\documentclass[10pt]{bmc_article}

\usepackage{cite} 
\usepackage{url}  
\usepackage{ifthen}  
\usepackage{multicol}   
\usepackage[utf8]{inputenc} 
\usepackage{amsmath}
\usepackage{amssymb}
\usepackage{amsthm}
\usepackage{todonotes}
\usepackage{lineno}
\usepackage{url}
\usepackage[colorlinks=true,final=true,linkcolor=black,urlcolor=black,citecolor=black,filecolor=black]{hyperref}
\usepackage{numprint}
\usepackage{comment}
\usepackage{SweaveSlides}
\usepackage[printonlyused]{acronym}

\usepackage{color}

\DeclareMathOperator{\Po}{Poisson} 
\DeclareMathOperator{\Be}{Be} 
\DeclareMathOperator{\Ga}{Ga} 
\DeclareMathOperator{\cn}{cn} 
\DeclareMathOperator{\ccn}{ccn} 
\DeclareMathOperator{\Var}{Var} 
\DeclareMathOperator{\E}{\mathsf{E}} 
\DeclareMathOperator{\p}{\mathsf{p}} 
 
\usepackage{graphicx}

\newboolean{publ}

\newenvironment{bmcformat}{\begin{raggedright}\baselineskip20pt\sloppy\setboolean{publ}{false}}{\end{raggedright}\baselineskip20pt\sloppy}

\begin{document}
\begin{bmcformat}


\title{BayMeth: Improved DNA methylation quantification for affinity
capture sequencing data using a flexible Bayesian approach }



\author{Andrea Riebler$^{1,2,3,}$\correspondingauthor
\email{
Andrea Riebler\correspondingauthor - andrea.riebler@math.ntnu.no},
Mirco Menigatti$^{4}$\email{Mirco Menigatti - menigatti@imcr.uzh.ch},
Jenny Z. Song$^5$\email{Jenny Z. Song - j.song@garvan.org.au},
Aaron L. Statham$^5$	\email{Aaron L. Statham- a.statham@garvan.org.au},
Clare Stirzaker$^{5,6}$\email{Clare Stirzaker - c.stirzaker@garvan.org.au},
Nadiya Mahmud$^{7}$\email{Nadiya Mahmud - n.mahmud@qmul.ac.uk},
Charles A. Mein$^{7}$\email{Charles A. Mein - c.a.mein@qmul.ac.uk},
Susan J. Clark$^{5,6}$\email{Susan J. Clark - s.clark@garvan.org.au}
\and Mark D.~Robinson$^{1,8,}$\correspondingauthor%
\email{Mark D.~Robinson\correspondingauthor - mark.robinson@imls.uzh.ch}%
      }
      

\address{%
    \iid(1)Institute of Molecular Life Sciences, University of Zurich,
	Winterthurerstrasse 190, CH-8057 Zurich, Switzerland\\
    \iid(2)Institute of Social- and Preventive Medicine,
	University of Zurich, Hirschengraben 84, CH-8001 Zurich, Switzerland\\
	\iid(3)Department of Mathematical Sciences, Norwegian University of Science and Technology,
	N-7491 Trondheim, Norway\\
	\iid(4)Institute of Molecular Cancer Research, University of Zurich,
	Winterthurerstrasse 190, CH-8057 Zurich, Switzerland\\
    \iid(5)Epigenetics Laboratory, Cancer Research Program,
	Garvan Institute of Medical Research, Sydney 2010, New South Wales, Australia\\
    \iid(6)St Vincent's Clinical School, University of NSW, Sydney 2052, NSW, Australia\\
	\iid(7)Genome Centre, Barts and the London, Queen Mary,
	University of London, Charterhouse Square, London EC1M 6BQ, United Kingdom\\
	\iid(8)SIB Swiss Institute of Bioinformatics,
	University of Zurich, Zurich, Switzerland
}%

\maketitle


\begin{abstract}
\paragraph*{Background:} \ac{DNAme} is a critical component of the
epigenetic regulatory machinery and aberrations in \ac{DNAme}
patterns occur in many diseases, such as cancer.
Mapping and understanding \ac{DNAme} profiles offers considerable
promise for reversing the aberrant states.  There are several
approaches to analyze \ac{DNAme}, which vary widely in cost,
resolution and coverage. Affinity capture and high-throughput
sequencing of methylated DNA  (e.g.~MeDIP-seq or MBD-seq) strike
a good balance between the high cost of \ac{WGBS} and the low
coverage of methylation arrays. However, existing methods cannot
adequately differentiate between hypomethylation patterns and low
capture efficiency and do not offer flexibility to integrate \ac{CNV}.
Furthermore, no uncertainty estimates are provided, which may prove 
useful for combining data from multiple protocols or propagating into 
downstream analysis.

\paragraph*{Results:} We propose an empirical Bayes framework that
uses a fully methylated (i.e.~SssI treated) control sample to
transform observed read densities into regional methylation estimates.
In our model, inefficient capture can readily be distinguished from
low methylation levels by means of larger posterior variances.
Furthermore, we can integrate \ac{CNV} by introducing a multiplicative
offset into our Poisson model framework.  Notably, our model offers
analytic expressions for the mean and variance of the methylation
level and thus is fast to compute.  Our algorithm outperforms
existing approaches in terms of bias, mean-squared error and coverage
probabilities as illustrated on multiple reference datasets. Although
our method provides advantages even without the SssI-control, considerable
improvement is achieved by its incorporation.

\paragraph*{Conclusions:} Our model not only improves on existing
methods, but allows explicit modeling of \ac{CNV}, context-specific
prior information and offers a computationally-efficient analytic
estimator.  Our method can be applied to methylated DNA affinity
enrichment assays (e.g MBD-seq, MeDIP-seq) and a software
implementation is freely available in the Bioconductor
\texttt{Repitools} package.
\end{abstract}

\acresetall

\ifthenelse{\boolean{publ}}{\begin{multicols}{2}}{}


\section*{Background}
\ac{DNAme} is a critical component in the regulation of gene
expression, is precisely controlled in development and is known to
be aberrantly distributed in many diseases, such as cancer and
diabetes \cite{jones.baylin2007, slomko.etal2012}.
In differentiated cells, \ac{DNAme} occurs primarily in the CpG
dinucleotide context.  For CpG-island-associated promoters,
increases in \ac{DNAme} (i.e. hypermethylation) induce repression of
transcription, while hypomethylated promoters are generally
transcriptionally active.
In cancer, tumor suppressor gene promoters are frequently hypermethylated,
and therefore silenced, while hypomethylation can activate oncogenes,
which collectively can drive disease progression
\cite{clark-melki-2002, esteller-2007}.  
The detection and profiling of such abnormalities across cell types
and patient cohorts is of great medical relevance, to both our basic
understanding of how the disease manifests but also for the opportunities of translating this knowledge to the clinic \cite{ziller-etal-2013}.  Epigenetic patterns
can be used as diagnostic markers, predictors of response to
chemotherapy and for understanding mechanisms of disease progression
\cite{ruike-etal-2010, stein2011, baylin-jones-2011, jones-2012}.
Acquired epigenetic changes are potentially reversible, which provides
important therapeutic opportunities; notably, the US Food and Drug
Administration has approved at least four epigenetic drugs and others
are in late-stage clinical trials \cite{baylin-jones-2011}. \\

Four classes of methods are available to profile \ac{DNAme}
genome-wide: chemical conversion, endonuclease digestion, direct
sequencing and affinity enrichment; combinations of techniques are
also in use (e.g. \ac{RRBS} \cite{gu-etal-2010}).  For recent
reviews of the available platforms, see
\cite{laird-2010,lister-ecker-2009, kerick.etal2012}.  Treatment of
DNA with \ac{BS} is the gold standard, giving a single-base readout
that preserves methylated cytosines while unmethylated cytosines
are converted to uracil \cite{clark-etal-1994}.  This approach can
be coupled with high-throughput sequencing, e.g.~\ac{WGBS},  or a
``genotyping'' microarray (e.g.~Illumina Human Methylation 450k
array \cite{bibikova-etal-2011}).  Because \ac{WGBS} is genome-wide,
it inefficiently reveals methylation status for low CpG-density
regions \cite{robinson-etal-2010-3} and is cost-limiting for larger
cohorts; however, recent statistical frameworks allow trading
coverage for replication \cite{hansen-etal-2011} and sequencing targeted regions may be a plausible way to increase efficiency \cite{lee-etal-2011, lee-etal-2012}.  Meanwhile, Illumina
arrays cover less than 2\% of genomic CpG sites and are only available
for profiling human DNA, while enzymatic digestion approaches are
limited by the location of specific sequences.  There is considerable
excitement surrounding third generation sequencing technologies that
directly infer methylation status, but these are not yet readily
available and generally offer lower throughput
\cite{clarke-etal-2009, flusberg-etal-2010}.

An attractive alternative that provides a good tradeoff between cost
and coverage, albeit at lower resolution, is affinity capture of
methylated DNA in combination with high-throughput sequencing
(e.g.~MeDIP-seq \cite{down-etal-2008, ruike-etal-2010}). Using
affinity capture with antibodies to 5-methylcytosine or methyl-CpG
binding domain-based (MBDCap) proteins, subpopulations of methylated
DNA are captured, prepared, sequenced and mapped to a reference
genome (see Laird et al. \cite{laird-2010}).
\AA berg et al.~\cite{aberg.etal2012} studied the use of MBD-seq for
methylome-wide association studies on 1500 case–control samples, and
proved the potential of MBD-seq as a cost-effective tool in large-scale
disease studies.
A recent comparative
study highlighted that affinity capture methods can uncover a
significantly larger fraction of differentially methylated regions
than the Illumina 450k array \cite{clark.etal2012}. With
appropriate normalization, the density of mapped reads can be
transformed to a quantitative readout of the regional methylation
level.  However, the capability of these procedures to interrogate a
given genomic region is largely related to CpG-density, which
influences the efficiency of capture and can differ from protocol to
protocol \cite{deMeyer-etal-2013, robinson-etal-2010-3, nair-etal-2011}.
Thus, statistical approaches are needed.

Several methods have been proposed to estimate \ac{DNAme} from
affinity-based \ac{DNAme} data. For example, \ac{MBD}-isolated
Genome Sequencing, a variant of \ac{MBD}-seq, assumed a constant
rate of reads genome-wide and used a single threshold
to binarize as methylated or not \cite{serre-etal-2010}.  
State-of-the-art methods, such as Batman \cite{down-etal-2008} and
MEDIPS \cite{chavez-etal-2010},  build a linear model relating read
density and CpG-density, which is then used to normalize the observed
read densities. For MeDIP-seq data, both algorithms showed similar
estimation performance \cite{chavez-etal-2010}, while MEDIPS is
considerably more time-efficient. Recently, a tool called BALM used
deep sequencing of \ac{MBD}-captured populations and a
bi-asymmetric-Laplace model to provide CpG-specific methylation
estimates \cite{lan-etal-2011}.  All methods, however, suffer from
the same limitations: i) low capture efficiency cannot easily be
distinguished from low methylation level; ii) other factors that
directly affect read density, such as \ac{CNV}, are not easily taken
into account.
For \ac{CNV} correction, a few possibilities have emerged, such as omitting
known regions of amplification \cite{ruike-etal-2010} or adjusting
read densities manually \cite{feber-etal-2011}; or, iii) adjust using read
density from an input sample \cite{lan-etal-2011}.  Very recently, a method based on combining profiles from MeDIP/MBD-seq and methylation-sensitive restriction enzyme sequencing (MRE-seq) on the same samples with a computational approach using conditional random fields appears promising \cite{stevens-etal-2013}.


We present a novel empirical Bayes model called BayMeth, based on the
Poisson distribution, that explicitly models (affinity capture) read
densities of a fully methylated control (e.g. SssI-treated DNA)
together with those from a sample of interest. Here, SssI data
provide the model an awareness of where in the genome the assay can
detect \ac{DNAme} and the model allows integration of \ac{CNV} and
potentially other estimable factors that affect read density. 
Notably, we derive an analytic expression for the mean methylation
level and also for the variance. Interval estimates, such as credible intervals, can be computed
using numerical integration of the analytical posterior marginal
distributions. Using \ac{MBD}-seq on \ac{IMR-90} DNA, where ``true''
methylation levels are available from \ac{WGBS}, we show favorable
performance against existing approaches in terms of bias, mean-squared
error, Spearman correlation and coverage probabilities. Notably, we show
that improved performance can even be observed when ignoring SssI-data.
Model-based SssI correction, however, does not only lead to better
performance, but allows, in addition, to compare more easily
data originating from different capture platforms by propagating the platform-specific uncertainty. In an application to  \ac{MBD}-seq
data on \ac{LNCaP} cells, we show that directly integrating \ac{CNV}
data provides additional performance gains. The performance on
historical data, where no matched SssI sample is available, is
demonstrated using data on embryonic stem cells, and colon tumor
and normal samples presented in \cite{bock-etal-2010}.
 
\section*{Results}

\subsection*{BayMeth: A Bayesian framework for translating read densities into methylation levels}
\ac{DNAme} data is obtained by \ac{MBD}-seq or a similar affinity
enrichment assay.  Let $y_{iS}$ and $y_{iC}$ denote the observed
number of (uniquely) mapped reads for genomic regions $i=1, \ldots, n$
for the sample of interest and the SssI control, respectively.
Throughout this paper, we use non-overlapping 100bp regions that
have at least 75\% mappable bases  
(see Methods).
Let
\begin{align}
  y_{iS}|\mu_i, \lambda_i &\sim \Po\left(f \times \frac{\cn_i}{\ccn}
	\times \mu_i \times \lambda_i\right) \text{, and} \label{eq:sI} \\
  y_{iC}|\lambda_i &\sim \Po(\lambda_i), \label{eq:SssI}
\end{align}
with $\lambda_i > 0$, $0 < \mu_i < 1$
Here, $\lambda_i$ denotes the region-specific read density at full
methylation, $\mu_i$ the regional methylation level and $f > 0$
represents the (effective) relative sequencing depth between libraries
(i.e.~a normalization offset). An approximately linear
relationship between the copy number state and \ac{MBD}-seq read
density was established \cite{robinson-etal-2012}. Hence, if needed,
we include a multiplicative offset
$\frac{\cn_i}{\ccn}$ into our model formulation, where $\cn_i$
denotes the copy number state at region $i$ and $\ccn$ is cell's
most prominent \ac{CNV} state (e.g.~two in normal cells).

\subsubsection*{Closed-form posterior methylation quantities}

In a Bayesian framework, prior distributions are assigned to all
parameters. The methylation level ($\mu_i$) has support from zero
to one. Potential priors include mixtures of beta distributions
or a Dirac-Beta-Dirac mixture. In the latter, a beta distribution 
is combined with point masses placed on zero and on one. 
The mixture weights can be either unknown or fixed. 
By default, BayMeth assumes a uniform prior distribution
(i.e., a beta distribution with both parameters set to 1)
for $\mu_i$.
For the region-specific density, we assume a gamma distribution,
i.e.~$\lambda_i \sim \Ga(\alpha, \beta)$ using
shape $\alpha > 0$ and rate $\beta > 0$ hyperparameters, which are
determined in a CpG-dependent manner (see next Section). To make
inferences for the regional methylation levels, $\mu_i$, we integrate
out $\lambda_i$ from the joint posterior distribution:
\begin{linenomath*}
 \begin{equation*}\label{eq:post-marg}
 \begin{split}
 	\p(\mu_i|y_{i1}, y_{i2}) & = \int_0^\infty
		\p(\lambda_i, \mu_i|y_{i1}, y_{i2}) d\lambda_i \\
 	& = \int_0^\infty \frac{\p(y_{i1}|\lambda_i,\mu_{i})
		\p(y_{i2}|\lambda_i)\p(\lambda_i) \p(\mu_i)}{\p(y_{i1},y_{i2})}  d\lambda_i.
  \end{split}
\end{equation*}
\end{linenomath*}
Notably, $\p(y_{i1},y_{i2})$ can be calculated analytically
\cite{fader-hardie-2000}, so that the marginal posterior distribution
\begin{linenomath}
\begin{equation}\label{eq:riebler-etal-postMarg}
	\p(\mu_i|y_{i1}, y_{i2})=  \frac{\mu_{i}^{y_{i1}}}{W}
	\left( 1- \frac{E (1-\mu_i)}{\beta + 1 + E}\right)^{-(\alpha+y_{i1}+y_{i2})},
\end{equation}
\end{linenomath}
is given in closed form with $E=f \cdot \frac{\cn_i}{\ccn}$ and
\begin{linenomath}
\begin{equation*}\label{eq:riebler-etal-W}
	W = \frac{1}{y_{i1} + 1} \times\,  _2F_1\left(y_{i1} + y_{i2} +
		\alpha, 1; y_{i1} + 2; \frac{E}{\beta + 1 + E}\right).
\end{equation*}
\end{linenomath}
where $_2F_1(\:)$ is the Gauss hypergeometric function \cite[page 558]{Abramowitz-stegun-1972}.
The posterior mean and the variance are analytically available
(see Additional file~1) and therefore efficient to compute;
credible intervals (quantile-based or HPD) can be computed
from Equation \eqref{eq:riebler-etal-postMarg}. Wald credible
intervals are computed on the logit scale, where $\text{logit}(\mu_i)=\log(\mu_i/(1-\mu_i)$), 
and then transformed back.
These intervals are based on assuming asymptotic normality of the logit methylation
estimate. The $95\%$ Wald interval on logit scale is computed by 
$\text{logit}(\hat\mu_i) \pm 1.96 \cdot \hat{\sigma}_i$, where
$\hat{\sigma}_i$ is the standard error estimate of $\text{logit}(\hat\mu_i)$.
For detailed statistical derivations, also including more general prior distributions
for $\mu_i$, we refer to the Additional file~1.

\subsubsection*{Empirical Bayes for prior hyperparameter specification} 
Our method takes advantage of the relationship between CpG-density
and read depth to formulate a CpG-density-dependent prior distribution
for $\lambda_i$ (and possibly unknown parameters in the prior distribution 
of $\mu_i$). Taking CpG-density into account the prior should
stabilize the methylation estimation procedure for low counts and
in the presence of sampling variability. All unknown hyperparameters 
are determined in a
CpG-density-dependent manner using empirical Bayes. For each genomic bin
of predetermined size, e.g., 100bp, we determine the weighted
number of CpG dinucleotides within an enlarged window, say 700bp,
around the center of the bin (see Methods and MEDME\cite{pelizzola-etal-2008}).
Each region is classified based on its CpG-density into one of
$K(=100)$ non-overlapping CpG-density intervals (see x-axis tick marks
in Figure~S1 of Additional file~2). 

For each class separately, we derive the values for the hyperparameters
under an empirical Bayes framework using maximum likelihood.
Both read depths, from the SssI control and the sample of
interest, are thereby taken into account, since $\lambda_i$ is a joint
parameter affecting both.  We end up with $K$ parameter sets. To illustrate 
the (known)
relationship between SssI read count and CpG density, we considered
only the SssI Poisson model (Equation~\eqref{eq:SssI}) and derived 
the prior predictive
distribution by integrating $\lambda_i$ out; this results in a
negative binomial distribution for each CpG class (see Figure 1
using SssI data from \cite{robinson-etal-2010-2} which are later used
in the analysis of the IMR-90 cell line).

\subsubsection*{SssI-free BayMeth}

Although we recommend collecting at least a single SssI sample under 
the same protocol as the data of interest, BayMeth can, in principle, be 
run without a SssI-control sample. The statistical framework then only 
involves the Poisson model for the sample of interest (Equation \eqref{eq:sI}) 
and no longer borrows 
strength from the information included in the SssI-control sample 
(Equation \eqref{eq:SssI}). The same model is used in the analysis of underreported 
count data in economics \cite{schmittlein1985, winkelmann-1996, fader-hardie-2000}, 
where it is assumed that the number of registered purchase events underreports
the actual purchase rate. 
According to Fader and Hardie 
\cite{fader-hardie-2000} the parameters $\lambda_i$ and 
$\mu_i$ are identifiable assuming that the gamma and beta prior distributions
are able to capture unobserved heterogeneity in the read density rate and 
the methylation level. As in the framework with SssI data, 
parameters for the gamma prior distributions of the 
region-specific read density $\bf{\lambda}$ can still be determined in 
dependence on the CpG density, however, no information can be borrowed 
from the fully methylated control.
Furthermore, the determination of the normalizing offset $f$ gets
more involved. Interpretation moves from the (effective)
relative sequencing depth between libraries to the number of bins
potentially ``under risk'' to be methylated in the sample of interest.
Here, we fix $f$ at the $99^{th}$ quantile of the number of reads.
The results for the posterior mean and variance of the methylation
level change accordingly (see Additional file~1).

\subsection*{Analysis of affinity capture methylation data with a matched SssI sample}

In the following, we apply BayMeth to affinity capture methylation data
where we collected a SssI-control sample under the same conditions
(e.g. same elution protocol) used for the samples of interest. Hence, both
data components are matched.

\subsubsection*{BayMeth improves estimation and provides realistic variability estimates}

To take advantage of the Lister et al. \cite{lister-etal-2009}
single-base-resolution high-coverage methylome obtained by \ac{WGBS},
we generated IMR-90 \ac{MBD}-seq data under the same protocol as our previously published SssI MBD-seq dataset \cite{robinson-etal-2010-2} i.e.~using a single fraction with high salt
elution buffer (MethylMiner\texttrademark). We applied
BayMeth to chromosome 7 consisting of $\numprint{1588214}$ non-overlapping bins
of width 100bp. Only bins with at least 75\% mappable bases were included, which
leads to the analysis of $\numprint{1221753}$ ($\sim 77\%$) bins.
We run BayMeth in two configurations: 1) incorporating SssI information and
assuming a uniform prior between zero and one for the methylation parameter;
2) ignoring SssI information and assuming a Dirac-Beta-Dirac 
mixture prior distribution for the methylation parameter. 
That means we set a point mass on zero
and on one, giving each a prior weight of $10\%$. The parameters of the central
beta component are thereby assumed to be unknown.
The normalizing offset $f=0.581$ for configuration 1) is found based on 
calculating a scaling factor
between highly methylated regions in \ac{IMR-90} relative to the
SssI control (see Methods and Figure~S2 of Additional file~2). 
The prior parameters for the gamma distributions, and the parameters of the
beta distribution in configuration 2), are determined by empirical
Bayes, as discussed above (see also further details in Methods). 
We compared the results of BayMeth, both ignoring and taking advantage of 
the SssI control, to those obtained by Batman \cite{down-etal-2008}, MEDIPS
\cite{chavez-etal-2010} and BALM \cite{lan-etal-2011}.
To provide plausible uncertainty estimates with Batman, we increased the default number of
generated samples from $100$ to $500$. The \ac{WGBS} data,
here considered to be the ``truth'' (at suitable depth), and the CpG-specific BALM
methylation estimates are collapsed into 100bp bin estimates
(see Methods) to match the estimates from MEDIPS, Batman and our
approach.  For about $53\%$ ($\numprint{645451}$) of the analyzed
bins, no WGBS data are available (largely due to lack of CpG sites).
For $17259$ bins, no methylation estimates are provided by Batman, so that
in total, algorithm comparisons are conducted on the remaining
$559043$ bins.


The behavior of BayMeth (including SssI-information) and Batman is illustrated using an example
region of chromosome~7 (see Figure~2A). \ac{WGBS} levels, CpG-density
and read counts per 100bp region of \ac{MBD}-seq SssI and \ac{IMR-90}
sample are shown.  As expected, the number of reads in the SssI
control is related to the CpG-density, whereas the read density in
(\ac{MBD}-seq) IMR-90 is modulated by both the region-specific density
and the \ac{DNAme} level. Regions lacking both \ac{IMR-90} and SssI
reads suggest inefficient \ac{MBD}-based affinity capture
(e.g. region `a'). Figure~2B shows posterior samples from Batman
and inferred posterior distributions from BayMeth.  For region `a',
Batman's posterior samples are concentrated between $0.7$ and $1$
(mean equal to $0.85$).  In contrast, BayMeth returns a mean
methylation level of $0.49$ together with a large $95\%$ \ac{HPD}
interval $(0, 0.94)$, reflecting the uncertainty from having no SssI
reads sampled. The credible interval covers nearly the entire interval,
reflecting that no reliable estimate can be made for this bin due to
inefficient capture.  For regions with no \ac{IMR-90} reads
but efficient capture (e.g.~region `b'), both BayMeth and Batman
provide sensible posterior marginal distributions and low \ac{DNAme}
estimates. If there are a small number of reads for \ac{IMR-90} with
efficient capture (e.g.~region `c'), the BayMeth posterior marginal
is more disperse than Batman's, while both are close to zero. Region
`d' has a high number of reads for both samples and a true methylation
level around $0.95$. This level is covered by the $95\%$ \ac{HPD}
region of BayMeth, while it lies outside of the density mass obtained
by Batman overestimating this region.

Table~1 summarizes the estimation performance for chromosome 7 by
means of mean bias (difference between the posterior mean $\hat{\mu_i}$
and the true value $\mu_i$),  MSE (mean of squared differences),
Spearman correlation and compares it to a BayMeth version ignoring
SssI-information, Batman, MEDIPS and BALM.
To account for uncertainty
present in the \ac{WGBS} estimates, we applied a threshold on the
depth; we assess the performance using bins with at least $33$ \ac{WGBS}
reads (unmethylated and methylated) corresponding to the $25\%$
quantile of depth in the truth, which results in $414352$ bins.
Results are stratified into five groups according to depth in the
SssI control, which should represent a surrogate of the capture
efficiency. The first group $[0,4]$ encompasses primarily low-CpG
regions that are not well captured in \ac{MBD} experiments, while
the high $(27, 168]$ group represents primarily CpG island regions.
On average, Batman tends to overestimate \ac{DNAme} while MEDIPS and
BALM tend to underestimate. BayMeth, in contrast, is almost unbiased.
The smaller bias in the point estimates obtained by BayMeth is also
reflected in the MSE. For all methods, the MSE decreases with higher
SssI depth, as expected due to the efficiency of capture. For all
depth groups, BayMeth has the highest correlation with the \ac{WGBS}
estimates, which increases with higher SssI depth. The SssI-free
version of BayMeth performs comparable to the other approaches,
with slightly smaller bias and MSE, however, smaller correlation for
bins with low SssI depth.
A smoothed density representation of regional methylation estimates
for the highest SssI depth group, namely $(27, 168]$, plotted for
all methods against the ``true'' \ac{WGBS} methylation levels
are shown in Figure~3; overall, BayMeth provides the most accurate
point estimates. The overestimation
of Batman and underestimation of MEDIPS and BALM is obvious, while
the BayMeth errors vary almost symmetrically. Comparing BALM CpG-wise
to \ac{WGBS} lead to similar conclusions as in the bin-specific
setting (results not shown). Notably, the pattern of the SssI-free BayMeth estimation (i.e. overestimation) is similar to Batman, which may be expected given that no information is drawn from the SssI sample.

To assess calibration, we computed coverage probabilities
(frequency that the ``true'' methylation value is captured within a credible interval).
Stratified by the ``true'' \ac{WGBS} methylation level, Figure~4 shows
coverage probabilities at  $95\%$ level for regions deemed
to be inside or outside a CpG island (Figure~S1 of Additional file~2).
\ac{HPD} intervals, quantile-based and Wald-based credible intervals (CI) are
computed for BayMeth while only quantile-based CIs are available for
Batman; coverage probabilities are not possible from MEDIPS and BALM
output. As mentioned, Batman tends to underestimate the variance,
resulting in lower coverage probabilities of the \ac{WGBS} values;
in contrast, BayMeth's coverage probabilities are much closer to the
nominal levels and seem to be stable across the stratification. For the
SssI-free BayMeth quantile-based  credible intervals are computed which
are generally better than those provided by Batman (see Table~1 and Figure~4),
indicating a more realistic methylation estimation. Table~S1 in Additional file~2)
shows for the same stratification the mean bias for BayMeth, Batman, MEDIPS
and BALM. While the latter two provide low mean bias for bins where the truth
lies within $[0,0.2]$, Batman performs best for highly methylated bins. 
BayMeth shows good performance for bins where the true methylation level is 
intermediate or high. Similar to Batman reasonable estimates are obtained over 
the whole range of methylation states when considering bins in CpG islands. 
Interpreting the mean bias the uncertainty around the obtained
estimates should taken into account and hence the results should be set 
into context with Figure~4. Combining bias and calibration BayMeth shows 
good performance and seems to improve on existing approaches.

\subsubsection*{CNV-aware BayMeth improves \ac{DNAme} estimation for prostate cancer cells}

In the following, we illustrate the benefits of directly integrating
\ac{CNV} information into a cancer \ac{MBD}-seq dataset. We apply
our methodology to the autosomes of the \ac{LNCaP} cell line.  To
motivate such an adjustment, Figure~5 shows the estimated copy number
across chromosome~13 (with many non-neutral regions), together with
tiled \ac{MBD}-seq read counts. Copy number estimates were derived
using the PICNIC algorithm on Affymetrix genotyping arrays
(see Methods). Although read densities at a specific genomic
region (again, 100bp non-overlapping bins) are influenced by a
combination of effects (e.g. \ac{DNAme}, CpG-density), a relationship
between \ac{CNV} and number of reads is clearly visible. In
particular, a difference in read counts between regions with four
copies and those with smaller copy numbers is apparent. We adjust
for this bias through a multiplicative offset $\frac{\cn_i}{\ccn}$,
where the prominent state is four copies, i.e.,~$\ccn=4$ in Equation
\eqref{eq:sI} (see Figure~S3 of Additional file~2); note, this also assumes the 
SssI sample originates for a ``normal'' copy genome. In addition, regions
from this state ($\cn_i=4$) are used to determine the normalizing
offset $f$ (here, estimated to be $0.712$). The read depth stratified
by copy number state together with mean and median estimates is shown
in Figure~S4 of Additional file~2. In particular, for the three most frequent
\ac{CNV} states (2--4), read densities scale approximately linearly
(with a slope of 1) with \ac{CNV}, which justifies the structure of
our multiplicative offset; copy-number offsets are given in Table~2.
Figure~6 shows the bias of \ac{DNAme} point estimates of the
different methods by integer \ac{CNV} state (2--5); here, we used
the Illumina Human Methylation 450k array as the ``true'' methylation
(see Methods), since methylation status should be unaffected by
\ac{CNV} \cite{houseman-etal-2009}.  Because \ac{CNV} only affects
\ac{MBD} capture for methylated regions, we restrict this comparison
to bins where the true methylation state is larger than $0.5$ and we
applied a threshold of $13$ (median after excluding bins with a low
depth of $[0,4]$) to the number of reads in the SssI-control to
select for regions where \ac{MBD}-seq has good performance. Similar
to the \ac{IMR-90} data, MEDIPS and BALM tend to underestimate,
while Batman tends to overestimate. For BayMeth we show four different
approaches neglecting CNV and/or SssI-information. As previously, 
we use a uniform prior for the methylation level when taking advantage
of the SssI sample, and a Dirac-Beta-Dirac mixture with fixed weights (0.1, 0.8, 0.1)
but unknown beta parameters in the SssI-free case. In the SssI-free
version the normalization offset $f$ is determined as the $99\%$
percentile of the number or reads for the sample of interest having
copy number state $4$, while the reads of all bins are used when
neglecting additionally the CNV information. Without the additional
multiplicative offset (i.e. without $\cn_i \equiv \ccn$) to account
for CNV, BayMeth provides biased estimates, predictably by
\ac{CNV} state.  After including the copy-number-specific offset,
these copy number specific biases almost disappear, whereby the
SssI-free version still shows slight overestimation.
A smoothed scatterplot illustrating the benefits of including
the copy-number-specific offset is shown in Figure~7 for copy
number state two. In particular, bins that have been falsely
underestimated (due to two copies instead of four) are corrected (see top-right panel).
Due to overestimation in the SssI-free version (bottom-left) the methylation
estimates for copy number state two do not show such a strong bias.
Adjusting for CNV in this case slightly increases the bias (bottom-right).
Table~3 shows mean bias, MSE and Spearman correlation for the
different approaches stratified by copy number state. In all
measures, the \ac{CNV}-aware standard (including SssI) version
of BayMeth performs best. While the differences in the correlation
estimates are small, clear advantages can be seen in terms of bias
and MSE when compared to Batman, MEDIPS or BALM. In contrast to
the other approaches, the bias/MSE performance estimates stay almost constant
over the different copy number states and are close to zero.

\subsubsection*{Improved correlation across methylation kits on IMR-90 DNA}

One potential advantage of the proposed model-based SssI correction
is that data originating from different capture platforms can be more
easily compared.  In this situation, propagation of the uncertainty
becomes important, since methods to capture methylated DNA have
different CpG-dependent affinity and therefore different estimation precision.  To demonstrate this, we captured
methylated DNA from IMR-90 and SssI DNA using six approaches: low,
medium and high salt elutions from MethylCap Kit\texttrademark,
500nM and 1000nM salt fractions from MethylMiner\texttrademark and
MeDIP. Autosomes were analyzed
with BayMeth using specific SssI data for each kit. The derived
MA-plots together with the obtained normalizing offsets $f$ 
for each sample are shown in Figure~S5 of Additional file~2. 
Unusual high counts were excluded in the derivation of the prior 
parameters \cite{pickrell.etal2011}, but methylation
estimates are derived for all bins. For bins where
the estimated credible interval width (HPD) is smaller than $0.4$,
Figure~8 compares the unnormalized read density between the six kits
(upper triangular panel), and the obtained methylation estimates (lower 
triangular panels). Clearly, capture affinities across
the six kits vary drastically, while the SssI-based correction
makes the comparison much clearer.  In addition,
the SssI data from this collection of platforms may be useful for the community to pair
with their in-house data, assuming similar procedures have been
followed (see Discussion), in order to benefit from the use of
SssI-based read density correction from BayMeth.

\subsection*{Analysis of affinity capture methylation data WITHOUT a matched SssI sample}

Next, we applied (default) BayMeth to the MethylCap sequencing data of \cite{bock-etal-2010}, provided at
\url{http://www.broadinstitute.org/labs/meissner/mirror/papers/meth-benchmark/index.html}, and 
denoted as the ``Bock'' data below. 
Absolute read densities are available for four samples: HUES6 ES cell line, 
HUES8 ES cell line, colon tumor tissue, colon normal tissue 
(same donor as for colon tumor tissue), and given for (non-overlapping) 50bp bins. 
There is no matched SssI sample available for these data. To take 
advantage of BayMeth in analyzing these data,  we use a non-matching 
SssI sample, but one chosen to be maximally compatible to the preparation 
conditions of Bock data\cite{bock-etal-2010} 
(i.e. MethylCap at low salt concentration: 200mM NaCl). 
Regions from the data available were lifted over to hg19 coordinates 
(see Additional file~3 for details). 
Although there are still slight differences in the preparation 
of the samples of interest and the SssI sample, which arise from 
different used read length (36bp versus 75bp, respectively) and 
different read extensions (300bp versus 150bp, respectively) before 
calculating the read frequencies, we regard the SssI sample as a 
reasonably suitable control for running BayMeth.
We analyzed all autosomes after removing bins that have no read 
depth in any of the four samples, leading to $\numprint{42955764}$
bins. As in the previous analyses, 
we restrict our focus on bins that have at least 75\% mappable
bases, which means $37013409$. That is $86\%$ of all bins. 
A detailed description of all data preparation steps and the 
data analysis using BayMeth based on the R-package {\tt  Repitools}
is given in Additional file~3.
To assess the methylation estimates obtained with BayMeth, 
we compare them to RRBS data available from the Bock study\cite{bock-etal-2010}. 
As in the methylation kit analysis we masked unusual high counts in the derivation of the prior parameters 
as they sometimes cause problems in the numerical optimization routine, however,
methylation estimates are derived for more than $99.5\%$ of these masked bins. 
Interestingly, several high count regions are explained by 
unannotated high copy number regions, see Pickrell et al. \cite{pickrell.etal2011}.

Methylation estimates are obtained for about $37$ million bins each of 
width 50bp, while RRBS estimates are only 
available for approximately $4\%$ of these bins. We assess the performance of BayMeth 
using bins where the depth in the RRBS is larger than $20$.
Furthermore, we focus on bins where we believe in the SssI control, that means 
where the read depth is at least $10$. 
Figure~9 shows regional methylation estimates obtained by BayMeth compared to 
RRBS derived methylation levels for all four samples of interest where the
corresponding posterior standard deviation is smaller than $0.15$. 
In particular, low methylation levels are well predicted for all samples. 
While high methylation levels are partly underestimated 
by BayMeth for the human embryonic stem cell 
line HUES8, estimates for HUES 6, colon tumor and color normal 
tissue reproduce the true methylation for all levels. Although, in the latter
two slight overestimation is visible. This is partly caused by bins for
which we observe low read depth in SssI, but extreme depth in the
sample of interest. BayMeth predicts these bins comprehensibly with high precision 
(low standard deviation), which may, however, not coincide with the RRBS estimates.
Figure 10 shows regional posterior variances obtained by BayMeth compared
to SssI depth for bins where the depth in the RRBS is larger than $20$.
The posterior variance decreases with increasing SssI 
depth. However, the range of posterior variances for low SssI depth is
large. The red square contains the bins illustrated in Figure 9.
Of note, comparisons to other methods are not possible for the Bock data, since we do 
not have access to the raw reads.

\section*{Discussion and conclusions}

DNA methylation plays a crucial role in various biological processes
and is known to be aberrant in several human diseases, such as cancer.
There are now a multitude of methylation profiling platforms, each
with inherent advantages and disadvantages. Bisulfite-based
approaches are considered the gold standard since they allow
quantification at single-base resolution. However, applied
genome-wide, this technique can be inefficient and expensive,
in terms of CpGs covered per read or base sequenced
\cite{ziller-etal-2013,robinson-etal-2010-3}. On the other hand, affinity capture
based approaches, such as \ac{MBD} or MeDIP, combined with sequencing
seem to provide a good compromise between cost and coverage, albeit
at lower resolution.  Thus, we consider MBD-seq and its variants to
be an attractive alternative and have developed an efficient data
analytic approach to facilitate their use. In addition, MBD-seq has
recently been demonstrated using only hundreds of nanograms of
starting DNA, thus making it applicable to a wider range of studies,
such as clinical samples \cite{taiwo-etal-2012}.
%


The key to our proposed method is the use of methylated DNA captured
from a fully methylated SssI control; to facilitate accurate
transformation of read counts into methylation, we recommend such a
sample should be collected under the same conditions
(e.g. same elution) used for the samples of interest. In our analyses,
we used commercially available SssI-treated DNA
\cite{robinson-etal-2010-2, nair-etal-2011} for the \ac{MBD}-seq
experiments and verified with the 450k platform that the overwhelming
majority of CpG sites are indeed methylated
(see Figure~S6 of Additional file~2); similarly, such a sample can be
constructed directly and inexpensively \cite{carvalho-etal-2012}.
Our proposed method, BayMeth, is a flexible empirical Bayes approach
that transforms read densities into regional methylation estimates.
Our model is based on a Poisson distribution and takes advantage of
SssI control data in two ways: i) we model SssI data jointly with
data from a sample of interest to preserve the linearity of the
methylation estimation; ii) we explicitly get information about the
region-specific read density as a function of CpG-density.  Our method
is similar in principle to  MEDME, which was applied to fully
methylated MeDIP microarray intensities \cite{pelizzola-etal-2008}.
However, our approach necessarily modifies assumptions for count data
(i.e. read densities versus probe intensities) and is effectively a
{\em moderation} between the global fit that MEDME implements and a
region-specific correction.  We showed that BayMeth delivers improved
performance against state-of-the-art techniques for \ac{MBD}-seq data,
using multiple datasets where independent ``true'' methylation levels
are available from \ac{WGBS} or bisulphite-based methylation arrays.
In general, MEDIPS and BALM underestimate the methylation levels and
do not offer variability estimates.  Batman performs reasonably well,
but our results suggest that variability estimates are generally
underestimated and the method is very computationally demanding.
Our model performs best in point estimation and is the only method
that provides reasonable interval estimates.  Notably, BayMeth offers
analytic expressions for the posterior marginal distribution and the
posterior  mean and variance, avoiding computationally-expensive
sampling algorithms. Furthermore, we can explicitly integrate existing
\ac{CNV} data, which offers improvement when applied to cancer
datasets.  \ac{CNV} adjustments may be possible with existing
approaches such as Batman or MEDIPS, based on ad-hoc transformations
of the read counts (e.g. see \cite{feber-etal-2011}), but are not
included within the model formulation.  In contrast, our model
preserves the count nature of the data.  To adjust the modeled
mean for effects arising due to library composition or \ac{CNV},
we introduced a normalization offset. This strategy is quite general
and could be extended beyond composition and \ac{CNV}
(e.g. see \cite{hansen-etal-2012,robinson-etal-2012}).


%

 



 
A conceptual similar Bayesian hierarchical model, which involves
MCMC sampling, has been proposed in the context of Methyl-Seq
experiments, where methylation levels are derived based on enzymatic
digestion using two enzymes \cite{wu-etal-2011}; a separate Poisson
model is assumed for the tag counts of each enzyme. The models are
linked through a shared parameter while one Poisson model contains a
methylation level parameter $\mu$, assumed to be uniformly
distributed {\em a priori}; our model may have applications
in this domain.  In the applications presented here, a uniform
prior distribution for the methylation level was observed to perform
best when taking SssI information into account, while a mixture
prior of a point mass at zero and at one, combined with a beta distribution,
performed best when ignoring SssI information.
The analytical expressions for the mean, variance and posterior marginal
distribution are also available when using a mixture of beta distributions (see Methods). 
Therefore, context-specific information, such as CpG-density or the position
relative to transcriptional features, could be incorporated in the prior
distribution for the methylation level. We have tried various weighted
mixtures of two or three beta distributions that build in contextual
information; however, these did not outperform the uniform prior when borrowing
strength from the SssI sample. The reason lies probably in the fact
that there is only one data point for each methylation parameter. Hence, using an
informative prior distribution for the methylation level, it is very difficult
for the data to overcome this prior guess.  

It is well known that methylation levels are dependent within
neighboring regions. Thus, a potential improvement may involve
modeling correlation between neighboring genomic bins. One approach
might be Gaussian Markov random fields \cite{GMRFbook}; however,
the analytical summaries are lost, so the the gain in performance
may not justify the more complex model and associated computational cost.

BayMeth may also be regarded as pre-processing step for performing
differential methylation analysis. The uncertainty in methylation estimates
obtained by BayMeth could be propagated to downstream analysis, which
may lead to improved inference on differential methylation.
  
\section*{Methods}\label{sec:methods}

\subsection*{\ac{MBD}-seq on IMR-90, LNCaP and SssI DNA}

We used LNCaP and SssI \ac{MBD}-seq data and Affymetrix genotyping
array data (LNCaP only) from Robinson et al.
\cite{robinson-etal-2010-2}. The data can be found at
\url{http://www.ncbi.nlm.nih.gov/geo} under accession number GSE24546.
Similarly, IMR-90 \ac{MBD}-seq is available from GSE38679.
Details of the DNA capture, preparation and sequencing can be
found in Robinson et al. \cite{robinson-etal-2010-2, nair-etal-2011}.

\subsection*{\ac{MBD}-seq and Methylated DNA Immunoprecipitation sequencing (MeDIP-seq)
for comparing data from different methylation kits}

For comparing data obtained by different methylation kits we captured 
methylated DNA from IMR-90 and SssI DNA as follows.
Genomic DNA was sheared to 150-200bp using the Covaris S220 sonicator.
MBD capture was performed using the MethylMiner Methylated DNA Enrichment Kit (Invitrogen) 
and the MethylCap Kit (Diagenode) following the manufactures recommended protocols. 
The bound fractions were eluted at 500mM and 1M NaCl for MethylMiner and with 
buffers at different salt concentrations (low, medium, high) for the MethylCap. 
Sequencing libraries were prepared with the SOLiD Fragment Library Construction Kit (Applied Biosystems)
MeDIP-seq methylation immunocapture and library preparation were performed using 
the MeDIP Kit (Active Motif) following the manufactures recommended protocol.

\subsection*{Calculation of CpG-density}

CpG-density is defined to be a weighted count of CpG sites in a
predefined region. We used the function \texttt{cpgDensityCalc}
provided by the R-package {\tt Repitools} \cite{statham-etal-2010} to get
bin-specific CpG-density estimates using a linear weighting
function and a window size of 700bp (since we expect fragments
around 300bp).

\subsection*{Calculation of mappability}

Using Bowtie, all possible 36bp reads of the genome were mapped back
against the hg18 reference, with no mismatches. At each base, a read
can either unambiguously map or not. A mappability estimate gives the
proportion of reads that can be mapped to a specific regions. To get
bin-specific mappability estimates we used the function
\texttt{mappabilityCalc} in the {\tt Repitools} package
\cite{statham-etal-2010}. In our analysis, a window of 500bp was used
(250bp upstream and downstream from the center from each 100bp bin)
and the percentage of mappable bases was computed.

For the methylation kits analysis we used mappability estimates for hg19 provided
by ENCODE on \url{http://hgdownload.cse.ucsc.edu/goldenPath/hg19/encodeDCC/wgEncodeMapability/
wgEncodeCrgMapabilityAlign100mer.bigWig},
from which we derived a weighted mean based on the window size.
Analogously, we used \url{http://hgdownload.cse.ucsc.edu/goldenPath/
hg19/encodeDCC/wgEncodeMapability/wgEncodeCrgMapabilityAlign50mer.bigWig} for
the Bock data analysis.

  \subsection*{Derivation of region-specific methylation estimates from WGBS}
In the Lister et al. \ac{IMR-90} WGBS data, the number of reads
$r_j^+$ and $r_j^-$ overlaying a cytosine $j$ in the positive $(+)$
and negative strand $(-)$, respectively, is available. Furthermore,
the number of these reads, $m_j^+$ and $m_j^-$, that contain a
methylated cytosine, is known. A single-base methylation estimate
can be obtained by $(m_j^+ + m_j^-)/(r_j^+ + r_j^-)$. To get a
bin-specific methylation estimate all cytosines lying within a bin
of interest $\mathcal{B}$ are taken into account:
\begin{linenomath*}
\begin{equation*}
	\mu_{\mathcal{B}} = \frac{\sum_{j \in \mathcal{B}}(m_j^+ + m_j^-)}{\sum_{j \in \mathcal{B}} (r_j^+ + r_j^-)}.
\end{equation*}
\end{linenomath*}
Here, $\sum_{j \in \mathcal{B}} (r_j^+ + r_j^-)$ is termed depth.

  \subsection*{Derivation of region-specific methylation estimates from 450K arrays}
First, the Illumina HumanMethylation450 methylation array was
preprocessed using default parameter of the \texttt{minfi} package
\cite{hansen-aryee}; for each sample, a vector of {\em beta values},
one for each targeted CpG site representing methylation estimates
are produced. To obtain (100bp) bin-specific methylation profiles,
we averaged beta values from all CpG sites within 100bp (upstream and
downstream; total window of 200bp) from the center of our 100bp bins.

\subsection*{Derivation of region-specific methylation estimates from RRBS data}

For the Bock data analysis, information on RRBS data were available from 
\url{http://www.broadinstitute.org/labs/meissner/mirror/papers/meth-benchmark/index.html}, 
which we considered as gold standard. Both, the number of reads that overlay a cytosine (T) 
and the number of cytosines that stay a cytosine (M), i.e. are methylated, 
are given. Note, that for one CpG site there is only information from one strand available.
To get smooth methylation estimates, we used 150bp bins (overlapping by 100bp). 
The methylation level for one 150bp bin $i$ was derived as:
\begin{equation*}
	m_i= \frac{\sum M_{\in i}}{\sum T_{\in i}}.
\end{equation*}
That means using information for all CpG sites that fall into bin $i$.

\subsection*{Determining the normalizing offset}

The composition of a library influences the resulting read densities
\cite{robinson-oshlack-2010}.  For example, the SssI control
represents a more diverse set of DNA fragments since it captures the
vast majority of CpG rich regions in the genome.  Therefore, if the
total sequencing depth were to be fixed, one would expect a relative
undersampling of regions in SssI, compared to a sample of interest
that is presumably largely unmethylated. To adjust the modeled mean
(in the Poisson model) for these composition effects, we estimate a
normalizing factor $f$ that accounts simultaneously for overall
sequencing depth and composition. 
Figure~S2 of Additional file~2 shows an $M$ (log-ratio) versus $A$
(average-log-count) plot at 50,000 randomly chosen (100bp) bins for
\ac{IMR-90} compared to the fully methylated control. A clear offset
from zero is visible, where the distribution of $M$ values is skewed
in the negative direction. The normalization offset $f$ is estimated
as $f=2^{median(M_{A > q})}$, with $q$ corresponding to a high
(here, $0.998$; more than $35000$ points in both applications)
quantile of $A$.
In cancer samples where \ac{CNV} are common, the normalization
factor $f$ is calculated from bins that originate from the most
prominent copy number state (e.g., $\ccn=4$ in \ac{LNCaP} cells).

\subsection*{Estimation of copy number}
Copy number estimates were estimated from Affymetrix SNP6.0 genotyping
array data by PICNIC \cite{greenman-etal-2010}, using default
parameters. PICNIC is an algorithm based on a hidden Markov model
to produce absolute allelic copy number segmentation.

\subsection*{Details on BayMeth methodology}

The methodology of BayMeth is roughly divided into two steps:
\begin{enumerate}
\item An empirical Bayes procedure to derive sensible prior parameters for all
parameters in the model.
\item The analytical derivation of the posterior marginal distribution, 
posterior expectation and variance for the methylation levels. Credible intervals
are derived numerically from the posterior marginal distribution.
\end{enumerate}
The details for both steps are provided in Additional file~1. 
In practice BayMeth can be used almost as a black box within the
Bioconductor package {\tt Repitools} \cite{statham-etal-2010}.

\subsection*{Details on Batman specifications}

Batman is an algorithm implemented in JAVA and run from the command
prompt. The original Batman can be downloaded from
\url{http://td-blade.gurdon.cam.ac.uk/software/batman/}; we used an
unreleased version ``20090617'' received directly from Thomas Down
that had MeDIP-seq-specific enhancements; the commands used to run
Batman are given at the Supplementary website.

\subsection*{Details on MEDIPS specifications}

We used the R-Bioconductor MEDIPS version 1.4.0 and followed the
available tutorial (\url{medips.molgen.mpg.de/MEDIPS.1.0.0/MEDIPS.pdf}
from October 18, 2010); the detailed command sequence is given at
the Supplementary website.
MEDIPS returns methylation estimates in the range from zero to $1000$,
which we rescaled to the interval $[0,1]$. In our comparison, we used
the absolute methylation score (AMS) provided by MEDIPS.

%
%

\subsection*{Details on BALM specifications}

BALM is an algorithm implemented in C and C++ and run from the
command prompt. The original BALM can be downloaded from
\url{http://motif.bmi.ohio-state.edu/BALM/}; We used the version 1.01.
The detailed command sequence is given at the Supplementary website.
BALM returns a vector of methylation estimates, one for each targeted
CpG site. To obtain (100bp) bin-specific methylation profiles,
we averaged the methylation estimates from all CpG sites within 100bp
(upstream and downstream; total window of 200bp) from the center of
our 100bp bins. For the IMR-90 data set BALM was run without an input
control. To assess the effect of the missing input control, we run
BALM using a sample from a normal human prostate epithelial cell line
(PrEC) as input control which lead to almost identical performance
results.

\subsection*{Software}

BayMeth is fully integrated into the R-package {\tt Repitools} and available
from the Bioconductor web page \url{http://www.bioconductor.org/packages/release/bioc/html/Repitools.html}. Data (semi-processed), R Code for all figures and analyses are
provided on \url{http://imlspenticton.uzh.ch/robinson_lab/BayMeth/index.html}. 

\section*{Acronyms}
\begin{acronym}[DNAme]
\acro{DNAme}{DNA methylation}
 \acro{RRBS}{reduced representation bisulphite sequencing}
 \acro{CNV}{copy number variation}
 \acro{WGBS}{whole genome bisulphite sequencing}
 \acro{BS}{sodium bisulphite}
 \acro{MBD}{methyl binding domain}
 \acro{IMR-90}{human lung fibroblast}
 \acro{LNCaP}{human prostate carcinoma}
 \acro{DSS}{David-Sebastiani score}
 \acro{HPD}{highest posterior density}
 \end{acronym}

\section*{Authors contributions}
    The statistical approach was conceived and developed by AR and
MDR, with biological and technical insight from ALS, CS, MM and SJC.
Implementation and data analyses were done by AR with contributions
from MDR. Data was collected by JZS, ALS, NM, CM and MM.  AR and MDR
wrote the manuscript with input from all authors. All authors read
and approved the final manuscript.

\section*{Additional Files}

\subsection*{Additional file 1 --- Statistical details of BayMeth}

This document describes all details of the BayMeth methodology. Two different 
prior distributions for the methylation level are presented, namely, a mixture 
of beta distributions, and a mixture of a point mass at zero, a beta distribution 
and a point mass at one (Dirac-beta-Dirac prior). An empirical Bayes procedure 
is outlined to derive prior parameters. Analytical derivation of the posterior 
marginal distribution and parameter estimation is described for both priors.
We outline the derivations for the standard BayMeth version, i.e.~taking advantage of 
SssI information, and for the SssI-free version.

\subsection*{Additional file 2 --- Supplementary figures and tables}

This document contains six supplementary figures and one supplementary table.
Detailed descriptions are provided within the file.

\subsection*{Additional file 3 --- BayMeth analysis of ``Bock'' data}

This document outlines all data preparation steps performed and presents
detailed R-code for the BayMeth analysis conducted 
using the Bioconductor package {\tt Repitools}.

\section*{Acknowledgements}
\ifthenelse{\boolean{publ}}{\small}{}
AR gratefully acknowledges funding of the ``Forschungskredit'' and
the URPP (University Research Priority Program in
Systems Biology/Functional Genomics) grant of the University of
Zurich. MM  acknowledges grant funding from the Swiss Cancer League (KFS-02739-02-2011). 
SJC gratefully acknowledges grant funding from NHMRC and NBCF.  
MDR acknowledges financial support from SNSF project grant (143883) and from the European Commission through the 7th Framework Collaborative Project RADIANT (Grant Agreement Number: 305626).
We thank Elena Zotenko, Marcel Coolen, Giancarlo Marra,
Mattia Pelizzola, Mark van de Wiel and Leonhard Held for useful discussions on the
experimental, computational and statistical strategies.


{\ifthenelse{\boolean{publ}}{\footnotesize}{\small}
 \bibliographystyle{bmc_article}  
  \bibliography{literature} }     


\begin{thebibliography}{10}
\providecommand{\url}[1]{[#1]}
\providecommand{\urlprefix}{}

\bibitem{jones.baylin2007}
Jones PA, Baylin SB: \textbf{{The epigenomics of cancer}}. \emph{Cell} 2007,
  \textbf{128}(4):683--692.

\bibitem{slomko.etal2012}
Slomko H, Heo HJ, Einstein FH: \textbf{{M}inireview: epigenetics of obesity and
  diabetes in humans}. \emph{Endocrinology} 2012, \textbf{153}(3):1025--1030.

\bibitem{clark-melki-2002}
Clark SJ, Melki J: \textbf{{DNA} methylation and gene silencing in cancer:
  which is the guilty party?} \emph{Oncogene} 2002, \textbf{21}(35):5380--5387.

\bibitem{esteller-2007}
Esteller M: \textbf{{C}ancer epigenomics: {DNA} methylomes and
  histone-modification maps}. \emph{Nature Reviews Genetics} 2007,
  \textbf{8}(4):286--298.

\bibitem{ziller-etal-2013}
Ziller MJ, Gu H, M{\"u}ller F, Donaghey J, Tsai LTY, Kohlbacher O, De~Jager PL,
  Rosen ED, Bennett DA, Bernstein BE, Gnirke A, Meissner A: \textbf{Charting a
  dynamic DNA methylation landscape of the human genome}. \emph{Nature} 2013,
  \textbf{500}(7463):477--481.

\bibitem{ruike-etal-2010}
Ruike Y, Imanaka Y, Sato F, Shimizu K, Tsujimoto G: \textbf{{G}enome-wide
  analysis of aberrant methylation in human breast cancer cells using
  methyl-{DNA} immunoprecipitation combined with high-throughput sequencing}.
  \emph{BMC Genomics} 2010, \textbf{11}:137.

\bibitem{stein2011}
Stein RA: \textbf{{E}pigenetics---{T}he link between infectious diseases and
  cancer}. \emph{Journal of the American Medical Association} 2011,
  \textbf{305}(14):1484--1485.

\bibitem{baylin-jones-2011}
Baylin SB, Jones PA: \textbf{{A} decade of exploring the cancer
  epigenome---biological and translational implications}. \emph{Nature Reviews
  Cancer} 2011, \textbf{11}(10):726--734.

\bibitem{jones-2012}
Jones PA: \textbf{{Functions of DNA methylation: islands, start sites, gene
  bodies and beyond}}. \emph{Nature Reviews Genetics} 2012,
  \urlprefix\url{[http://dx.doi.org/10.1038/nrg3230]}.

\bibitem{gu-etal-2010}
Gu H, Bock C, Mikkelsen TS, Jager N, Smith ZD, Tomazou E, Gnirke A, Lander ES,
  Meissner A: \textbf{{Genome-scale DNA methylation mapping of clinical samples
  at single-nucleotide resolution}}. \emph{Nature Methods} 2010,
  \textbf{7}(2):133--136.

\bibitem{laird-2010}
Laird PW: \textbf{{P}rinciples and challenges of genome-wide {DNA} methylation
  analysis}. \emph{Nature Reviews Genetics} 2010, \textbf{11}(3):191--203.

\bibitem{lister-ecker-2009}
Lister R, Ecker JR: \textbf{{F}inding the fifth base: genome-wide sequencing of
  cytosine methylation}. \emph{Genome Research} 2009, \textbf{19}(6):959--966.

\bibitem{kerick.etal2012}
Kerick M, Fischer A, Schweiger MR: \textbf{{G}eneration and {A}nalysis of
  {G}enome-{W}ide {DNA} {M}ethylation {M}aps}. In \emph{Bioinformatics for High
  Throughput Sequencing}. Edited by Rodr{\'i}guez-Ezpeleta N, Hackenberg M,
  Aransay AM, Springer New York 2012:151--167.

\bibitem{clark-etal-1994}
Clark SJ, Harrison J, Paul CL, Frommer M: \textbf{{H}igh sensitivity mapping of
  methylated cytosines}. \emph{Nucleic Acids Research} 1994,
  \textbf{22}(15):2990--2997.

\bibitem{bibikova-etal-2011}
Bibikova M, Barnes B, Tsan C, Ho V, Klotzle B, Le JM, Delano D, Zhang L,
  Schroth GP, Gunderson KL, Fan JB, Shen R: \textbf{{High} density {DNA}
  methylation array with single {CpG} site resolution}. \emph{Genomics} 2011,
  \textbf{98}(4):288--295.

\bibitem{robinson-etal-2010-3}
Robinson MD, Statham AL, Speed TP, Clark SJ: \textbf{{P}rotocol matters: which
  metylome are you actually studying?} \emph{Epigenomics} 2010,
  \textbf{2}(4):587--598.

\bibitem{hansen-etal-2011}
Hansen KD, Timp W, Bravo HC, Sabunciyan S, Langmead B, McDonald OG, Wen B, Wu
  H, Liu Y, Diep D, Briem E, Zhang K, Irizarry RA, Feinberg AP:
  \textbf{{Increased methylation variation in epigenetic domains across cancer
  types}}. \emph{Nat. Genet.} 2011, \textbf{43}(8):768--775.

\bibitem{lee-etal-2011}
Lee EJ, Pei L, Srivastava G, Joshi T, Kushwaha G, Choi JH, Robertson KD, Wang
  X, Colbourne JK, Zhang L, Schroth GP, Xu D, Zhang K, Shi H: \textbf{Targeted
  bisulfite sequencing by solution hybrid selection and massively parallel
  sequencing}. \emph{Nucleic Acids Research} 2011, \textbf{39}(19):e127.

\bibitem{lee-etal-2012}
Lee EJ, Luo J, Wilson JM, Shi H: \textbf{Analyzing the cancer methylome through
  targeted bisulfite sequencing}. \emph{Cancer letters} 2012.  in press.

\bibitem{clarke-etal-2009}
Clarke J, Wu HC, Jayasinghe L, Patel A, Reid S, Bayley H: \textbf{{Continuous
  base identification for single-molecule nanopore DNA sequencing}}.
  \emph{Nature Nanotechnology} 2009, \textbf{4}(4):265--270.

\bibitem{flusberg-etal-2010}
Flusberg BA, Webster DR, Lee JH, Travers KJ, Olivares EC, Clark TA, Korlach J,
  Turner SW: \textbf{{Direct detection of DNA methylation during
  single-molecule, real-time sequencing}}. \emph{Nature Methods} 2010,
  \textbf{7}(6):461--465.

\bibitem{down-etal-2008}
Down TA, Rakyan VK, Turner DJ, Flicek P, Li H, Kulesha E, Gr{\"a}f S, Johnson
  N, Herrero J, Tomazou EM, Thorne NP, B{\"a}ckdahl L, Herberth M, Howe KL,
  Jackson DK, Miretti MM, Marioni JC, Birney E, Hubbard TJP, Durbin R,
  Tavar{\'e} S, Beck S: \textbf{{A} {Bayesian} deconvolution strategy for
  immunoprecipitation-based {DNA} methylome analysis}. \emph{Nature
  Biotechnology} 2008, \textbf{26}(7):779--785.

\bibitem{aberg.etal2012}
Aberg KA, McClay JL, Nerella S, Xie LY, Clark SL, Hudson AD, Buksz{\'a}r J,
  Adkins D, Consortium SS, Hultman CM, et~al.: \textbf{{MBD-seq as a
  cost-effective approach for methylome-wide association studies: demonstration
  in 1500 case-control samples}}. \emph{Epigenomics} 2012,
  \textbf{4}(6):605--621.

\bibitem{clark.etal2012}
Clark C, Palta P, Joyce CJ, Scott C, Grundberg E, Deloukas P, Palotie A, Coffey
  AJ: \textbf{{A comparison of the whole genome approach of MeDIP-seq to the
  targeted approach of the Infinium HumanMethylation450 BeadChip(®) for
  methylome profiling}}. \emph{PLoS ONE} 2012, \textbf{7}(11):e50233.

\bibitem{deMeyer-etal-2013}
De~Meyer T, Mampaey E, Vlemmix M, Denil S, Trooskens G, Renard JP, De~Keulenaer
  S, Dehan P, Menschaert G, Van~Criekinge W: \textbf{{Quality evaluation of
  methyl binding domain based kits for enrichment DNA-methylation sequencing}}.
  \emph{PLoS ONE} 2013, \textbf{8}(3):e59068.

\bibitem{nair-etal-2011}
Nair SS, Coolen MW, Stirzaker C, Song JZ, Statham AL, Strbenac D, Robinson MD,
  Clark SJ: \textbf{{Comparison of methyl-DNA immunoprecipitation (MeDIP) and
  methyl-CpG binding domain (MBD) protein capture for genome-wide DNA
  methylation analysis reveal CpG sequence coverage bias}}. \emph{Epigenetics}
  2011, \textbf{6}:34--44.

\bibitem{serre-etal-2010}
Serre D, Lee BH, Ting AH: \textbf{{MBD}-isolated genome sequencing provides a
  high-throughput and comprehensive survey of {DNA} methylation in the human
  genome}. \emph{Nucleic Acids Research} 2010, \textbf{38}(2):391--399.

\bibitem{chavez-etal-2010}
Chavez L, Jozefczuk J, Grimm C, Dietrich J, Timmermann B, Lehrach H, Herwig R,
  Adjaye J: \textbf{{C}omputational analysis of genome-wide {DNA} methylation
  during the differentiation of human embryonic stem cells along the endodermal
  lineage}. \emph{Genome Research} 2010, \textbf{20}(10):1441--1450.

\bibitem{lan-etal-2011}
Lan X, Adams C, Landers M, Dudas M, Krissinger D, Marnellos G, Bonneville R, Xu
  M, Wang J, Huang THM, Meredith G, Jin VX: \textbf{{H}igh resolution detection
  and analysis of {C}p{G} dinucleotides methylation using {MBD}-seq
  technology}. \emph{PLoS ONE} 2011, \textbf{6}(7):e22226.

\bibitem{feber-etal-2011}
Feber A, Wilson GA, Zhang L, Presneau N, Idowu B, Down TA, Rakyan VK, Noon LA,
  Lloyd AC, Stupka E, Schiza V, Teschendorff AE, Schroth GP, Flanagan A, Beck
  S: \textbf{{Comparative methylome analysis of benign and malignant peripheral
  nerve sheath tumors}}. \emph{Genome Research} 2011, \textbf{21}(4):515--524.

\bibitem{stevens-etal-2013}
Stevens M, Cheng JB, Li D, Xie M, Hong C, Maire CL, Ligon KL, Hirst M, Marra
  MA, Costello JF, Wang T: \textbf{Estimating absolute methylation levels at
  single-CpG resolution from methylation enrichment and restriction enzyme
  sequencing methods}. \emph{Genome Research} 2013, \textbf{23}(9):1541--1553.

\bibitem{bock-etal-2010}
Bock C, Tomazou EM, Brinkman A, M{\"u}ller F, Simmer F, Gu H, J{\"a}ger N,
  Gnirke A, Stunnenberg HG, Meissner A: \textbf{{G}enome-wide mapping of {DNA}
  methylation: a quantitative technology comparison}. \emph{Nature
  Biotechnology} 2010, \textbf{28}:1106--1114.

\bibitem{robinson-etal-2012}
Robinson MD, Strbenac D, Stirzaker C, Statham AL, Song JZ, Speed TP, Clark SJ:
  \textbf{{Copy-number-aware differential analysis of quantitative DNA
  sequencing data}}. \emph{Genome Research} 2012,
  \urlprefix\url{[http://dx.doi.org/10.1101/gr.139055.112]}.

\bibitem{fader-hardie-2000}
Fader PS, Hardie BGS: \textbf{{A} note on modelling underreported {P}oisson
  counts}. \emph{Journal of Applied Statistics} 2000, \textbf{27}(8):953--964.

\bibitem{Abramowitz-stegun-1972}
Abramowitz M, Stegun IA: \emph{{H}andbook of {M}athematical functions with
  {F}ormulas, {G}raphs and {M}athematical {T}ables}. New York: Dover
  Publications 1972.

\bibitem{pelizzola-etal-2008}
Pelizzola M, Koga Y, Urban AE, Krauthammer M, Weissman S, Halaban R, Molinaro
  AM: \textbf{{MEDME}: an experimental and analytical methodology for the
  estimation of {DNA} methylation levels based on microarray derived
  {MeDIP}-enrichment}. \emph{Genome Research} 2008, \textbf{18}(10):1652--1659.

\bibitem{robinson-etal-2010-2}
Robinson MD, Stirzaker C, Statham AL, Coolen MW, Song JZ, Nair SS, Strbenac D,
  Speed TP, Clark SJ: \textbf{{E}valuation of affinity-based genome-wide {DNA}
  methylation data: effects of {CpG} density, amplification bias, and copy
  number variation}. \emph{Genome Research} 2010, \textbf{20}(12):1719--1729.

\bibitem{schmittlein1985}
Schmittlein DC, Bemmaor AC, Morrison DG: \textbf{Why does the NBD model work?
  Robustness in representing product purchases, brand purchases and imperfectly
  recorded purchases}. \emph{Marketing Science} 1985, \textbf{4}:255--266.

\bibitem{winkelmann-1996}
Winkelmann R: \textbf{Markov chain Monte Carlo analysis of underreported count
  data with an application to worker absenteeism}. \emph{Empirical Economics}
  1996, \textbf{21}(4):575--587.

\bibitem{lister-etal-2009}
Lister R, Pelizzola M, Dowen RH, Hawkins RD, Hon G, Tonti-Filippini J, Nery JR,
  Lee L, Ye Z, Ngo QM, Edsall L, Antosiewicz-Bourget J, Stewart R, Ruotti V,
  Millar AH, Thomson JA, Ren B, Ecker JR: \textbf{{H}uman {DNA} methylomes at
  base resolution show widespread epigenomic differences}. \emph{Nature} 2009,
  \textbf{462}(7271):315--322.

\bibitem{houseman-etal-2009}
Houseman EA, Christensen BC, Karagas MR, Wrensch MR, Nelson HH, Wiemels JL,
  Zheng S, Wiencke JK, Kelsey KT, Marsit CJ: \textbf{{Copy number variation has
  little impact on bead-array-based measures of DNA methylation}}.
  \emph{Bioinformatics} 2009, \textbf{25}(16):1999--2005.

\bibitem{pickrell.etal2011}
Pickrell J, Gaffney D, Gilad Y, Pritchard J: \textbf{{F}alse positive peaks in
  {C}h{IP}-seq and other sequencing-based functional assays caused by
  unannotated high copy number regions}. \emph{Bioinformatics} 2011,
  \textbf{27}(15):2144--2146.

\bibitem{taiwo-etal-2012}
Taiwo O, Wilson GA, Morris T, Seisenberger S, Reik W, Pearce D, Beck S, Butcher
  LM: \textbf{{Methylome analysis using MeDIP-seq with low DNA
  concentrations}}. \emph{Nature Protocols} 2012, \textbf{7}(4):617--636.

\bibitem{carvalho-etal-2012}
Carvalho RH, Haberle V, Hou J, van Gent T, Thongjuea S, van Ijcken W, Kockx C,
  Brouwer R, Rijkers E, Sieuwerts A, Foekens J, van Vroonhoven M, Aerts J,
  Grosveld F, Lenhard B, Philipsen S: \textbf{{Genome-wide DNA methylation
  profiling of non-small cell lung carcinomas}}. \emph{Epigenetics Chromatin}
  2012, \textbf{5}:9.

\bibitem{hansen-etal-2012}
Hansen KD, Irizarry RA, Wu Z: \textbf{{Removing technical variability in
  RNA-seq data using conditional quantile normalization}}. \emph{Biostatistics}
  2012, \textbf{13}(2):204--216.

\bibitem{wu-etal-2011}
Wu G, Yi N, Absher D, Zhi D: \textbf{{S}tatistical quantification of
  methylation levels by next-generation sequencing}. \emph{PLoS ONE} 2011,
  \textbf{6}(6):e21034.

\bibitem{GMRFbook}
Rue H, Held L: \emph{{G}aussian {M}arkov {R}andom {F}ields: {T}heory and
  {A}pplications}. London: Chapman \& Hall/CRC Press 2005.

\bibitem{statham-etal-2010}
Statham AL, Strbenac D, Coolen MW, Stirzaker C, Clark SJ, Robinson MD:
  \textbf{{R}epitools: an {R} package for the analysis of enrichment-based
  epigenomic data}. \emph{Bioinformatics} 2010, \textbf{26}(13):1662--1663.

\bibitem{hansen-aryee}
Hansen KD, Aryee M: \emph{minfi: {A}nalyze {I}llumina's 450k methylation
  arrays}. [R package version 1.3.3].

\bibitem{robinson-oshlack-2010}
Robinson MD, Oshlack A: \textbf{{A} scaling normalization method for
  differential expression analysis of {RNA-seq} data}. \emph{Genome Biology}
  2010, \textbf{11}(3):R25.

\bibitem{greenman-etal-2010}
Greenman CD, Bignell G, Butler A, Edkins S, Hinton J, Beare D, Swamy S,
  Santarius T, Chen L, Widaa S, Futreal PA, Stratton MR: \textbf{{PICNIC}: an
  algorithm to predict absolute allelic copy number variation with microarray
  cancer data}. \emph{Biostatistics} 2010, \textbf{11}:164--175.

\end{thebibliography}

\newcommand{\BMCxmlcomment}[1]{}

\BMCxmlcomment{

<refgrp>

<bibl id="B1">
  <title><p>{The epigenomics of cancer}</p></title>
  <aug>
    <au><snm>Jones</snm><fnm>P. A.</fnm></au>
    <au><snm>Baylin</snm><fnm>S. B.</fnm></au>
  </aug>
  <source>Cell</source>
  <pubdate>2007</pubdate>
  <volume>128</volume>
  <issue>4</issue>
  <fpage>683</fpage>
  <lpage>-692</lpage>
</bibl>

<bibl id="B2">
  <title><p>{M}inireview: epigenetics of obesity and diabetes in
  humans</p></title>
  <aug>
    <au><snm>Slomko</snm><fnm>H</fnm></au>
    <au><snm>Heo</snm><fnm>HJ</fnm></au>
    <au><snm>Einstein</snm><fnm>FH</fnm></au>
  </aug>
  <source>Endocrinology</source>
  <publisher>Endocrine Soc</publisher>
  <pubdate>2012</pubdate>
  <volume>153</volume>
  <issue>3</issue>
  <fpage>1025</fpage>
  <lpage>-1030</lpage>
</bibl>

<bibl id="B3">
  <title><p>{DNA} methylation and gene silencing in cancer: which is the guilty
  party?</p></title>
  <aug>
    <au><snm>Clark</snm><fnm>SJ</fnm></au>
    <au><snm>Melki</snm><fnm>J</fnm></au>
  </aug>
  <source>Oncogene</source>
  <pubdate>2002</pubdate>
  <volume>21</volume>
  <issue>35</issue>
  <fpage>5380</fpage>
  <lpage>-5387</lpage>
</bibl>

<bibl id="B4">
  <title><p>{C}ancer epigenomics: {DNA} methylomes and histone-modification
  maps</p></title>
  <aug>
    <au><snm>Esteller</snm><fnm>M</fnm></au>
  </aug>
  <source>Nature Reviews Genetics</source>
  <pubdate>2007</pubdate>
  <volume>8</volume>
  <issue>4</issue>
  <fpage>286</fpage>
  <lpage>-298</lpage>
</bibl>

<bibl id="B5">
  <title><p>Charting a dynamic DNA methylation landscape of the human
  genome</p></title>
  <aug>
    <au><snm>Ziller</snm><fnm>MJ</fnm></au>
    <au><snm>Gu</snm><fnm>H</fnm></au>
    <au><snm>M{\"u}ller</snm><fnm>F</fnm></au>
    <au><snm>Donaghey</snm><fnm>J</fnm></au>
    <au><snm>Tsai</snm><fnm>LTY</fnm></au>
    <au><snm>Kohlbacher</snm><fnm>O</fnm></au>
    <au><snm>De Jager</snm><fnm>PL</fnm></au>
    <au><snm>Rosen</snm><fnm>ED</fnm></au>
    <au><snm>Bennett</snm><fnm>DA</fnm></au>
    <au><snm>Bernstein</snm><fnm>BE</fnm></au>
    <au><snm>Gnirke</snm><fnm>A</fnm></au>
    <au><snm>Meissner</snm><fnm>A</fnm></au>
  </aug>
  <source>Nature</source>
  <publisher>Nature Publishing Group</publisher>
  <pubdate>2013</pubdate>
  <volume>500</volume>
  <issue>7463</issue>
  <fpage>477</fpage>
  <lpage>-481</lpage>
</bibl>

<bibl id="B6">
  <title><p>{G}enome-wide analysis of aberrant methylation in human breast
  cancer cells using methyl-{DNA} immunoprecipitation combined with
  high-throughput sequencing</p></title>
  <aug>
    <au><snm>Ruike</snm><fnm>Y</fnm></au>
    <au><snm>Imanaka</snm><fnm>Y</fnm></au>
    <au><snm>Sato</snm><fnm>F</fnm></au>
    <au><snm>Shimizu</snm><fnm>K</fnm></au>
    <au><snm>Tsujimoto</snm><fnm>G</fnm></au>
  </aug>
  <source>BMC Genomics</source>
  <pubdate>2010</pubdate>
  <volume>11</volume>
  <issue>1</issue>
  <fpage>137</fpage>
</bibl>

<bibl id="B7">
  <title><p>{E}pigenetics---{T}he link between infectious diseases and
  cancer</p></title>
  <aug>
    <au><snm>Stein</snm><fnm>RA</fnm></au>
  </aug>
  <source>Journal of the American Medical Association</source>
  <pubdate>2011</pubdate>
  <volume>305</volume>
  <issue>14</issue>
  <fpage>1484</fpage>
  <lpage>-1485</lpage>
</bibl>

<bibl id="B8">
  <title><p>{A} decade of exploring the cancer epigenome---biological and
  translational implications</p></title>
  <aug>
    <au><snm>Baylin</snm><fnm>S B</fnm></au>
    <au><snm>Jones</snm><fnm>P A</fnm></au>
  </aug>
  <source>Nature Reviews Cancer</source>
  <pubdate>2011</pubdate>
  <volume>11</volume>
  <issue>10</issue>
  <fpage>726</fpage>
  <lpage>-734</lpage>
</bibl>

<bibl id="B9">
  <title><p>{Functions of DNA methylation: islands, start sites, gene bodies
  and beyond}</p></title>
  <aug>
    <au><snm>Jones</snm><fnm>P A</fnm></au>
  </aug>
  <source>Nature Reviews Genetics</source>
  <pubdate>2012</pubdate>
  <url>http://dx.doi.org/10.1038/nrg3230</url>
</bibl>

<bibl id="B10">
  <title><p>{Genome-scale DNA methylation mapping of clinical samples at
  single-nucleotide resolution}</p></title>
  <aug>
    <au><snm>Gu</snm><fnm>H.</fnm></au>
    <au><snm>Bock</snm><fnm>C.</fnm></au>
    <au><snm>Mikkelsen</snm><fnm>T. S.</fnm></au>
    <au><snm>Jager</snm><fnm>N.</fnm></au>
    <au><snm>Smith</snm><fnm>Z. D.</fnm></au>
    <au><snm>Tomazou</snm><fnm>E.</fnm></au>
    <au><snm>Gnirke</snm><fnm>A.</fnm></au>
    <au><snm>Lander</snm><fnm>E. S.</fnm></au>
    <au><snm>Meissner</snm><fnm>A.</fnm></au>
  </aug>
  <source>Nature Methods</source>
  <pubdate>2010</pubdate>
  <volume>7</volume>
  <issue>2</issue>
  <fpage>133</fpage>
  <lpage>-136</lpage>
</bibl>

<bibl id="B11">
  <title><p>{P}rinciples and challenges of genome-wide {DNA} methylation
  analysis</p></title>
  <aug>
    <au><snm>Laird</snm><fnm>P W</fnm></au>
  </aug>
  <source>Nature Reviews Genetics</source>
  <pubdate>2010</pubdate>
  <volume>11</volume>
  <issue>3</issue>
  <fpage>191</fpage>
  <lpage>-203</lpage>
</bibl>

<bibl id="B12">
  <title><p>{F}inding the fifth base: genome-wide sequencing of cytosine
  methylation</p></title>
  <aug>
    <au><snm>Lister</snm><fnm>R</fnm></au>
    <au><snm>Ecker</snm><fnm>JR</fnm></au>
  </aug>
  <source>Genome Research</source>
  <pubdate>2009</pubdate>
  <volume>19</volume>
  <issue>6</issue>
  <fpage>959</fpage>
  <lpage>-966</lpage>
</bibl>

<bibl id="B13">
  <title><p>{G}eneration and {A}nalysis of {G}enome-{W}ide {DNA} {M}ethylation
  {M}aps</p></title>
  <aug>
    <au><snm>Kerick</snm><fnm>M</fnm></au>
    <au><snm>Fischer</snm><fnm>A</fnm></au>
    <au><snm>Schweiger</snm><fnm>MR</fnm></au>
  </aug>
  <source>Bioinformatics for High Throughput Sequencing</source>
  <publisher>Springer New York</publisher>
  <editor>Rodr{\'i}guez-Ezpeleta, Naiara and Hackenberg, Michael and Aransay,
  Ana M.</editor>
  <pubdate>2012</pubdate>
  <fpage>151</fpage>
  <lpage>-167</lpage>
</bibl>

<bibl id="B14">
  <title><p>{H}igh sensitivity mapping of methylated cytosines</p></title>
  <aug>
    <au><snm>Clark</snm><fnm>S. J.</fnm></au>
    <au><snm>Harrison</snm><fnm>J.</fnm></au>
    <au><snm>Paul</snm><fnm>C. L.</fnm></au>
    <au><snm>Frommer</snm><fnm>M.</fnm></au>
  </aug>
  <source>Nucleic Acids Research</source>
  <pubdate>1994</pubdate>
  <volume>22</volume>
  <issue>15</issue>
  <fpage>2990</fpage>
  <lpage>-2997</lpage>
</bibl>

<bibl id="B15">
  <title><p>{High} density {DNA} methylation array with single {CpG} site
  resolution</p></title>
  <aug>
    <au><snm>Bibikova</snm><fnm>M</fnm></au>
    <au><snm>Barnes</snm><fnm>B</fnm></au>
    <au><snm>Tsan</snm><fnm>C</fnm></au>
    <au><snm>Ho</snm><fnm>V</fnm></au>
    <au><snm>Klotzle</snm><fnm>B</fnm></au>
    <au><snm>Le</snm><fnm>JM</fnm></au>
    <au><snm>Delano</snm><fnm>D</fnm></au>
    <au><snm>Zhang</snm><fnm>L</fnm></au>
    <au><snm>Schroth</snm><fnm>GP</fnm></au>
    <au><snm>Gunderson</snm><fnm>KL</fnm></au>
    <au><snm>Fan</snm><fnm>JB</fnm></au>
    <au><snm>Shen</snm><fnm>R</fnm></au>
  </aug>
  <source>Genomics</source>
  <pubdate>2011</pubdate>
  <volume>98</volume>
  <issue>4</issue>
  <fpage>288</fpage>
  <lpage>295</lpage>
</bibl>

<bibl id="B16">
  <title><p>{P}rotocol matters: which metylome are you actually
  studying?</p></title>
  <aug>
    <au><snm>Robinson</snm><fnm>M D</fnm></au>
    <au><snm>Statham</snm><fnm>A L</fnm></au>
    <au><snm>Speed</snm><fnm>T P</fnm></au>
    <au><snm>Clark</snm><fnm>S J</fnm></au>
  </aug>
  <source>Epigenomics</source>
  <pubdate>2010</pubdate>
  <volume>2</volume>
  <issue>4</issue>
  <fpage>587</fpage>
  <lpage>-598</lpage>
</bibl>

<bibl id="B17">
  <title><p>{Increased methylation variation in epigenetic domains across
  cancer types}</p></title>
  <aug>
    <au><snm>Hansen</snm><fnm>K. D.</fnm></au>
    <au><snm>Timp</snm><fnm>W.</fnm></au>
    <au><snm>Bravo</snm><fnm>H. C.</fnm></au>
    <au><snm>Sabunciyan</snm><fnm>S.</fnm></au>
    <au><snm>Langmead</snm><fnm>B.</fnm></au>
    <au><snm>McDonald</snm><fnm>O. G.</fnm></au>
    <au><snm>Wen</snm><fnm>B.</fnm></au>
    <au><snm>Wu</snm><fnm>H.</fnm></au>
    <au><snm>Liu</snm><fnm>Y.</fnm></au>
    <au><snm>Diep</snm><fnm>D.</fnm></au>
    <au><snm>Briem</snm><fnm>E.</fnm></au>
    <au><snm>Zhang</snm><fnm>K.</fnm></au>
    <au><snm>Irizarry</snm><fnm>R. A.</fnm></au>
    <au><snm>Feinberg</snm><fnm>A. P.</fnm></au>
  </aug>
  <source>Nat. Genet.</source>
  <pubdate>2011</pubdate>
  <volume>43</volume>
  <issue>8</issue>
  <fpage>768</fpage>
  <lpage>-775</lpage>
</bibl>

<bibl id="B18">
  <title><p>Targeted bisulfite sequencing by solution hybrid selection and
  massively parallel sequencing</p></title>
  <aug>
    <au><snm>Lee</snm><fnm>EJ</fnm></au>
    <au><snm>Pei</snm><fnm>L</fnm></au>
    <au><snm>Srivastava</snm><fnm>G</fnm></au>
    <au><snm>Joshi</snm><fnm>T</fnm></au>
    <au><snm>Kushwaha</snm><fnm>G</fnm></au>
    <au><snm>Choi</snm><fnm>JH</fnm></au>
    <au><snm>Robertson</snm><fnm>KD</fnm></au>
    <au><snm>Wang</snm><fnm>X</fnm></au>
    <au><snm>Colbourne</snm><fnm>JK</fnm></au>
    <au><snm>Zhang</snm><fnm>L</fnm></au>
    <au><snm>Schroth</snm><fnm>GP</fnm></au>
    <au><snm>Xu</snm><fnm>D</fnm></au>
    <au><snm>Zhang</snm><fnm>K</fnm></au>
    <au><snm>Shi</snm><fnm>H</fnm></au>
  </aug>
  <source>Nucleic Acids Research</source>
  <pubdate>2011</pubdate>
  <volume>39</volume>
  <issue>19</issue>
  <fpage>e127</fpage>
</bibl>

<bibl id="B19">
  <title><p>Analyzing the cancer methylome through targeted bisulfite
  sequencing</p></title>
  <aug>
    <au><snm>Lee</snm><fnm>EJ</fnm></au>
    <au><snm>Luo</snm><fnm>J</fnm></au>
    <au><snm>Wilson</snm><fnm>JM</fnm></au>
    <au><snm>Shi</snm><fnm>H</fnm></au>
  </aug>
  <source>Cancer letters</source>
  <publisher>Elsevier</publisher>
  <pubdate>2012</pubdate>
  <inpress />
</bibl>

<bibl id="B20">
  <title><p>{Continuous base identification for single-molecule nanopore DNA
  sequencing}</p></title>
  <aug>
    <au><snm>Clarke</snm><fnm>J.</fnm></au>
    <au><snm>Wu</snm><fnm>H. C.</fnm></au>
    <au><snm>Jayasinghe</snm><fnm>L.</fnm></au>
    <au><snm>Patel</snm><fnm>A.</fnm></au>
    <au><snm>Reid</snm><fnm>S.</fnm></au>
    <au><snm>Bayley</snm><fnm>H.</fnm></au>
  </aug>
  <source>Nature Nanotechnology</source>
  <pubdate>2009</pubdate>
  <volume>4</volume>
  <issue>4</issue>
  <fpage>265</fpage>
  <lpage>-270</lpage>
</bibl>

<bibl id="B21">
  <title><p>{Direct detection of DNA methylation during single-molecule,
  real-time sequencing}</p></title>
  <aug>
    <au><snm>Flusberg</snm><fnm>B. A.</fnm></au>
    <au><snm>Webster</snm><fnm>D. R.</fnm></au>
    <au><snm>Lee</snm><fnm>J. H.</fnm></au>
    <au><snm>Travers</snm><fnm>K. J.</fnm></au>
    <au><snm>Olivares</snm><fnm>E. C.</fnm></au>
    <au><snm>Clark</snm><fnm>T. A.</fnm></au>
    <au><snm>Korlach</snm><fnm>J.</fnm></au>
    <au><snm>Turner</snm><fnm>S. W.</fnm></au>
  </aug>
  <source>Nature Methods</source>
  <pubdate>2010</pubdate>
  <volume>7</volume>
  <issue>6</issue>
  <fpage>461</fpage>
  <lpage>-465</lpage>
</bibl>

<bibl id="B22">
  <title><p>{A} {Bayesian} deconvolution strategy for immunoprecipitation-based
  {DNA} methylome analysis</p></title>
  <aug>
    <au><snm>Down</snm><fnm>TA</fnm></au>
    <au><snm>Rakyan</snm><fnm>VK</fnm></au>
    <au><snm>Turner</snm><fnm>DJ</fnm></au>
    <au><snm>Flicek</snm><fnm>P</fnm></au>
    <au><snm>Li</snm><fnm>H</fnm></au>
    <au><snm>Kulesha</snm><fnm>E</fnm></au>
    <au><snm>Gr{\"a}f</snm><fnm>S</fnm></au>
    <au><snm>Johnson</snm><fnm>N</fnm></au>
    <au><snm>Herrero</snm><fnm>J</fnm></au>
    <au><snm>Tomazou</snm><fnm>EM</fnm></au>
    <au><snm>Thorne</snm><fnm>NP</fnm></au>
    <au><snm>B{\"a}ckdahl</snm><fnm>L</fnm></au>
    <au><snm>Herberth</snm><fnm>M</fnm></au>
    <au><snm>Howe</snm><fnm>KL</fnm></au>
    <au><snm>Jackson</snm><fnm>DK</fnm></au>
    <au><snm>Miretti</snm><fnm>MM</fnm></au>
    <au><snm>Marioni</snm><fnm>JC</fnm></au>
    <au><snm>Birney</snm><fnm>E</fnm></au>
    <au><snm>Hubbard</snm><fnm>TJP</fnm></au>
    <au><snm>Durbin</snm><fnm>R</fnm></au>
    <au><snm>Tavar{\'e}</snm><fnm>S</fnm></au>
    <au><snm>Beck</snm><fnm>S</fnm></au>
  </aug>
  <source>Nature Biotechnology</source>
  <pubdate>2008</pubdate>
  <volume>26</volume>
  <issue>7</issue>
  <fpage>779</fpage>
  <lpage>-785</lpage>
</bibl>

<bibl id="B23">
  <title><p>{MBD-seq as a cost-effective approach for methylome-wide
  association studies: demonstration in 1500 case-control samples}</p></title>
  <aug>
    <au><snm>Aberg</snm><fnm>KA</fnm></au>
    <au><snm>McClay</snm><fnm>JL</fnm></au>
    <au><snm>Nerella</snm><fnm>S</fnm></au>
    <au><snm>Xie</snm><fnm>LY</fnm></au>
    <au><snm>Clark</snm><fnm>SL</fnm></au>
    <au><snm>Hudson</snm><fnm>AD</fnm></au>
    <au><snm>Buksz{\'a}r</snm><fnm>J</fnm></au>
    <au><snm>Adkins</snm><fnm>D</fnm></au>
    <au><snm>Consortium</snm><fnm>SS</fnm></au>
    <au><snm>Hultman</snm><fnm>CM</fnm></au>
    <au><cnm>others</cnm></au>
  </aug>
  <source>Epigenomics</source>
  <publisher>Future Medicine</publisher>
  <pubdate>2012</pubdate>
  <volume>4</volume>
  <issue>6</issue>
  <fpage>605</fpage>
  <lpage>-621</lpage>
</bibl>

<bibl id="B24">
  <title><p>{A comparison of the whole genome approach of MeDIP-seq to the
  targeted approach of the Infinium HumanMethylation450 BeadChip(®) for
  methylome profiling}</p></title>
  <aug>
    <au><snm>Clark</snm><fnm>C.</fnm></au>
    <au><snm>Palta</snm><fnm>P.</fnm></au>
    <au><snm>Joyce</snm><fnm>C. J.</fnm></au>
    <au><snm>Scott</snm><fnm>C.</fnm></au>
    <au><snm>Grundberg</snm><fnm>E.</fnm></au>
    <au><snm>Deloukas</snm><fnm>P.</fnm></au>
    <au><snm>Palotie</snm><fnm>A.</fnm></au>
    <au><snm>Coffey</snm><fnm>A. J.</fnm></au>
  </aug>
  <source>PLoS ONE</source>
  <pubdate>2012</pubdate>
  <volume>7</volume>
  <issue>11</issue>
  <fpage>e50233</fpage>
</bibl>

<bibl id="B25">
  <title><p>{Quality evaluation of methyl binding domain based kits for
  enrichment DNA-methylation sequencing}</p></title>
  <aug>
    <au><snm>De Meyer</snm><fnm>T.</fnm></au>
    <au><snm>Mampaey</snm><fnm>E.</fnm></au>
    <au><snm>Vlemmix</snm><fnm>M.</fnm></au>
    <au><snm>Denil</snm><fnm>S.</fnm></au>
    <au><snm>Trooskens</snm><fnm>G.</fnm></au>
    <au><snm>Renard</snm><fnm>J. P.</fnm></au>
    <au><snm>De Keulenaer</snm><fnm>S.</fnm></au>
    <au><snm>Dehan</snm><fnm>P.</fnm></au>
    <au><snm>Menschaert</snm><fnm>G.</fnm></au>
    <au><snm>Van Criekinge</snm><fnm>W.</fnm></au>
  </aug>
  <source>PLoS ONE</source>
  <pubdate>2013</pubdate>
  <volume>8</volume>
  <issue>3</issue>
  <fpage>e59068</fpage>
</bibl>

<bibl id="B26">
  <title><p>{Comparison of methyl-DNA immunoprecipitation (MeDIP) and
  methyl-CpG binding domain (MBD) protein capture for genome-wide DNA
  methylation analysis reveal CpG sequence coverage bias}</p></title>
  <aug>
    <au><snm>Nair</snm><fnm>S S</fnm></au>
    <au><snm>Coolen</snm><fnm>M W</fnm></au>
    <au><snm>Stirzaker</snm><fnm>C</fnm></au>
    <au><snm>Song</snm><fnm>J Z</fnm></au>
    <au><snm>Statham</snm><fnm>A L</fnm></au>
    <au><snm>Strbenac</snm><fnm>D</fnm></au>
    <au><snm>Robinson</snm><fnm>M D</fnm></au>
    <au><snm>Clark</snm><fnm>S J</fnm></au>
  </aug>
  <source>Epigenetics</source>
  <pubdate>2011</pubdate>
  <volume>6</volume>
  <issue>1</issue>
  <fpage>34</fpage>
  <lpage>-44</lpage>
</bibl>

<bibl id="B27">
  <title><p>{MBD}-isolated genome sequencing provides a high-throughput and
  comprehensive survey of {DNA} methylation in the human genome</p></title>
  <aug>
    <au><snm>Serre</snm><fnm>D</fnm></au>
    <au><snm>Lee</snm><fnm>BH</fnm></au>
    <au><snm>Ting</snm><fnm>AH</fnm></au>
  </aug>
  <source>Nucleic Acids Research</source>
  <pubdate>2010</pubdate>
  <volume>38</volume>
  <issue>2</issue>
  <fpage>391</fpage>
  <lpage>-399</lpage>
</bibl>

<bibl id="B28">
  <title><p>{C}omputational analysis of genome-wide {DNA} methylation during
  the differentiation of human embryonic stem cells along the endodermal
  lineage</p></title>
  <aug>
    <au><snm>Chavez</snm><fnm>L</fnm></au>
    <au><snm>Jozefczuk</snm><fnm>J</fnm></au>
    <au><snm>Grimm</snm><fnm>C</fnm></au>
    <au><snm>Dietrich</snm><fnm>J</fnm></au>
    <au><snm>Timmermann</snm><fnm>B</fnm></au>
    <au><snm>Lehrach</snm><fnm>H</fnm></au>
    <au><snm>Herwig</snm><fnm>R</fnm></au>
    <au><snm>Adjaye</snm><fnm>J</fnm></au>
  </aug>
  <source>Genome Research</source>
  <pubdate>2010</pubdate>
  <volume>20</volume>
  <issue>10</issue>
  <fpage>1441</fpage>
  <lpage>-1450</lpage>
</bibl>

<bibl id="B29">
  <title><p>{H}igh resolution detection and analysis of {C}p{G} dinucleotides
  methylation using {MBD}-seq technology</p></title>
  <aug>
    <au><snm>Lan</snm><fnm>X</fnm></au>
    <au><snm>Adams</snm><fnm>C</fnm></au>
    <au><snm>Landers</snm><fnm>M</fnm></au>
    <au><snm>Dudas</snm><fnm>M</fnm></au>
    <au><snm>Krissinger</snm><fnm>D</fnm></au>
    <au><snm>Marnellos</snm><fnm>G</fnm></au>
    <au><snm>Bonneville</snm><fnm>R</fnm></au>
    <au><snm>Xu</snm><fnm>M</fnm></au>
    <au><snm>Wang</snm><fnm>J</fnm></au>
    <au><snm>Huang</snm><fnm>THM</fnm></au>
    <au><snm>Meredith</snm><fnm>G</fnm></au>
    <au><snm>Jin</snm><fnm>VX</fnm></au>
  </aug>
  <source>PLoS ONE</source>
  <publisher>Public Library of Science</publisher>
  <pubdate>2011</pubdate>
  <volume>6</volume>
  <issue>7</issue>
  <fpage>e22226</fpage>
</bibl>

<bibl id="B30">
  <title><p>{Comparative methylome analysis of benign and malignant peripheral
  nerve sheath tumors}</p></title>
  <aug>
    <au><snm>Feber</snm><fnm>A.</fnm></au>
    <au><snm>Wilson</snm><fnm>G. A.</fnm></au>
    <au><snm>Zhang</snm><fnm>L.</fnm></au>
    <au><snm>Presneau</snm><fnm>N.</fnm></au>
    <au><snm>Idowu</snm><fnm>B.</fnm></au>
    <au><snm>Down</snm><fnm>T. A.</fnm></au>
    <au><snm>Rakyan</snm><fnm>V. K.</fnm></au>
    <au><snm>Noon</snm><fnm>L. A.</fnm></au>
    <au><snm>Lloyd</snm><fnm>A. C.</fnm></au>
    <au><snm>Stupka</snm><fnm>E.</fnm></au>
    <au><snm>Schiza</snm><fnm>V.</fnm></au>
    <au><snm>Teschendorff</snm><fnm>A. E.</fnm></au>
    <au><snm>Schroth</snm><fnm>G. P.</fnm></au>
    <au><snm>Flanagan</snm><fnm>A.</fnm></au>
    <au><snm>Beck</snm><fnm>S.</fnm></au>
  </aug>
  <source>Genome Research</source>
  <pubdate>2011</pubdate>
  <volume>21</volume>
  <issue>4</issue>
  <fpage>515</fpage>
  <lpage>-524</lpage>
</bibl>

<bibl id="B31">
  <title><p>Estimating absolute methylation levels at single-CpG resolution
  from methylation enrichment and restriction enzyme sequencing
  methods</p></title>
  <aug>
    <au><snm>Stevens</snm><fnm>M</fnm></au>
    <au><snm>Cheng</snm><fnm>JB</fnm></au>
    <au><snm>Li</snm><fnm>D</fnm></au>
    <au><snm>Xie</snm><fnm>M</fnm></au>
    <au><snm>Hong</snm><fnm>C</fnm></au>
    <au><snm>Maire</snm><fnm>CL</fnm></au>
    <au><snm>Ligon</snm><fnm>KL</fnm></au>
    <au><snm>Hirst</snm><fnm>M</fnm></au>
    <au><snm>Marra</snm><fnm>MA</fnm></au>
    <au><snm>Costello</snm><fnm>JF</fnm></au>
    <au><snm>Wang</snm><fnm>T</fnm></au>
  </aug>
  <source>Genome Research</source>
  <publisher>Cold Spring Harbor Lab</publisher>
  <pubdate>2013</pubdate>
  <volume>23</volume>
  <issue>9</issue>
  <fpage>1541</fpage>
  <lpage>-1553</lpage>
</bibl>

<bibl id="B32">
  <title><p>{G}enome-wide mapping of {DNA} methylation: a quantitative
  technology comparison</p></title>
  <aug>
    <au><snm>Bock</snm><fnm>C</fnm></au>
    <au><snm>Tomazou</snm><fnm>E M</fnm></au>
    <au><snm>Brinkman</snm><fnm>A</fnm></au>
    <au><snm>M{\"u}ller</snm><fnm>F</fnm></au>
    <au><snm>Simmer</snm><fnm>F</fnm></au>
    <au><snm>Gu</snm><fnm>H</fnm></au>
    <au><snm>J{\"a}ger</snm><fnm>N</fnm></au>
    <au><snm>Gnirke</snm><fnm>A</fnm></au>
    <au><snm>Stunnenberg</snm><fnm>H G</fnm></au>
    <au><snm>Meissner</snm><fnm>A</fnm></au>
  </aug>
  <source>Nature Biotechnology</source>
  <pubdate>2010</pubdate>
  <volume>28</volume>
  <fpage>1106</fpage>
  <lpage>1114</lpage>
</bibl>

<bibl id="B33">
  <title><p>{Copy-number-aware differential analysis of quantitative DNA
  sequencing data}</p></title>
  <aug>
    <au><snm>Robinson</snm><fnm>M D</fnm></au>
    <au><snm>Strbenac</snm><fnm>D</fnm></au>
    <au><snm>Stirzaker</snm><fnm>C</fnm></au>
    <au><snm>Statham</snm><fnm>A L</fnm></au>
    <au><snm>Song</snm><fnm>J Z</fnm></au>
    <au><snm>Speed</snm><fnm>T P</fnm></au>
    <au><snm>Clark</snm><fnm>S J</fnm></au>
  </aug>
  <source>Genome Research</source>
  <pubdate>2012</pubdate>
  <url>http://dx.doi.org/10.1101/gr.139055.112</url>
</bibl>

<bibl id="B34">
  <title><p>{A} note on modelling underreported {P}oisson counts</p></title>
  <aug>
    <au><snm>Fader</snm><fnm>PS</fnm></au>
    <au><snm>Hardie</snm><fnm>BGS</fnm></au>
  </aug>
  <source>Journal of Applied Statistics</source>
  <pubdate>2000</pubdate>
  <volume>27</volume>
  <issue>8</issue>
  <fpage>953</fpage>
  <lpage>964</lpage>
</bibl>

<bibl id="B35">
  <title><p>{H}andbook of {M}athematical functions with {F}ormulas, {G}raphs
  and {M}athematical {T}ables</p></title>
  <aug>
    <au><snm>Abramowitz</snm><fnm>M.</fnm></au>
    <au><snm>Stegun</snm><fnm>I. A.</fnm></au>
  </aug>
  <publisher>New York: Dover Publications</publisher>
  <editor>Abramowitz, M. and Stegun, I. A.</editor>
  <pubdate>1972</pubdate>
</bibl>

<bibl id="B36">
  <title><p>{MEDME}: an experimental and analytical methodology for the
  estimation of {DNA} methylation levels based on microarray derived
  {MeDIP}-enrichment</p></title>
  <aug>
    <au><snm>Pelizzola</snm><fnm>M</fnm></au>
    <au><snm>Koga</snm><fnm>Y</fnm></au>
    <au><snm>Urban</snm><fnm>AE</fnm></au>
    <au><snm>Krauthammer</snm><fnm>M</fnm></au>
    <au><snm>Weissman</snm><fnm>S</fnm></au>
    <au><snm>Halaban</snm><fnm>R</fnm></au>
    <au><snm>Molinaro</snm><fnm>AM</fnm></au>
  </aug>
  <source>Genome Research</source>
  <pubdate>2008</pubdate>
  <volume>18</volume>
  <issue>10</issue>
  <fpage>1652</fpage>
  <lpage>-1659</lpage>
</bibl>

<bibl id="B37">
  <title><p>{E}valuation of affinity-based genome-wide {DNA} methylation data:
  effects of {CpG} density, amplification bias, and copy number
  variation</p></title>
  <aug>
    <au><snm>Robinson</snm><fnm>MD</fnm></au>
    <au><snm>Stirzaker</snm><fnm>C</fnm></au>
    <au><snm>Statham</snm><fnm>AL</fnm></au>
    <au><snm>Coolen</snm><fnm>MW</fnm></au>
    <au><snm>Song</snm><fnm>JZ</fnm></au>
    <au><snm>Nair</snm><fnm>SS</fnm></au>
    <au><snm>Strbenac</snm><fnm>D</fnm></au>
    <au><snm>Speed</snm><fnm>TP</fnm></au>
    <au><snm>Clark</snm><fnm>SJ</fnm></au>
  </aug>
  <source>Genome Research</source>
  <pubdate>2010</pubdate>
  <volume>20</volume>
  <issue>12</issue>
  <fpage>1719</fpage>
  <lpage>-1729</lpage>
</bibl>

<bibl id="B38">
  <title><p>Why does the NBD model work? Robustness in representing product
  purchases, brand purchases and imperfectly recorded purchases</p></title>
  <aug>
    <au><snm>Schmittlein</snm><fnm>D. C.</fnm></au>
    <au><snm>Bemmaor</snm><fnm>A. C.</fnm></au>
    <au><snm>Morrison</snm><fnm>D. G.</fnm></au>
  </aug>
  <source>Marketing Science</source>
  <pubdate>1985</pubdate>
  <volume>4</volume>
  <fpage>255</fpage>
  <lpage>-266</lpage>
</bibl>

<bibl id="B39">
  <title><p>Markov chain Monte Carlo analysis of underreported count data with
  an application to worker absenteeism</p></title>
  <aug>
    <au><snm>Winkelmann</snm><fnm>R</fnm></au>
  </aug>
  <source>Empirical Economics</source>
  <publisher>Springer</publisher>
  <pubdate>1996</pubdate>
  <volume>21</volume>
  <issue>4</issue>
  <fpage>575</fpage>
  <lpage>-587</lpage>
</bibl>

<bibl id="B40">
  <title><p>{H}uman {DNA} methylomes at base resolution show widespread
  epigenomic differences</p></title>
  <aug>
    <au><snm>Lister</snm><fnm>R</fnm></au>
    <au><snm>Pelizzola</snm><fnm>M</fnm></au>
    <au><snm>Dowen</snm><fnm>RH</fnm></au>
    <au><snm>Hawkins</snm><fnm>RD</fnm></au>
    <au><snm>Hon</snm><fnm>G</fnm></au>
    <au><snm>Tonti Filippini</snm><fnm>J</fnm></au>
    <au><snm>Nery</snm><fnm>JR</fnm></au>
    <au><snm>Lee</snm><fnm>L</fnm></au>
    <au><snm>Ye</snm><fnm>Z</fnm></au>
    <au><snm>Ngo</snm><fnm>QM</fnm></au>
    <au><snm>Edsall</snm><fnm>L</fnm></au>
    <au><snm>Antosiewicz Bourget</snm><fnm>J</fnm></au>
    <au><snm>Stewart</snm><fnm>R</fnm></au>
    <au><snm>Ruotti</snm><fnm>V</fnm></au>
    <au><snm>Millar</snm><fnm>AH</fnm></au>
    <au><snm>Thomson</snm><fnm>JA</fnm></au>
    <au><snm>Ren</snm><fnm>B</fnm></au>
    <au><snm>Ecker</snm><fnm>JR</fnm></au>
  </aug>
  <source>Nature</source>
  <pubdate>2009</pubdate>
  <volume>462</volume>
  <issue>7271</issue>
  <fpage>315</fpage>
  <lpage>-322</lpage>
</bibl>

<bibl id="B41">
  <title><p>{Copy number variation has little impact on bead-array-based
  measures of DNA methylation}</p></title>
  <aug>
    <au><snm>Houseman</snm><fnm>E. A.</fnm></au>
    <au><snm>Christensen</snm><fnm>B. C.</fnm></au>
    <au><snm>Karagas</snm><fnm>M. R.</fnm></au>
    <au><snm>Wrensch</snm><fnm>M. R.</fnm></au>
    <au><snm>Nelson</snm><fnm>H. H.</fnm></au>
    <au><snm>Wiemels</snm><fnm>J. L.</fnm></au>
    <au><snm>Zheng</snm><fnm>S.</fnm></au>
    <au><snm>Wiencke</snm><fnm>J. K.</fnm></au>
    <au><snm>Kelsey</snm><fnm>K. T.</fnm></au>
    <au><snm>Marsit</snm><fnm>C. J.</fnm></au>
  </aug>
  <source>Bioinformatics</source>
  <pubdate>2009</pubdate>
  <volume>25</volume>
  <issue>16</issue>
  <fpage>1999</fpage>
  <lpage>-2005</lpage>
</bibl>

<bibl id="B42">
  <title><p>{F}alse positive peaks in {C}h{IP}-seq and other sequencing-based
  functional assays caused by unannotated high copy number regions</p></title>
  <aug>
    <au><snm>Pickrell</snm><fnm>J.K.</fnm></au>
    <au><snm>Gaffney</snm><fnm>D.J.</fnm></au>
    <au><snm>Gilad</snm><fnm>Y.</fnm></au>
    <au><snm>Pritchard</snm><fnm>J.K.</fnm></au>
  </aug>
  <source>Bioinformatics</source>
  <publisher>Oxford Univ Press</publisher>
  <pubdate>2011</pubdate>
  <volume>27</volume>
  <issue>15</issue>
  <fpage>2144</fpage>
  <lpage>-2146</lpage>
</bibl>

<bibl id="B43">
  <title><p>{Methylome analysis using MeDIP-seq with low DNA
  concentrations}</p></title>
  <aug>
    <au><snm>Taiwo</snm><fnm>O</fnm></au>
    <au><snm>Wilson</snm><fnm>G A</fnm></au>
    <au><snm>Morris</snm><fnm>T</fnm></au>
    <au><snm>Seisenberger</snm><fnm>S.</fnm></au>
    <au><snm>Reik</snm><fnm>W</fnm></au>
    <au><snm>Pearce</snm><fnm>D</fnm></au>
    <au><snm>Beck</snm><fnm>S</fnm></au>
    <au><snm>Butcher</snm><fnm>L M</fnm></au>
  </aug>
  <source>Nature Protocols</source>
  <pubdate>2012</pubdate>
  <volume>7</volume>
  <issue>4</issue>
  <fpage>617</fpage>
  <lpage>-636</lpage>
</bibl>

<bibl id="B44">
  <title><p>{Genome-wide DNA methylation profiling of non-small cell lung
  carcinomas}</p></title>
  <aug>
    <au><snm>Carvalho</snm><fnm>R. H.</fnm></au>
    <au><snm>Haberle</snm><fnm>V.</fnm></au>
    <au><snm>Hou</snm><fnm>J.</fnm></au>
    <au><snm>Gent</snm><fnm>T.</fnm></au>
    <au><snm>Thongjuea</snm><fnm>S.</fnm></au>
    <au><snm>Ijcken</snm><fnm>W.</fnm></au>
    <au><snm>Kockx</snm><fnm>C.</fnm></au>
    <au><snm>Brouwer</snm><fnm>R.</fnm></au>
    <au><snm>Rijkers</snm><fnm>E.</fnm></au>
    <au><snm>Sieuwerts</snm><fnm>A.</fnm></au>
    <au><snm>Foekens</snm><fnm>J.</fnm></au>
    <au><snm>Vroonhoven</snm><fnm>M.</fnm></au>
    <au><snm>Aerts</snm><fnm>J.</fnm></au>
    <au><snm>Grosveld</snm><fnm>F.</fnm></au>
    <au><snm>Lenhard</snm><fnm>B.</fnm></au>
    <au><snm>Philipsen</snm><fnm>S.</fnm></au>
  </aug>
  <source>Epigenetics Chromatin</source>
  <pubdate>2012</pubdate>
  <volume>5</volume>
  <issue>1</issue>
  <fpage>9</fpage>
</bibl>

<bibl id="B45">
  <title><p>{Removing technical variability in RNA-seq data using conditional
  quantile normalization}</p></title>
  <aug>
    <au><snm>Hansen</snm><fnm>K. D.</fnm></au>
    <au><snm>Irizarry</snm><fnm>R. A.</fnm></au>
    <au><snm>Wu</snm><fnm>Z.</fnm></au>
  </aug>
  <source>Biostatistics</source>
  <pubdate>2012</pubdate>
  <volume>13</volume>
  <issue>2</issue>
  <fpage>204</fpage>
  <lpage>-216</lpage>
</bibl>

<bibl id="B46">
  <title><p>{S}tatistical quantification of methylation levels by
  next-generation sequencing</p></title>
  <aug>
    <au><snm>Wu</snm><fnm>G</fnm></au>
    <au><snm>Yi</snm><fnm>N</fnm></au>
    <au><snm>Absher</snm><fnm>D</fnm></au>
    <au><snm>Zhi</snm><fnm>D</fnm></au>
  </aug>
  <source>PLoS ONE</source>
  <publisher>Public Library of Science</publisher>
  <pubdate>2011</pubdate>
  <volume>6</volume>
  <issue>6</issue>
  <fpage>e21034</fpage>
</bibl>

<bibl id="B47">
  <title><p>{G}aussian {M}arkov {R}andom {F}ields: {T}heory and
  {A}pplications</p></title>
  <aug>
    <au><snm>Rue</snm><fnm>H.</fnm></au>
    <au><snm>Held</snm><fnm>L.</fnm></au>
  </aug>
  <publisher>London: Chapman \& Hall/CRC Press</publisher>
  <pubdate>2005</pubdate>
</bibl>

<bibl id="B48">
  <title><p>{R}epitools: an {R} package for the analysis of enrichment-based
  epigenomic data</p></title>
  <aug>
    <au><snm>Statham</snm><fnm>AL</fnm></au>
    <au><snm>Strbenac</snm><fnm>D</fnm></au>
    <au><snm>Coolen</snm><fnm>MW</fnm></au>
    <au><snm>Stirzaker</snm><fnm>C</fnm></au>
    <au><snm>Clark</snm><fnm>SJ</fnm></au>
    <au><snm>Robinson</snm><fnm>MD</fnm></au>
  </aug>
  <source>Bioinformatics</source>
  <pubdate>2010</pubdate>
  <volume>26</volume>
  <issue>13</issue>
  <fpage>1662</fpage>
  <lpage>-1663</lpage>
</bibl>

<bibl id="B49">
  <title><p>minfi: {A}nalyze {I}llumina's 450k methylation arrays</p></title>
  <aug>
    <au><snm>Hansen</snm><fnm>KD</fnm></au>
    <au><snm>Aryee</snm><fnm>M</fnm></au>
  </aug>
  <note>R package version 1.3.3</note>
</bibl>

<bibl id="B50">
  <title><p>{A} scaling normalization method for differential expression
  analysis of {RNA-seq} data</p></title>
  <aug>
    <au><snm>Robinson</snm><fnm>M D</fnm></au>
    <au><snm>Oshlack</snm><fnm>A</fnm></au>
  </aug>
  <source>Genome Biology</source>
  <pubdate>2010</pubdate>
  <volume>11</volume>
  <issue>3</issue>
  <fpage>R25</fpage>
</bibl>

<bibl id="B51">
  <title><p>{PICNIC}: an algorithm to predict absolute allelic copy number
  variation with microarray cancer data</p></title>
  <aug>
    <au><snm>Greenman</snm><fnm>CD</fnm></au>
    <au><snm>Bignell</snm><fnm>G</fnm></au>
    <au><snm>Butler</snm><fnm>A</fnm></au>
    <au><snm>Edkins</snm><fnm>S</fnm></au>
    <au><snm>Hinton</snm><fnm>J</fnm></au>
    <au><snm>Beare</snm><fnm>D</fnm></au>
    <au><snm>Swamy</snm><fnm>S</fnm></au>
    <au><snm>Santarius</snm><fnm>T</fnm></au>
    <au><snm>Chen</snm><fnm>L</fnm></au>
    <au><snm>Widaa</snm><fnm>S</fnm></au>
    <au><snm>Futreal</snm><fnm>PA</fnm></au>
    <au><snm>Stratton</snm><fnm>MR</fnm></au>
  </aug>
  <source>Biostatistics</source>
  <pubdate>2010</pubdate>
  <volume>11</volume>
  <issue>1</issue>
  <fpage>164</fpage>
  <lpage>175</lpage>
</bibl>

</refgrp>
} 


\ifthenelse{\boolean{publ}}{\end{multicols}}{}


\clearpage
\section*{Figures}
\subsection*{Figure 1 - SssI read depth versus CpG-density together with prior predictive distribution}\label{fig:prior}
\begin{center}
\includegraphics[width=.5\textwidth]{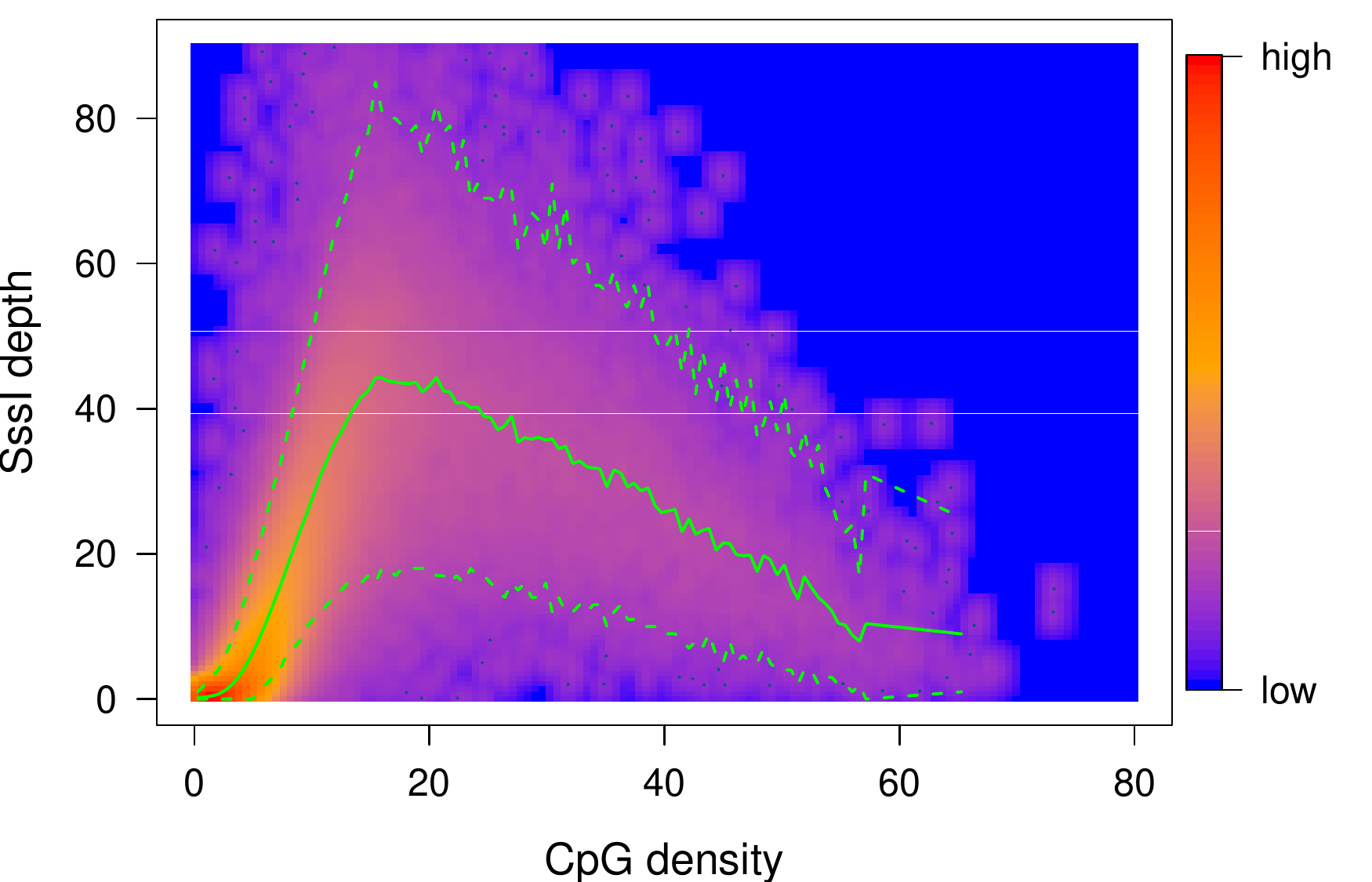}
\end{center}
Smoothed color density representation of SssI read depth versus
CpG-density together with mean (green solid line), $2.5\%$ and
$97.5\%$-quantile (green dashed lines) of the prior predictive
distribution for the SssI control sample. The parameters of this
negative binomial distribution are derived using an empirical Bayes
approach by maximizing the joint marginal distribution of the
\ac{IMR-90} and SssI control counts stratified into $100$
CpG-density groups. Only counts from bins with a mappability larger
than $0.75$ were considered.

\subsection*{Figure 2 - Example data tracks for IMR-90 chromosome 7}

\begin{center}
\includegraphics[width=\textwidth]{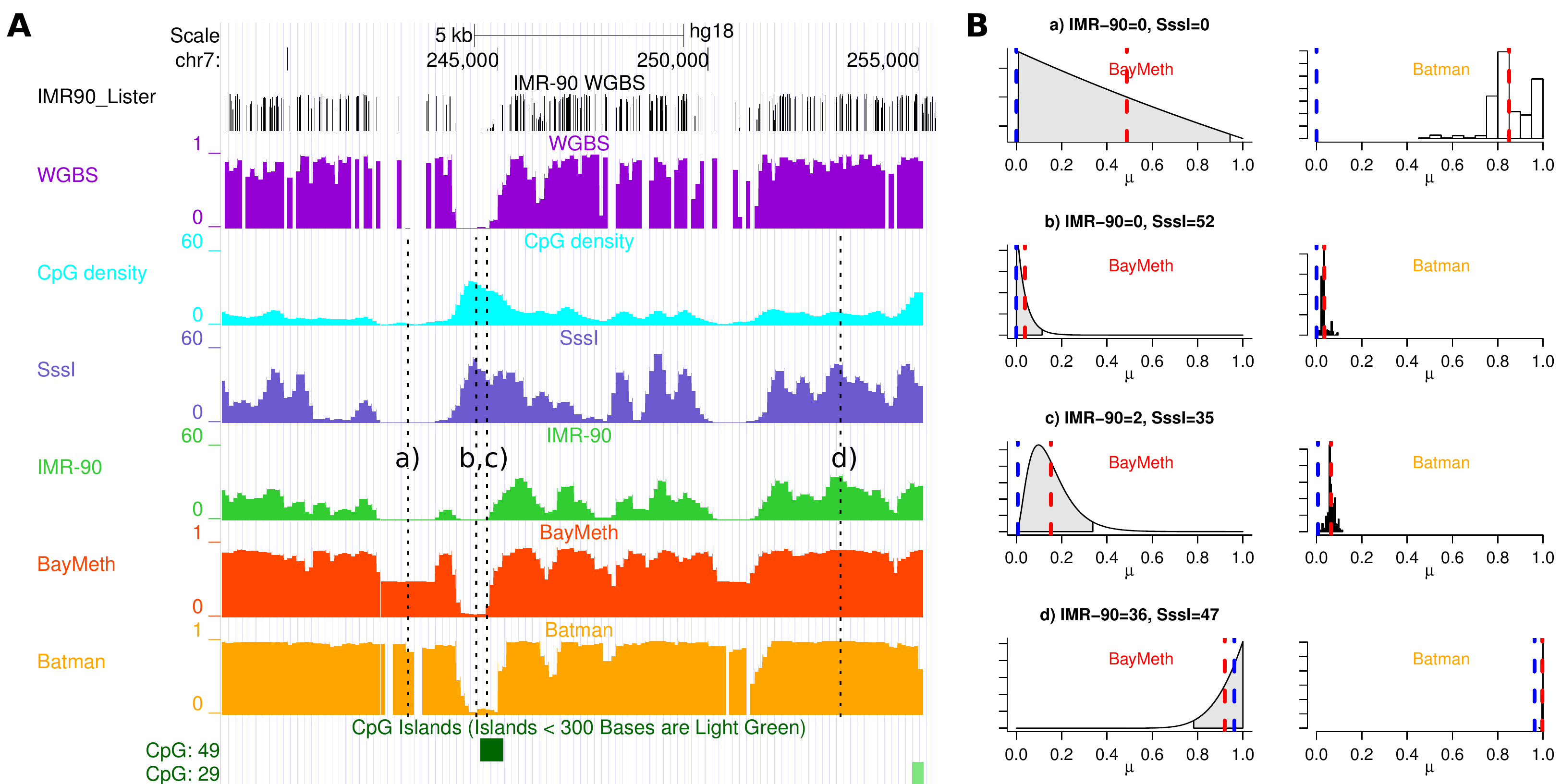}
\end{center}
Panel A: Shown are the \ac{WGBS} methylome (black) per CpG-site and
per 100bp bin (purple) as obtained by Lister and others
\cite{lister-etal-2009}. CpG-density (light blue), and read counts
for SssI-treated DNA (blue) and \ac{IMR-90} cells (green) obtained
by \ac{MBD}-seq  based on 100bp non-overlapping bins are shown.
Methylation estimates for BayMeth (red) and Batman (orange) are provided.
Panel B: For 4 specific bins of panel A (denoted a, b, c, d) detailed
posterior information of BayMeth and Batman is provided. For BayMeth
posterior marginals together with $95\%$ highest-posterior-density
(HPD) credible intervals 
(grey-shaded) are shown.
The posterior samples obtained by Batman are plotted as histograms. For both
approaches the posterior mean is indicated (red dashed line) together
with the ``true'' WGBS derived methylation estimate (blue dashed line).

\subsection*{Figure 3 - Regional methylation estimates for IMR-90 chromosome 7}
\begin{center}
\includegraphics[width=\textwidth]{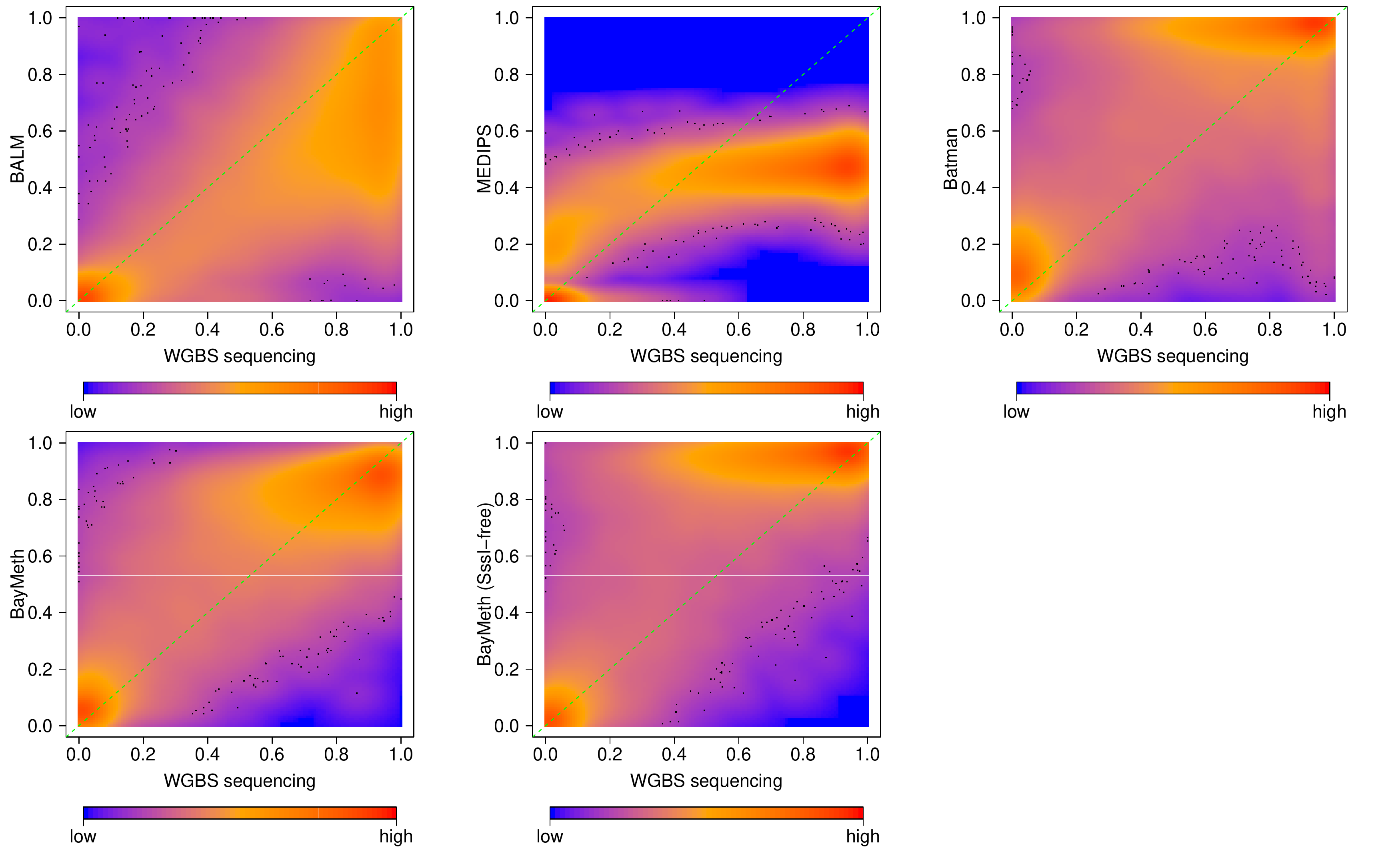}
\end{center}
Smoothed color density representation of regional \ac{DNAme} estimates
of BALM, MEDIPS, Batman, BayMeth and BayMeth ignoring SssI information, respectively, plotted against
WGBS methylation levels for the 75\% of bins with the largest depth
in the truth (cutoff are $33$ reads) where the depth in the SssI
control is $(27, 168]$. In addition the y equals x line
(green dashed line) is shown. Black points indicate outliers.

\subsection*{Figure 4 - Coverage probabilities stratified by CpG island status and true methylation level}
\begin{center}
\includegraphics[width=\textwidth]{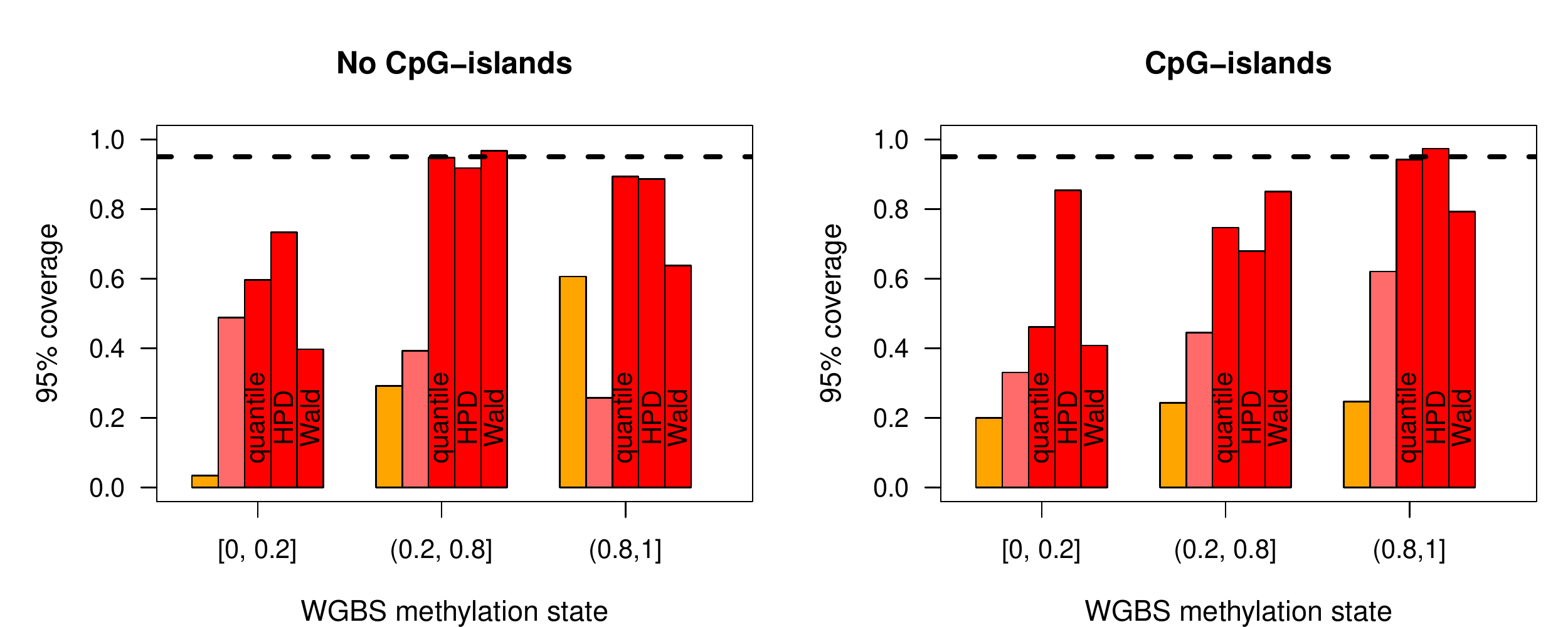}
\end{center}
Coverage probabilities (frequency in which the true value is within
a predefined credible interval) at 
$95\%$ level are shown for the 75\% of bins with the largest depth
in the truth (cutoff are $33$ reads) for Batman (orange), BayMeth ignoring
SssI control information (light red), and BayMeth (red). Three different
types (quantile-based, Wald, HPD) of credible intervals are shown for BayMeth,
while for Batman and the SssI-free version of BayMeth only quantile-based intervals
are available. MEDIPS and Balm do not return any uncertainty estimates.
The nominal coverage value is indicated (black dashed line)
as a reference. Genomic regions are stratified by CpG-density using
the threshold of $12.46$ which separates CpG islands from non-CpG
islands, compare Figure~S1 of Additional file~2. Further stratification by
the true methylation level as derived from \ac{WGBS}
\cite{lister-etal-2009} is provided.

\subsection*{Figure 5 - Relation between copy number state and regional affinity enrichment}\label{fig:cp_chr13}
\begin{center}
\includegraphics[width=\textwidth]{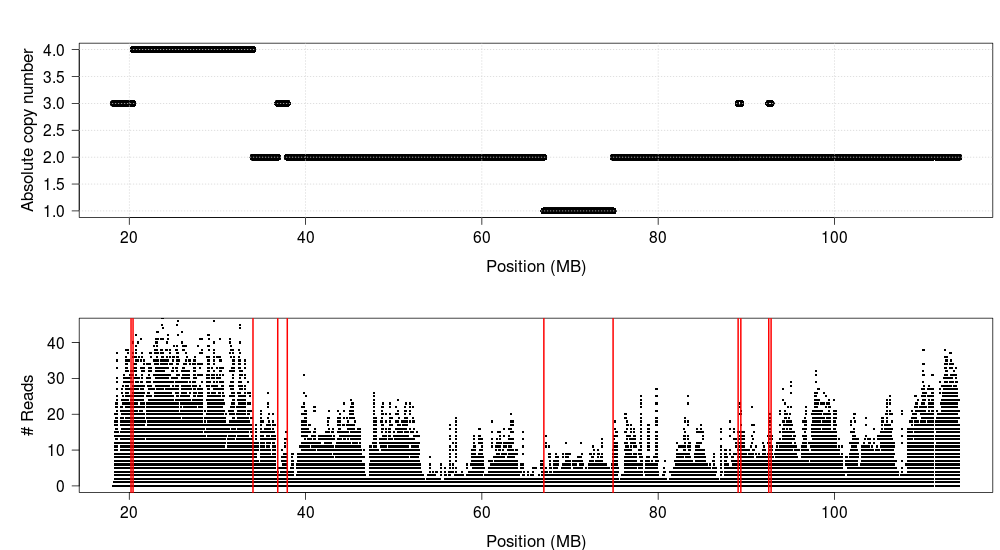}
\end{center}
Top: Copy number estimates of \ac{LNCaP} cell line obtained by the
PICNIC \cite{greenman-etal-2010} algorithm for 100bp bins across
human chromosome 13 with a mappability of at least 75\%.
Bottom: Read counts of affinity capture sequencing data for the
same bins.

\subsection*{Figure 6 - Bias of LNCaP methylation estimates compared to 450k array beta values}
	\begin{center}
	\includegraphics[width=\textwidth]{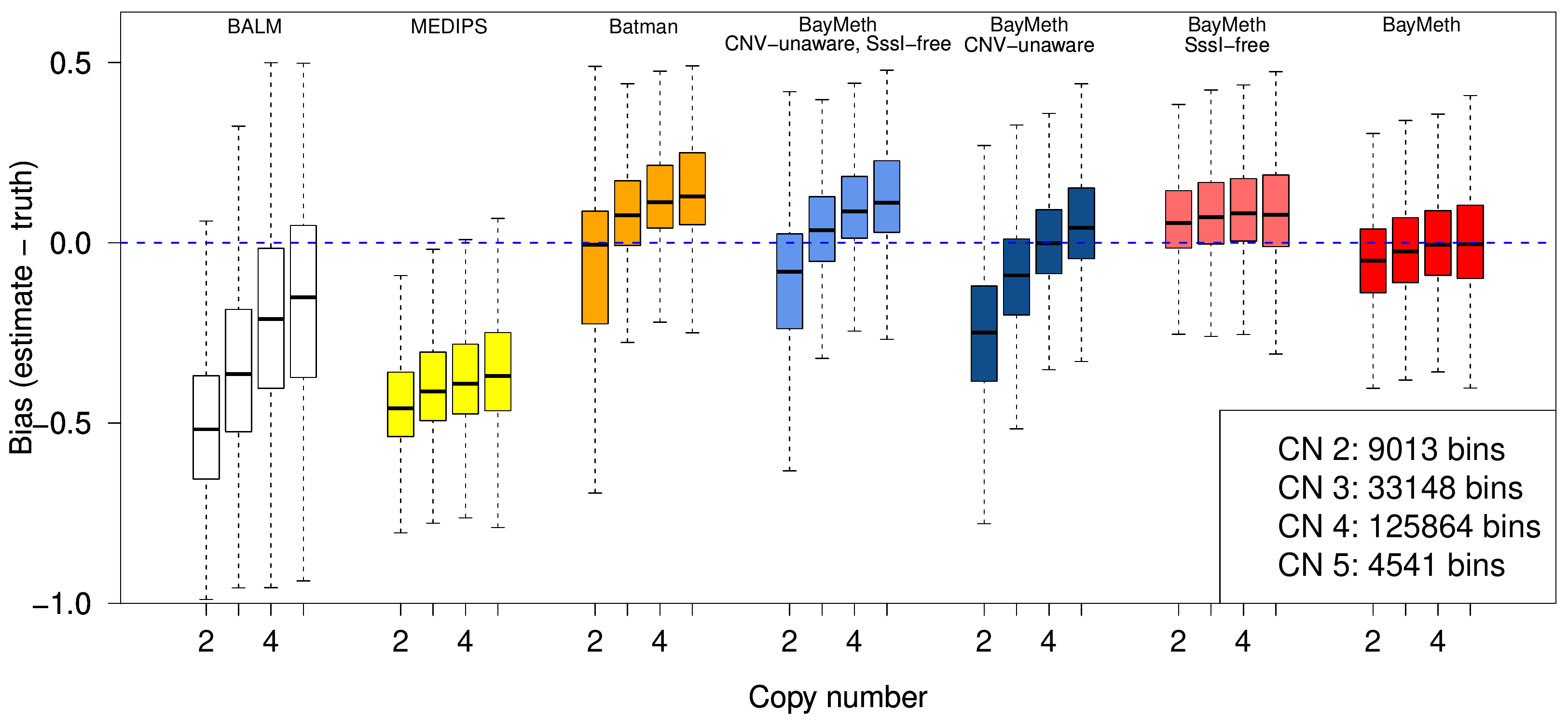}
	\end{center}
	
Boxplot of bias (Estimated methylation level - 450K array beta value)
for BALM (white), MEDIPS (yellow), Batman (orange), CNV-unaware and
SssI-free BayMeth (light blue), CNV-unaware BayMeth (dark blue),
SssI-free but CNV-aware BayMeth (light red) and CNV-aware BayMeth
(red) stratified by the most prominent copy number.
(Outliers are not shown.) Taking SssI information into account a
uniform prior for the methylation level was used, in the SssI-free
version a Dirac-Beta-Dirac mixture with weights fixed to
0.1, 0.8, 0.1 was used. The results are shown genome-wide for
100bp bins with at least $75\%$ mappability and where the true
methylation estimate is larger than $0.5$. A threshold of $13$ is
applied for the depth of SssI. The blue dashed line indicates a bias
of zero.

\subsection*{Figure 7 - Effect of adjusting for CNV in LNCaP cell line}

\begin{center}
	\includegraphics[width=\textwidth]{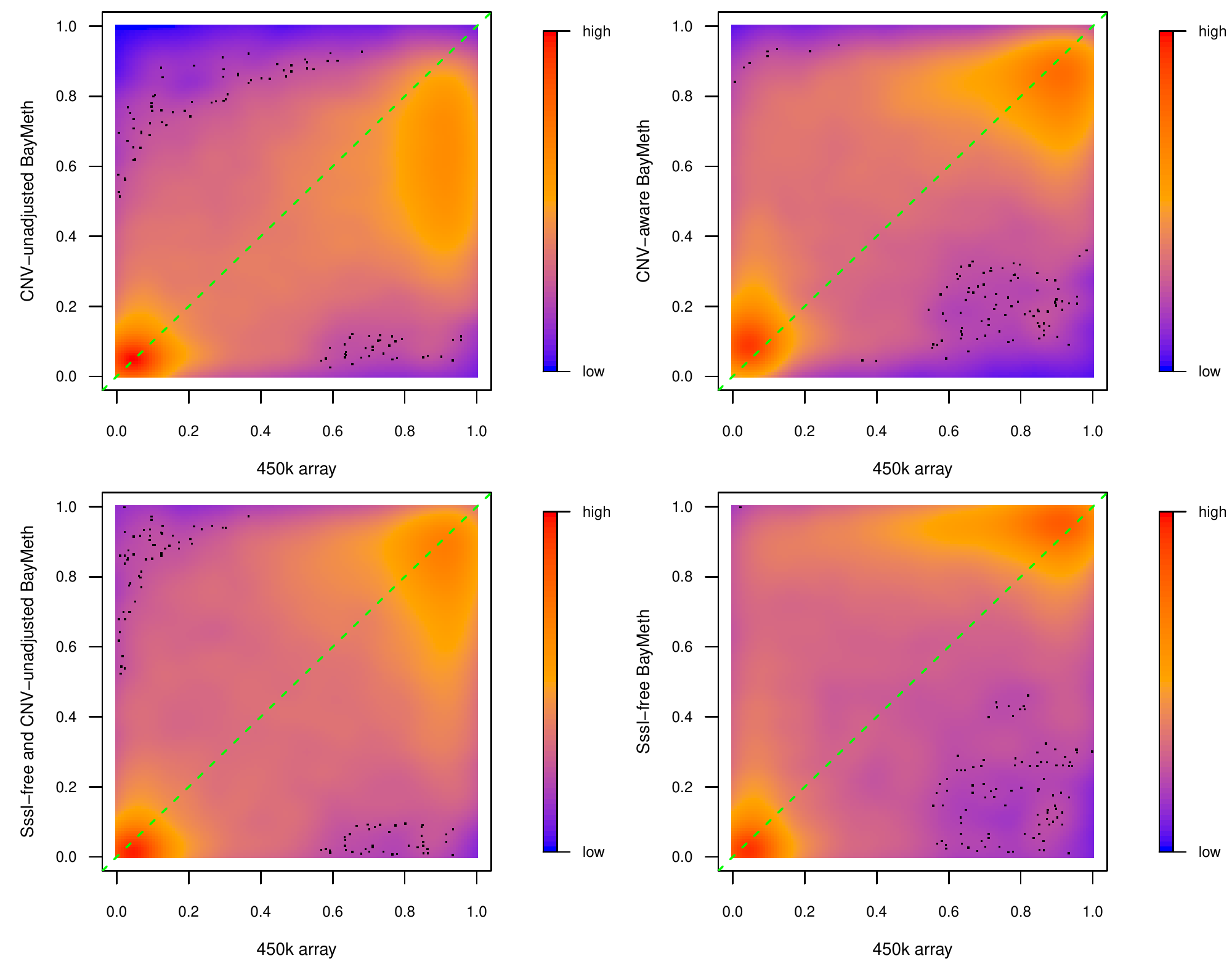}
\end{center}
Smoothed color density representation of methylation estimates for
copy number state two derived by BayMeth compared to 450k array beta
values.   A threshold of $13$ is applied for the depth of SssI, which
leads to $61969$ bins, of which we have for $18010$ 100bp-bins a beta
values and BayMeth estimate.  In addition the y equals x line
(green dashed line) is shown. Black points indicate outliers.
Top-Left: CNV-unaware BayMeth; Top-Right: CNV-aware BayMeth;
Bottom-Left: SssI-free and CNV-unaware BayMeth;
Bottom-Right: SssI-free BayMeth.

\subsection*{Figure 8 - Comparison of raw IMR-90 data and methylation estimates obtained by different methylation kits}
\begin{center}
 \includegraphics[width=\textwidth]{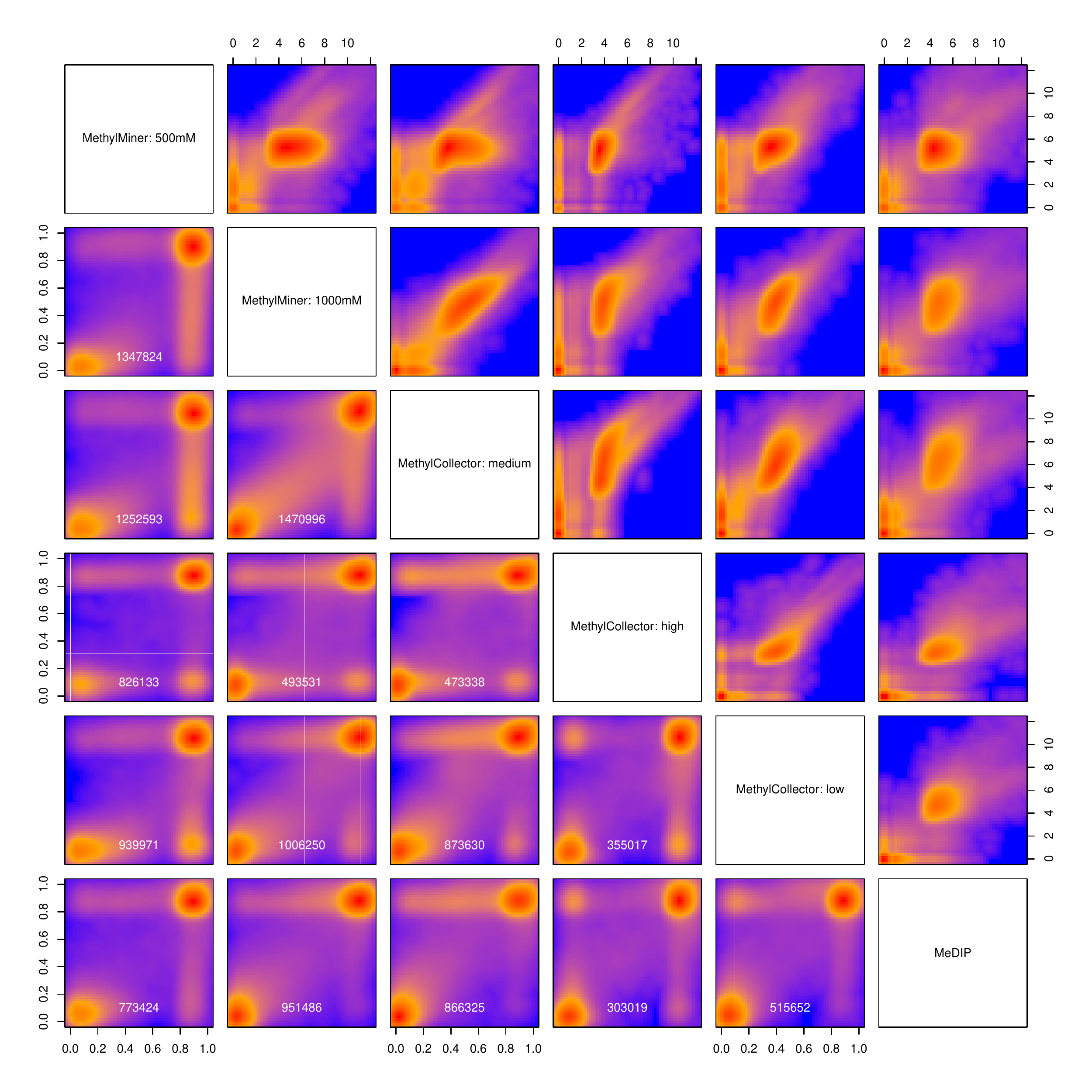}
\end{center}
Genomic bins (100bp), with a mappability larger than $75\%$, are selected for which the predicted HPD credible 
interval width is smaller than $0.4$. For these bins the upper triangular panels show a smoothed color density 
representation (going from blue representing low density to red for high density) of the raw counts and the 
lower triangular panels the estimated methylation levels obtained by different 
methylation kits against each other. The number of bins included is given in the panels of the lower triangular in white.


\subsection*{Figure 9 - Regional methylation estimates for samples of Bock data}

\begin{center}
	\includegraphics[width=\textwidth]{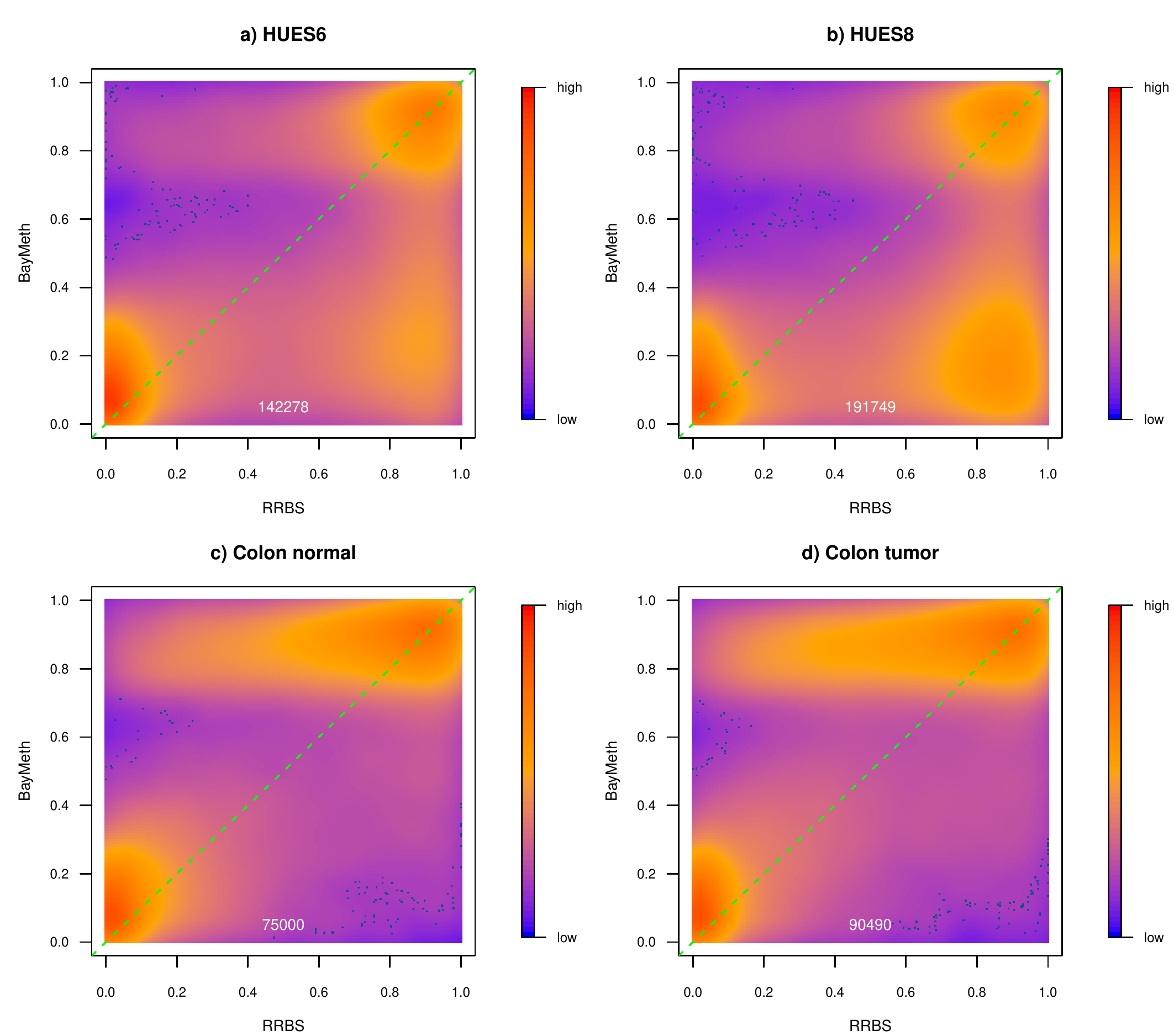}
\end{center}

Smoothed color density representation of regional DNAme estimates of BayMeth, plotted 
against RRBS methylation levels, where the estimated 
standard deviation of BayMeth is smaller than
$0.15$ for bins with more than 20 reads for RRBS and at least 
a depth of 10 in the SssI control. The number of bins included in the plot is shown at the bottom center
of the panels.

\subsection*{Figure 10 - Regional variance estimates versus SssI control for Bock data}

\begin{center}
	\includegraphics[width=\textwidth]{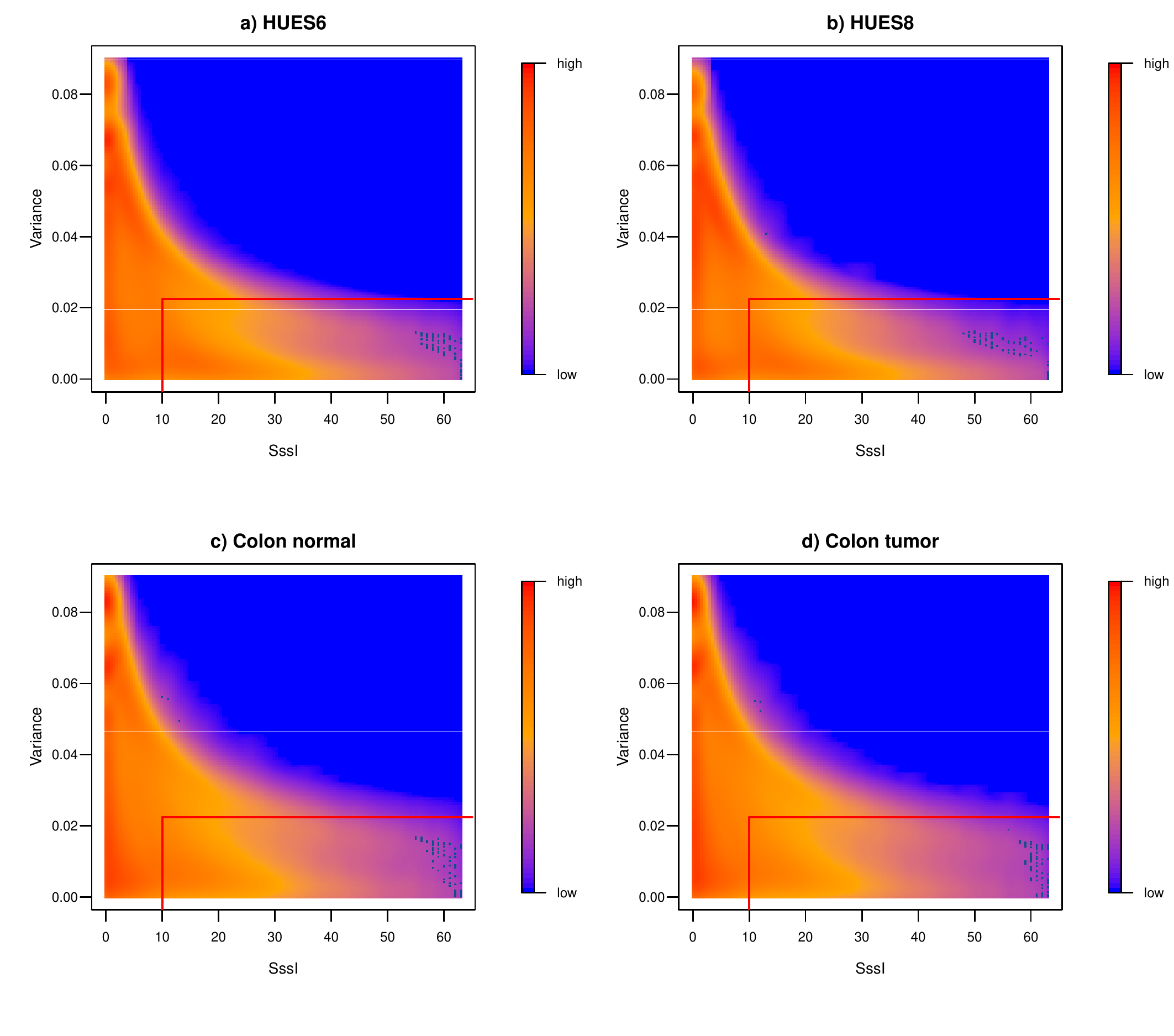}
\end{center}
Smooth color density representation of variance 
estimates obtained by BayMeth versus number of reads in the SssI control for a read depth 
larger than $20$ in RRBS. The red box contains the bins used in Figure 9 having at least
a depth of 10 in SssI and a standard deviation smaller than $0.15$, i.e. a variance smaller
than $0.025$.


\clearpage
 \section*{Tables}

\subsection*{Table 1 - Performance assessment for IMR-90 analysis (chromosome 7)}
Results are shown for  bins with a truth depth larger than the
$25\%$ quantile (cutoff are $33$ reads), stratified into
five groups by SssI depth. Shown are the number of bins per group,
mean bias,  MSE, Spearman correlation and coverage probabilities
at 
$95\%$ level.

    \par
    \mbox{
\begin{tabular}{rrrrrrrrr}
  SssI depth & \#Bins & Method & Bias & MSE & Cor & Wald & HPD & quantile \\ 
   \hline
$[0,4]$ & 305638 & BayMeth & -0.04 & 0.08 & 0.36 & 0.74 & 0.89 & 0.89 \\ 
   &  & BayMeth (SssI-free) & -0.19 & 0.20 & 0.23 & --- & --- & 0.24 \\ 
   &  & Batman & 0.22 & 0.14 & 0.31 & --- & --- & 0.43 \\ 
   &  & MEDIPS & -0.38 & 0.26 & 0.29 & --- & --- & --- \\ 
   &  & BALM & -0.48 & 0.33 & 0.32 & --- & --- & --- \\ 
   \hline
$(4,7]$ & 22196 & BayMeth & 0.05 & 0.05 & 0.65 & 0.84 & 0.88 & 0.87 \\ 
   &  & BayMeth (SssI-free) & -0.01 & 0.08 & 0.42 & --- & --- & 0.68 \\ 
   &  & Batman & 0.16 & 0.07 & 0.61 & --- & --- & 0.34 \\ 
   &  & MEDIPS & -0.23 & 0.11 & 0.45 & --- & --- & --- \\ 
   &  & BALM & -0.27 & 0.15 & 0.60 & --- & --- & --- \\ 
   \hline
$(7,14]$ & 28871 & BayMeth & 0.06 & 0.04 & 0.69 & 0.84 & 0.86 & 0.86 \\ 
   &  & BayMeth (SssI-free) & 0.02 & 0.05 & 0.57 & --- & --- & 0.79 \\ 
   &  & Batman & 0.16 & 0.07 & 0.65 & --- & --- & 0.28 \\ 
   &  & MEDIPS & -0.21 & 0.10 & 0.49 & --- & --- & --- \\ 
   &  & BALM & -0.21 & 0.11 & 0.66 & --- & --- & --- \\ 
   \hline
$(14,27]$ & 28928 & BayMeth & 0.05 & 0.03 & 0.76 & 0.81 & 0.85 & 0.82 \\ 
   &  & BayMeth (SssI-free) & 0.08 & 0.04 & 0.72 & --- & --- & 0.70 \\ 
   &  & Batman & 0.15 & 0.06 & 0.73 & --- & --- & 0.23 \\ 
   &  & MEDIPS & -0.20 & 0.09 & 0.59 & --- & --- & --- \\ 
   &  & BALM & -0.15 & 0.07 & 0.75 & --- & --- & --- \\ 
   \hline
$(27,168]$ & 28719 & BayMeth & 0.02 & 0.03 & 0.79 & 0.73 & 0.86 & 0.78 \\ 
   &  & BayMeth (SssI-free) & 0.11 & 0.04 & 0.77 & --- & --- & 0.48 \\ 
   &  & Batman & 0.11 & 0.05 & 0.75 & --- & --- & 0.20 \\ 
   &  & MEDIPS & -0.22 & 0.10 & 0.67 & --- & --- & --- \\ 
   &  & BALM & -0.14 & 0.06 & 0.76 & --- & --- & --- \\ 
   \hline
\end{tabular}
}

\subsection*{Table 2 - Copy number specific offset}
Copy number specific offsets defined as $f \times \frac{\cn_i}{\ccn}$
derived for 100bp non-overlapping bins of LNCaP autosomes, which have
a mappability of at least 75\%. Note, that $f$ is only derived based
on bins with the most common copy number state four.
 \par
    \mbox{
\begin{tabular}{rrrrrrrrr}
  \hline
 & 1 & 2 & 3 & 4 & 5 & 6 & 7 & 8 \\ 
  \hline
Combined offset & 0.178 & 0.356 & 0.534 & 0.712 & 0.889 & 1.067 & 1.245 & 1.423 \\ 
   \hline
\end{tabular}
}

\subsection*{Table 3 - Performance assessment for LNCaP analysis by copy number}

Results are shown for 100bp-bins with a mappability of at least $0.75$
stratified into the four most frequent copy number states. A threshold
of $13$ is applied for the depth of the SssI-control. Four BayMeth
four different variations are shown, depending on whether SssI-control
information is used and whether copy number information is integrated.
Taking SssI information into account a uniform prior for the
methylation level was used, in the SssI-free version a
Dirac-Beta-Dirac mixture with weights fixed to 0.1, 0.8, 0.1 was used.
Shown are the number of bins per copy number state, mean bias,  MSE
and Spearman correlation.
    \par
    \mbox{
\begin{tabular}{lrlrrr}
  Copy number & \#Bins & Method & Bias & MSE & Cor \\ 
   \hline
2 & 18010 & BayMeth & 0.04 & 0.04 & 0.78 \\ 
   &  & BayMeth (SssI-free) & 0.08 & 0.05 & 0.79 \\ 
   &  & BayMeth (CNV-unaware) & -0.11 & 0.06 & 0.78 \\ 
   &  & BayMeth (SssI-free, CNV-unaware) & -0.05 & 0.05 & 0.79 \\ 
   &  & Batman & 0.03 & 0.06 & 0.74 \\ 
   &  & MEDIPS & -0.23 & 0.11 & 0.76 \\ 
   &  & BALM & -0.29 & 0.16 & 0.78 \\ 
   \hline
3 & 65982 & BayMeth & 0.05 & 0.04 & 0.80 \\ 
   &  & BayMeth (SssI-free) & 0.09 & 0.05 & 0.80 \\ 
   &  & BayMeth (CNV-unaware) & -0.01 & 0.04 & 0.80 \\ 
   &  & BayMeth (SssI-free, CNV-unaware) & 0.05 & 0.04 & 0.80 \\ 
   &  & Batman & 0.11 & 0.06 & 0.77 \\ 
   &  & MEDIPS & -0.19 & 0.09 & 0.76 \\ 
   &  & BALM & -0.20 & 0.10 & 0.79 \\ 
   \hline
4 & 256074 & BayMeth & 0.05 & 0.04 & 0.81 \\ 
   &  & BayMeth (SssI-free) & 0.09 & 0.05 & 0.81 \\ 
   &  & BayMeth (CNV-unaware) & 0.05 & 0.04 & 0.81 \\ 
   &  & BayMeth (SssI-free, CNV-unaware) & 0.11 & 0.06 & 0.81 \\ 
   &  & Batman & 0.16 & 0.08 & 0.79 \\ 
   &  & MEDIPS & -0.17 & 0.09 & 0.76 \\ 
   &  & BALM & -0.12 & 0.07 & 0.80 \\ 
   \hline
5 & 11790 & BayMeth & 0.04 & 0.03 & 0.83 \\ 
   &  & BayMeth (SssI-free) & 0.07 & 0.05 & 0.82 \\ 
   &  & BayMeth (CNV-unaware) & 0.09 & 0.04 & 0.83 \\ 
   &  & BayMeth (SssI-free, CNV-unaware) & 0.12 & 0.06 & 0.82 \\ 
   &  & Batman & 0.18 & 0.08 & 0.80 \\ 
   &  & MEDIPS & -0.12 & 0.07 & 0.80 \\ 
   &  & BALM & -0.08 & 0.05 & 0.82 \\ 
   \hline
\end{tabular}
}

\clearpage

\section*{Additional file 1 --- Statistical details of BayMeth}

The methodology of BayMeth is roughly divided into two steps: 1) An empirical Bayes procedure to derive parameters
for the prior distributions of all parameters in the model. 2) The analytical derivation of the posterior marginal distribution,
posterior expectation and variance for the methylation levels. Credible intervals
are derived numerically from the posterior marginal distribution. Recall the model formulation 
provided in the main text:
\begin{align*}
  y_{iS}|\mu_i, \lambda_i &\sim \Po\left(f \times \frac{\cn_i}{\ccn}
	\times \mu_i \times \lambda_i\right) \text{, and}  \\
  y_{iC}|\lambda_i &\sim \Po(\lambda_i), 
\end{align*}

\section*{Prior specification}
For $\lambda_i$ we assume a gamma prior distribution with parameters $\alpha$ and $\beta$:
\begin{equation*}
  \lambda_i\mid \alpha, \beta = \frac{\beta^{\alpha}}{\Gamma(\alpha)}
  \lambda_i^{\alpha - 1} \exp(-\beta\lambda_i), \lambda_i > 0, \alpha, \beta > 0.
\end{equation*}
The methylation level $\mu_i$ has support from zero to one. We consider two groups of prior distributions:
\begin{itemize}
 \item a mixture of beta distributions, i.e., $\mu_i \sim \sum_{m=1}^M w_m \Be(a_m, b_m)$, where in its
 simplest form $M=1$. (The default configuration of BayMeth is $M=1$ and ($a=a_m=b=b_m=1$), i.e., a uniform
 distribution from zero to one.)
 \item a mixture of a point mass at zero, a beta distribution and a point mass at one. 
 We call this the Dirac-Beta-Dirac (DBD) prior distribution, which has the density
 \begin{equation}
    p(\mu_i) = w_0 \delta_0 + w_1 \Be(\mu_i; a, b) + w_2 \delta_1, \label{eq:DBD}
 \end{equation}
where
\begin{align*}
  \delta_0 &= \begin{cases} 0 & \text{if $\mu_i \ne 0$}\\
    1 & \text{if $\mu_i = 0$} \end{cases},&
  \delta_1 = \begin{cases} 0 & \text{if $\mu_i \ne 1$}\\
    1 & \text{if $\mu_i = 1$} \end{cases}
\end{align*}
and $w_0 + w_1 + w_2 = 1$.

\end{itemize}

\section*{Marginal distribution}

The empirical Bayes approach is based on the maximization of the marginal distribution.
For ease of readability let $ E= f\times \tfrac{\text{cn}_i}{\text{ccn}}$.
The joint marginal distribution of $y_{iS}, y_{iC}$ results as:
\begin{linenomath*}
\begin{equation*}
\begin{split}
  p&(y_{iS}, y_{iC}) = \int \int p(y_{iS} | \mu_i, \lambda_i) p(y_{iC}|\lambda_i) p(\lambda_i) p(\mu_i) d\lambda_i d\mu_i\\
    &= \int_0^1 p(\mu_i)  \left[\int_0^\infty p(y_{iS} | \mu_i, \lambda_i)p(y_{iC}|\lambda_i) p(\lambda_i) d\lambda_i\right]d\mu_i\\
    &= \int_0^1  p(\mu_i) \left[\int_0^\infty \frac{(E\mu_i  \lambda_i)^{y_{iS}}\lambda_i^{y_{iC}}}{y_{iS}!y_{iC}!}
    \exp\left(-E\mu_i\lambda_i\right) \times \exp(-\lambda_i) \times
      \frac{\beta^{\alpha}}{\Gamma(\alpha)} \lambda_i^{\alpha -1}\exp(-\beta \lambda_i)d\lambda_i\right]d\mu_i\\
    &= \int_0^1  p(\mu_i) \left[\frac{(E\mu_i)^{y_{iS}}}{y_{iS}!y_{iC}!}\frac{\beta^{\alpha}}{\Gamma(\alpha)}
    \int_0^\infty \lambda_i^{y_{iS}+y_{iC} + \alpha -1}\exp\left(-(E\mu_i + 1 + \beta) \lambda_i\right)d\lambda_i\right]d\mu_i\\
    &= \int_0^1  p(\mu_i) \left[\frac{(E\mu_i)^{y_{iS}}}{y_{iS}!y_{iC}!}\frac{\beta^{\alpha}}{\Gamma(\alpha)}
      \frac{\Gamma(y_{iS}+y_{iC} + \alpha)}{(E\mu_i + 1+ \beta)^{y_{iS}+y_{iC}+\alpha}}\right]d\mu_i\\
    &= \frac{E^{y_{iS}}}{y_{iS}!y_{iC}!}\frac{\beta^{\alpha}}{\Gamma(\alpha)}\Gamma(y_{iS} + y_{iC}+\alpha)
      \int_0^1 p(\mu_i)\frac{\mu_i^{y_{iS}}}{(E\mu_i + 1 +  \beta)^{y_{iS}+ y_{iC} + \alpha}}d\mu_i
\end{split}
\end{equation*}
\end{linenomath*}
What is left, is to choose a prior for $\mu_i$, that means to specify $p(\mu_i)$.

\section*{(Mixture of) beta distribution for the methylation level}

Consider the simple case where the number of mixture components
is one, so that $\mu_i \sim \Be(a,b)$,
i.e.~$p(\mu_i)=\frac{\Gamma(a+b)}{\Gamma(a)\Gamma(b)}\mu_i^{a-1}(1-\mu_i)^{b-1}$, $a, b > 0$. (For a uniform distribution
$a=b=1$). Then
\begin{equation}\label{eq:deriv_bb_nbd}
\begin{split}
  p&(y_{iS}, y_{iC}) = \frac{\Gamma(y_{iS} + y_{iC}+\alpha)}{\Gamma(\alpha)y_{iS}!y_{iC}!}\frac{\Gamma(a + b)}{\Gamma(a)\Gamma(b)}
      E^{y_{iS}}\beta^{\alpha}\int_0^1 \frac{\mu_i^{y_{iS} + a -1} (1-\mu_i)^{b-1}}{(E\mu_i + 1 +  \beta)^{y_{iS}+ y_{iC} + \alpha}}d\mu_i\\
  &= \frac{\Gamma(y_{iS} + y_{iC}+\alpha)}{\Gamma(\alpha)y_{iS}!y_{iC}!}\frac{\Gamma(a + b)}{\Gamma(a)\Gamma(b)}
      E^{y_{iS}}\frac{\beta^{\alpha}}{(E + 1+ \beta )^{y_{iS} + y_{iC} + \alpha}}
      \int_0^1 \frac{\mu_i^{y_{iS} + a -1} (1-\mu_i)^{b-1}}{\frac{(E\mu_i + 1 +  \beta)^{y_{iS}+ y_{iC} + \alpha}}{(E + 1 +  \beta)^{y_{iS}+ y_{iC} + \alpha}}}d\mu_i\\
  &= \frac{\Gamma(y_{iS} + y_{iC}+\alpha)}{\Gamma(\alpha)y_{iS}!y_{iC}!}\frac{\Gamma(a + b)}{\Gamma(a)\Gamma(b)}
      E^{y_{iS}}\frac{\beta^{\alpha}}{(\beta + 1 +E)^{y_{iS} + y_{iC} + \alpha}}
      \int_0^1 \frac{\mu_i^{y_{iS} + a -1} (1-\mu_i)^{b-1}}{\left(1-\frac{E}{E + 1 +  \beta} \cdot (1-\mu_i)\right)^{y_{iS}+ y_{iC} + \alpha}}d\mu_i\\
   &\stackrel{\star}{=} \frac{\Gamma(y_{iS} + y_{iC}+\alpha)}{\Gamma(\alpha)y_{iS}!y_{iC}!}\frac{\Gamma(a + b)}{\Gamma(a)\Gamma(b)}
       E^{y_{iS}}\frac{\beta^{\alpha}}{(\beta + 1 +E)^{y_{iS} + y_{iC} + \alpha}}
       \int_0^1 \frac{(1-t_i)^{y_{iS} + a -1} t_i^{b-1}}{\left(1-\frac{E}{E + 1 +  \beta} \cdot t_i\right)^{y_{iS}+ y_{iC} + \alpha}}dt_i\\
   &= \frac{\Gamma(y_{iS} + y_{iC}+\alpha)}{\Gamma(\alpha)y_{iS}!y_{iC}!}\left(\frac{\beta}{\beta + 1 +E}\right)^{\alpha}
 	\left(\frac{E}{\beta + 1 +E}\right)^{y_{iS}}\left(\frac{1}{\beta + 1 +E}\right)^{y_{iC}}
       \frac{\Gamma(a+b) \Gamma(y_{iS} + a)}{\Gamma(a)\Gamma(y_{iS}+a+b)}\times\\ &\hfill
 	 \hspace{3cm}_2F_1\left(y_{iS}+y_{iC}+\alpha, b; y_{iS} +a +b; \frac{E}{\beta + 1 +E}\right).
\end{split}
\end{equation}
 In the step marked with $^\star$ we substituted $(1-\mu_i)$ with
$t_i$, where $dt_i = -d\mu_i$,  to get the desired form of the Gauss
hypergeometric function (the limits of the integral stay thereby
unchanged), which is defined by:
\begin{linenomath*}
\begin{equation*}\label{eq:2F1}
  _2F_1(a,b;c;z) = \frac{\Gamma(c)}{\Gamma(b)\Gamma(c-b)}\int_0^1 t^{b-1}(1-t)^{c-b-1}(1-zt)^{-a}dt, \quad c > b > 0
\end{equation*}
\end{linenomath*}
where $|z| < 1$ is the radius of convergence
\cite[see page 558]{Abramowitz-stegun-1972}.
(Note,  $|z|=|E/(\beta + 1 + E)| < 1$ in \eqref{eq:deriv_bb_nbd},
so that convergence is granted).
Model \eqref{eq:deriv_bb_nbd} is similar to the beta binomial
(BB)/negative binomial (NB) model derived in \cite{schmittlein1985}
and \cite{fader-hardie-2000}.

Using a mixture of $M$ beta distributions as prior distribution for $\mu_i$, i.e.
$\mu_i \sim \sum_{m=1}^M w_m \Be(a_m, b_m)$, where
$0 \leq w_m \leq 1$, for all $m=1, \ldots, M$,  and $\sum_{m=1}^M w_m=1$ we get:

\begin{equation}
	p(y_{iS}, y_{iC}) = \frac{\Gamma(y_{iS} + y_{iC}+\alpha)}{\Gamma(\alpha)y_{iS}!y_{iC}!}\left(\frac{\beta}{\beta + 1 +E}\right)^{\alpha}
 	\left(\frac{E}{\beta + 1 +E}\right)^{y_{iS}}\left(\frac{1}{\beta + 1 +E}\right)^{y_{iC}} \times W\label{eq:bb_nbd}
\end{equation}
with
\begin{linenomath*}
\begin{equation*}
\begin{split}
W &= \sum_{m=1}^M\Biggl[
		w_m \cdot \frac{\Gamma(a_m+b_m) \Gamma(y_{iS} + a_m)}{\Gamma(a_m)\Gamma(y_{iS}+a_m+b_m)}\times\: _2F_1\left(y_{iS}+y_{iC}+\alpha, b_m; y_{iS} +a_m +b_m; \frac{E}{\beta + 1 +E}\right)	\Biggr].
\end{split}
\end{equation*}
\end{linenomath*}
Of note, ignoring the SssI information the marginal distribution changes to:
\begin{equation*}
\begin{split}
  p&(y_{iS}) = \int \int p(y_{iS} | \mu_i, \lambda_i) p(\lambda_i) p(\mu_i) d\lambda_i d\mu_i\\
   &=\frac{\Gamma(y_{iS} +\alpha)}{\Gamma(\alpha)y_{iS}!}\left(\frac{\beta}{\beta +E}\right)^{\alpha}
 	\left(\frac{E}{\beta +E}\right)^{y_{iS}}\times W
\end{split}
\end{equation*}
with 
\begin{equation}
\begin{split}
W &= \sum_{m=1}^M\Biggl[
	w_m \cdot \frac{\Gamma(a_m+b_m) \Gamma(y_{iS} + a_m)}{\Gamma(a_m)\Gamma(y_{iS}+a_m+b_m)}\times\: _2F_1\left(y_{iS}+\alpha, b_m; y_{iS} +a_m +b_m; \frac{E}{\beta +E}\right)	\Biggr].\label{eq:mixBetaW}
\end{split}
\end{equation}

\section*{Dirac-beta-Dirac distribution for the methylation level}

If we consider instead of a mixture beta distribution, the DBD prior
as given in Equation \eqref{eq:DBD}, we get the following
marginal distribution:
\begin{linenomath}
\begin{equation*}
\begin{split}
	p(y_{iS}, y_{iC}) = &\frac{\Gamma(y_{iS} + y_{iC}+\alpha)}{\Gamma(\alpha)y_{iS}!y_{iC}!}\left(\frac{\beta}{\beta + 1 +E}\right)^{\alpha}
 	\left(\frac{E}{\beta + 1 +E}\right)^{y_{iS}}\left(\frac{1}{\beta + 1 +E}\right)^{y_{iC}} \times W
\end{split}
\end{equation*}
\end{linenomath}
with 
\begin{linenomath}
\begin{equation*}
\begin{split}
W &= w_2 + w_1 \cdot \frac{\Gamma(a+b) \Gamma(y_{iS} + a)}{\Gamma(a)\Gamma(y_{iS}+a+b)}\times\: _2F_1\left(y_{iS}+y_{iC} + \alpha, b; y_{iS} +a +b; \frac{E}{\beta +E + 1}\right)	\Biggr].
\end{split}
\end{equation*}
\end{linenomath}
Ignoring the SssI information this marginal distribution changes to:
\begin{linenomath}
\begin{equation*}\label{eq:bb_nbd}
\begin{split}
	p(y_{iS}) = &\frac{\Gamma(y_{iS} +\alpha)}{\Gamma(\alpha)y_{iS}!}\left(\frac{\beta}{\beta +E}\right)^{\alpha}
 	\left(\frac{E}{\beta  +E}\right)^{y_{iS}} \times W
\end{split}
\end{equation*}
\end{linenomath}
with
\begin{equation*}
\begin{split}
W &= w_2 + w_1 \cdot \frac{\Gamma(a+b) \Gamma(y_{iS} + a)}{\Gamma(a)\Gamma(y_{iS}+a+b)}\times\: _2F_1\left(y_{iS} + \alpha, b; y_{iS} +a +b; \frac{E}{\beta +E}\right)	\Biggr].
\end{split}
\end{equation*}

\section*{Parameter estimation}

Independent of the prior choice for $\mu_i$, we have to determine
parameters $\alpha$ and $\beta$ of the gamma prior distribution for
$\lambda$. The default BayMeth assumes a uniform prior for $\mu_i$,
i.e. $M=1$ and $\mu_i \sim \Be(a=1, b=1)$,
and that SssI information is taken into account, therefore $\alpha$
and $\beta$ are the only parameters to determine. Under the empirical
Bayes approach, the parameters $\alpha$ and $\beta$ of Equation (2)
can be estimated using maximum likelihood. The parameters are thereby determined in a
CpG-density-dependent manner. Each 100bp bin is classified based on
its CpG-density into one of $K = 100$ non-overlapping CpG-density classes:
$\mathcal{C}_1, \ldots, \mathcal{C}_K$.
The class size $|\mathcal{C}_k|$, i.e.~the number of 100bp bins in
class $k$, is denoted by $n_k$. We derive for each class separately
the set of prior parameters using empirical Bayes leading finally to
$K$ parameter sets. The corresponding log likelihood function for
class $k$ is then given by
\begin{linenomath}
\begin{equation}\label{eq:loglik-eb}
  l(\alpha^{(k)}, \beta^{(k)} |\mathbf{y}^{(k)}_1, \mathbf{y}^{(k)}_2) =
	 \sum_{j=1}^{n_k} \log(p(y_{j1}^{(k)}, y_{j2}^{(k)} | \alpha^{(k)}, \beta^{(k)}).
\end{equation}
\end{linenomath}
Here $\mathbf{y}^{(k)}_S=(y_{1S}^{(k)}, \ldots, y_{n_kS}^{(k)})$ and $\mathbf{y}^{(k)}_C=(y_{1C}^{(k)}, \ldots, y_{n_kC}^{(k)})$ denote the read counts of the bins contained in class $\mathcal{C}_k$. Further   $\alpha^{(k)}$, $\beta^{(k)}$ denote the parameters for CpG-density class $k$.  In Equation \eqref{eq:loglik-eb}, we assume that genomic regions are independent. For a discussion of this assumption,
see the Discussion Section of the main paper. 
Considering a different prior distribution for $\mu_i$ the empirical Bayes approach extends
to the additional parameters appearing in the prior. They will be also
estimated in a CpG dependent manner. However, one should avoid including
too many parameters as this complicates the empirical Bayes procedure
and makes it more difficult to find the best parameters. In the case of the
DBD prior distribution we fixed the weights to $w_0=0.1$, $w_1=0.8$, $w_2=0.1$
and only estimated the parameters $a$ and $b$.

\section*{Derivation of the posterior marginal distribution}

\section*{Using a beta mixture prior for the methylation level}

Our main interest lies in the marginal posterior distribution of
the methylation level $\mu_i$

\begin{linenomath*}
\begin{equation*}
	p(\mu_i|y_{iS}, y_{iC}) = \int_0^\infty p(\lambda_i, \mu_i|y_{iS}, y_{iC}) d\lambda_i,
\end{equation*}
\end{linenomath*}

where

\begin{linenomath*}
\begin{equation*}
\begin{split}
	p(\lambda_i, \mu_i|y_{iS}, y_{iC}) &= \frac{p(y_{iS}, y_{iC}|\lambda_i, \mu_i) p(\lambda_i,\mu_i)}{p(y_{iS},y_{iC})}\\
	&\stackrel{\text{cond.indep}}{=}\frac{p(y_{iS}|\lambda_i,\mu_{i})p(y_{iC}|\lambda_i)p(\lambda_i) p(\mu_i)}{p(y_{iS},y_{iC})}\\
	&= \frac{\lambda_i^{y_{iS} + y_{iC} + \alpha-1}\exp(-(E\mu_i+1+\beta)\lambda_i)(\beta+1+E)^{\alpha +y_{iS} +y_{iC}}p(\mu_i)\mu_i^{y_{iS}}}{\Gamma(y_{iS} +y_{iC} + \alpha)\times W}.
\end{split}
\end{equation*}
\end{linenomath*}
Here, $W$ is as given in Equation \eqref{eq:mixBetaW}, and $\alpha$ and $\beta$ are the parameters for the gamma prior
distribution for $\lambda_i$ as determined by empirical Bayes (see above) for the
CpG-density class to which bin $i$ belongs.

Thus:
\begin{linenomath*}
\begin{equation*}
\begin{split}
p(\mu_i|y_{iS}, y_{iC}) &= \frac{\mu_i^{y_{iS}} p(\mu_i) (\beta+1 +E)^{\alpha+y_{iS} + y_{iC}}}{\Gamma(y_{iS} +y_{iC} +\alpha)\times W}
	\int_0^\infty \lambda_i^{y_{iS} +y_{iC} + \alpha -1}\exp(-(E\mu_i + 1 +\beta)\lambda_i)d\lambda_i\\
	&= \frac{\mu_i^{y_{iS}} p(\mu_i)}{W} \left(1- \frac{E(1-\mu_i)}{\beta +1 +E}\right)^{-(\alpha+y_{iS}+y_{iC})}.
\end{split}
\end{equation*}
\end{linenomath*}

The mean of the marginal posterior of $\mu_i$ is given by:
\begin{linenomath*}
\begin{equation*}
	\E(\mu_i|y_{iS}, y_{iC}) = \frac{A}{W}
\end{equation*}
\end{linenomath*}
with
\begin{linenomath*}
\begin{equation*}
\begin{split}
    A &= \sum_{m=1}^M\Biggl[
        w_m \cdot \frac{\Gamma(a_m+b_m) \Gamma(y_{iS} + a_m{+1})}{\Gamma(a_n)\Gamma(y_{iS}+a_m+b_m{+1})}\times\: _2F_1\left(y_{iS}+y_{iC}+\alpha, b_m; y_{iS} +a_m +b_m{+1}; \frac{E}{\beta + 1 +E}\right)   \Biggr].
\end{split}
\end{equation*}
\end{linenomath*}

\begin{proof}
\begin{linenomath*}
\begin{equation*}
\begin{split}
\E&(\mu_i|y_{iS}, y_{iC}) = \int_0^1 \mu_i p(\mu_i|y_{iS}, y_{iC})d\mu_i\\
	&=\frac{1}{W} \sum_{m=1}^M\Biggl[\int_0^1 \frac{w_m\frac{\Gamma(a_m+b_m)}{\Gamma(a_m)\Gamma(b_m)}\mu_i^{a_m+y_{iS}}(1-\mu_i)^{b_m-1}}{\left(1- \frac{E(1-\mu_i)}{\beta +1 +E}\right)^{\alpha+y_{iS}+y_{iC}}} d\mu_i\Biggr],
\end{split}
\end{equation*}
\end{linenomath*}

where each integral can again be written in terms of the Gauss
hypergeometric function:
\begin{linenomath*}
\begin{equation*}
\begin{split}
\int_0^1&\frac{w_m\frac{\Gamma(a_m+b_m)}{\Gamma(a_m)\Gamma(b_m)}\mu_i^{a_m+y_{iS}}(1-\mu_i)^{b_m-1}}{\left(1- \frac{E(1-\mu_i)}{\beta +1 +E}\right)^{\alpha+y_{iS}+y_{iC}}} d\mu_i \\
&=\frac{w_m\Gamma(a_m+b_m)}{\Gamma(a_m)\Gamma(b_m)} \int_0^1\frac{(1-t_i)^{a_m+y_{iS}}t_i^{b_m-1}}{\left(1- \frac{E}{\beta +1 +E}t_i\right)^{\alpha+y_{iS}+y_{iC}}} dt_i\\
&=\frac{w_m\Gamma(a_m+b_m)}{\Gamma(a_m)\Gamma(b_m)} \frac{\Gamma(b_m)\Gamma(y_{iS}+a_m+1)}{\Gamma(y_{iS}+a_m+b_m+1)}\: _2F_1\left(y_{iS}+y_{iC}+\alpha, b_m; y_{iS}+a_m+b_m+1;\frac{E}{\beta +1 +E}\right)\\
&=\frac{w_m\Gamma(a_m+b_m)\Gamma(y_{iS}+a_m+1)}{\Gamma(a_m)\Gamma(y_{iS}+a_m+b_m+1)}\: _2F_1\left(y_{iS}+y_{iC}+\alpha, b_m; y_{iS}+a_m+b_m+1;\frac{E}{\beta +1 +E}\right).
\end{split}
\end{equation*}
\end{linenomath*}
\end{proof}

The variance of the marginal posterior distribution of $\mu_i$ can
be computed using the computational formula for the variance
$\Var(\mu_i|y_{iS}, y_{iC}) = \E(\mu_i^2|y_{iS}, y_{iC})-(\E(\mu_i|y_{iS}, y_{iC}))^2$, where
\begin{linenomath*}
\begin{equation*}
\E(\mu_i^2|y_{iS}, y_{iC}) = \frac{B}{W}
\end{equation*}
\end{linenomath*}
with
\begin{linenomath*}
\begin{equation*}
\begin{split}
    B &= \sum_{m=1}^M\Biggl[
        w_m \cdot \frac{\Gamma(a_m+b_m) \Gamma(y_{iS} + a_n+2)}{\Gamma(a_m)\Gamma(y_{iS}+a_m+b_m+2)}\times\: _2F_1\left(y_{iS}+y_{iC}+\alpha, b_m; y_{iS} +a_m +b_m+2; \frac{E}{\beta + 1 +E}\right)   \Biggr],
\end{split}
\end{equation*}
\end{linenomath*}
so that
\begin{linenomath*}
\begin{equation*}
	\Var(\mu_i|y_{iS}, y_{iC}) = \frac{B}{W} - \left(\frac{A}{W}\right)^2.
\end{equation*}
\end{linenomath*}

Running BayMeth without a fully methylated control sample, we get
\begin{align*}
  \E(\mu_i|y_{iS}) = \frac{A}{W}
  \Var(\mu_i|y_{iS}) = \frac{B}{W} - \left(\frac{A}{W}\right)^2.\\
\end{align*}
$A$, $B$ and $W$ are:
\begin{linenomath*}
\begin{align*}
A &= \sum_{m=1}^M\Biggl[
	w_m \cdot \frac{\Gamma(a_m+b_m) \Gamma(y_{iS} + a_m + 1)}{\Gamma(a_m)\Gamma(y_{iS}+a_m+b_m + 1)}\times\: _2F_1\left(y_{iS}+\alpha, b_m; y_{iS} +a_m +b_m + 1; \frac{E}{\beta +E}\right)	\Biggr],\\
B &= \sum_{m=1}^M\Biggl[
	w_m \cdot \frac{\Gamma(a_m+b_m) \Gamma(y_{iS} + a_m + 2)}{\Gamma(a_m)\Gamma(y_{iS}+a_m+b_m + 2)}\times\: _2F_1\left(y_{iS}+\alpha, b_m; y_{iS} +a_m +b_m + 2; \frac{E}{\beta +E}\right)	\Biggr],\\
W &= \sum_{m=1}^M\Biggl[
	w_m \cdot \frac{\Gamma(a_m+b_m) \Gamma(y_{iS} + a_m)}{\Gamma(a_m)\Gamma(y_{iS}+a_m+b_m)}\times\: _2F_1\left(y_{iS}+\alpha, b_m; y_{iS} +a_m +b_m; \frac{E}{\beta +E}\right)	\Biggr].
\end{align*}
\end{linenomath*}

\section*{Using a DBD prior for the methylation level}

The posterior mean and variance can be derived analogously to the previous section.
Borrowing strength from a SssI sample, posterior mean and variance are given by:
\begin{align*}
  \E(\mu_i|y_{iS}, y_{iC}) &= \frac{A}{W}\\
  \Var(\mu_i|y_{iS}, y_{iC}) &= \frac{B}{W} - \left(\frac{A}{W}\right)^2.\\
\end{align*}
with 
\begin{linenomath*}
\begin{align*}
A &= w_2 + w_1 \cdot \frac{\Gamma(a+b) \Gamma(y_{iS} + a + 1)}{\Gamma(a)\Gamma(y_{iS}+a+b + 1)}\times\: _2F_1\left(y_{iS}+y_{iC} + \alpha, b; y_{iS} +a +b +1; \frac{E}{\beta +E + 1}\right)	\Biggr],\\
B &= w_2 + w_1 \cdot \frac{\Gamma(a+b) \Gamma(y_{iS} + a + 2)}{\Gamma(a)\Gamma(y_{iS}+a+b +2 )}\times\: _2F_1\left(y_{iS}+y_{iC} + \alpha, b; y_{iS} +a +b +2; \frac{E}{\beta +E + 1}\right)	\Biggr],\\
W &= w_2 + w_1 \cdot \frac{\Gamma(a+b) \Gamma(y_{iS}+ a)}{\Gamma(a)\Gamma(y_{iS}+a+b)}\times\: _2F_1\left(y_{iS}+y_{iC} + \alpha, b; y_{iS} +a +b; \frac{E}{\beta +E + 1}\right)	\Biggr].
\end{align*}
\end{linenomath*}
Assuming that no SssI sample is available, then 
\begin{align*}
  \E(\mu_i|y_{iS}) &= \frac{A}{W}\\
  \Var(\mu_i|y_{iS}) &= \frac{B}{W} - \left(\frac{A}{W}\right)^2.\\
\end{align*}
with 
\begin{linenomath*}
\begin{align*}
A &= w_2 + w_1 \cdot \frac{\Gamma(a+b) \Gamma(y_{iS} + a + 1)}{\Gamma(a)\Gamma(y_{iS}+a+b + 1)}\times\: _2F_1\left(y_{iS} + \alpha, b; y_{iS} +a +b +1; \frac{E}{\beta +E }\right)	\Biggr],\\
B &= w_2 + w_1 \cdot \frac{\Gamma(a+b) \Gamma(y_{iS} + a + 2)}{\Gamma(a)\Gamma(y_{iS}+a+b +2 )}\times\: _2F_1\left(y_{iS} + \alpha, b; y_{iS} +a +b +2; \frac{E}{\beta +E }\right)	\Biggr],\\
W &= w_2 + w_1 \cdot \frac{\Gamma(a+b) \Gamma(y_{iS}+ a)}{\Gamma(a)\Gamma(y_{iS}+a+b)}\times\: _2F_1\left(y_{iS} + \alpha, b; y_{iS} +a +b; \frac{E}{\beta +E}\right)	\Biggr].
\end{align*}
\end{linenomath*}

%
%

\clearpage

\section*{Supplementary Figures}
\subsection*{Figure S1 - CpG-density stratified by CpG island status}
\begin{center}
\includegraphics[width=.5\textwidth]{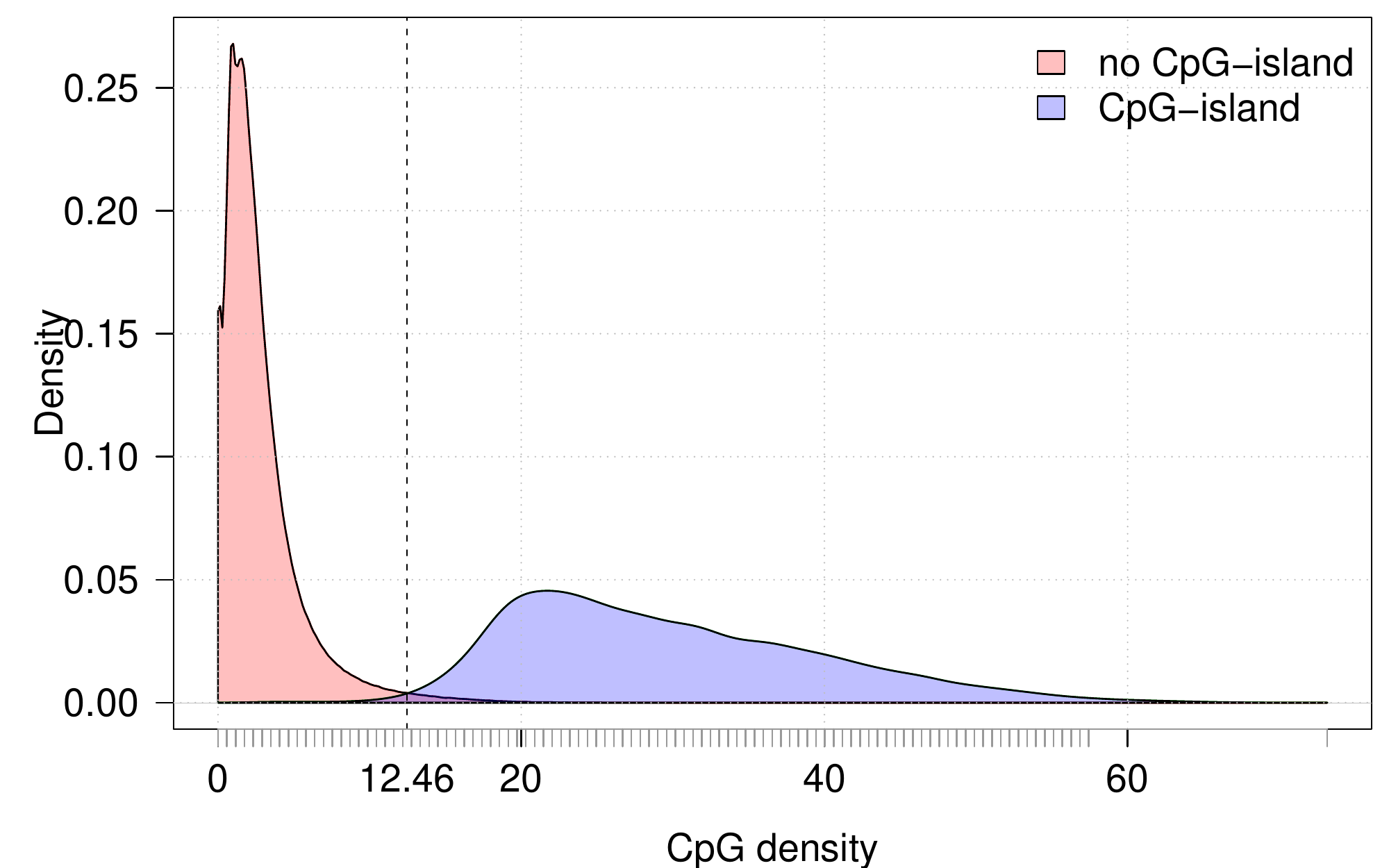}
\end{center}
Genome-wide CpG-density for bins with a mappability larger than 75\% stratified by
CpG island status as extracted from the cpgIslandExt-table of
the UCSC genome browser. The vertical line marks the intersection of both densities.
The grey tick-marks along the x-axis illustrate the CpG-density classes used for
the empirical Bayes approach in the IMR-90 application. 

\subsection*{Figure S2 - Normalizing offset}
\begin{center}
\includegraphics[width=0.5\textwidth]{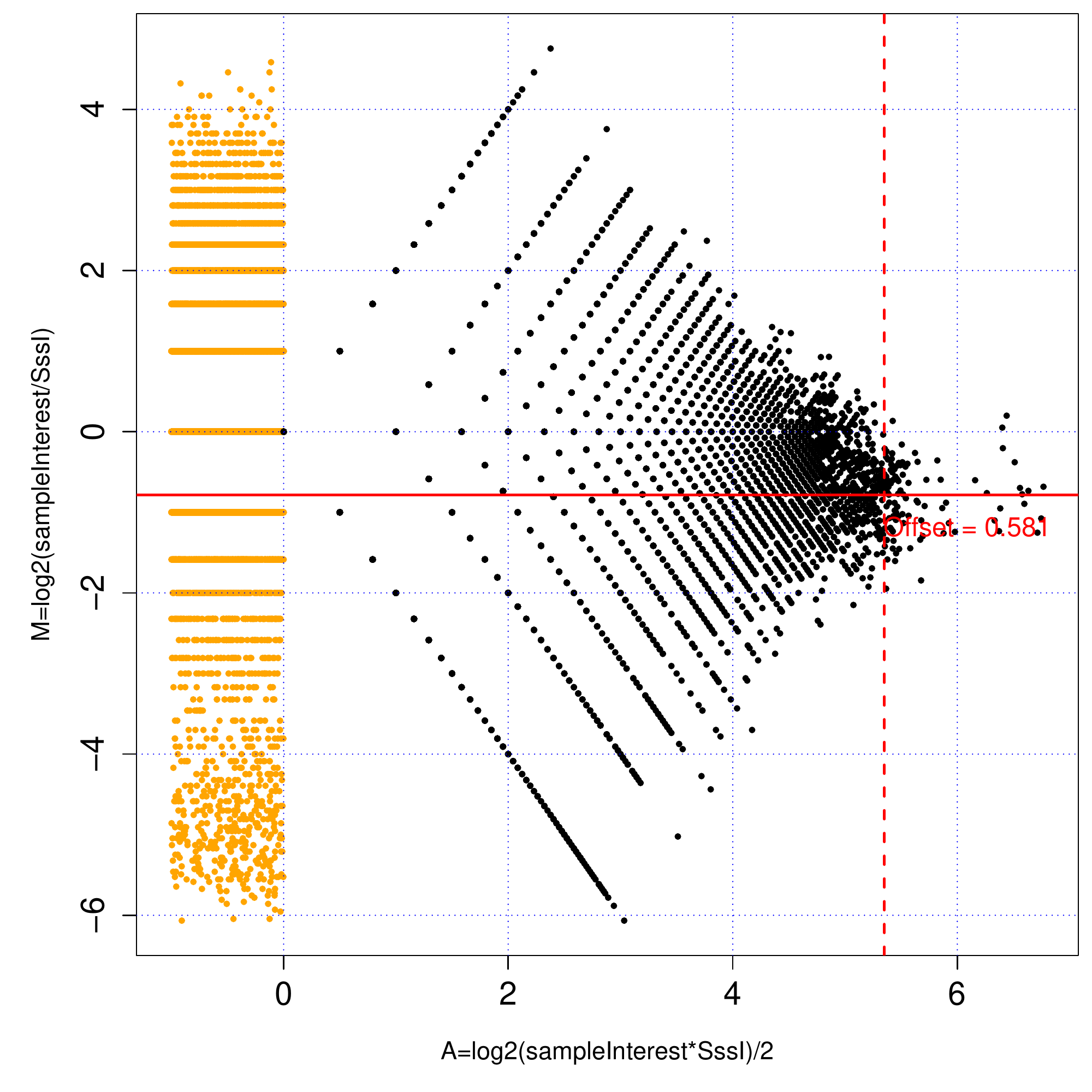}
\end{center}
Log-fold change ($M$) versus log-concentration ($A$) illustrated for
$50000$ randomly chosen bins. The red dotted line shows the $0.998$
quantile $q$ of $A$ determined from all bins. The red straight line
shows the estimated normalization offset $f=2^{median(M_{A > q})}$.
A 'smear' of yellow points at a low $A$ value represents counts that
are low in either of the two samples.

\subsection*{Figure S3 - Copy number frequencies for LNCaP}
\begin{center}
\includegraphics[width=0.5\textwidth]{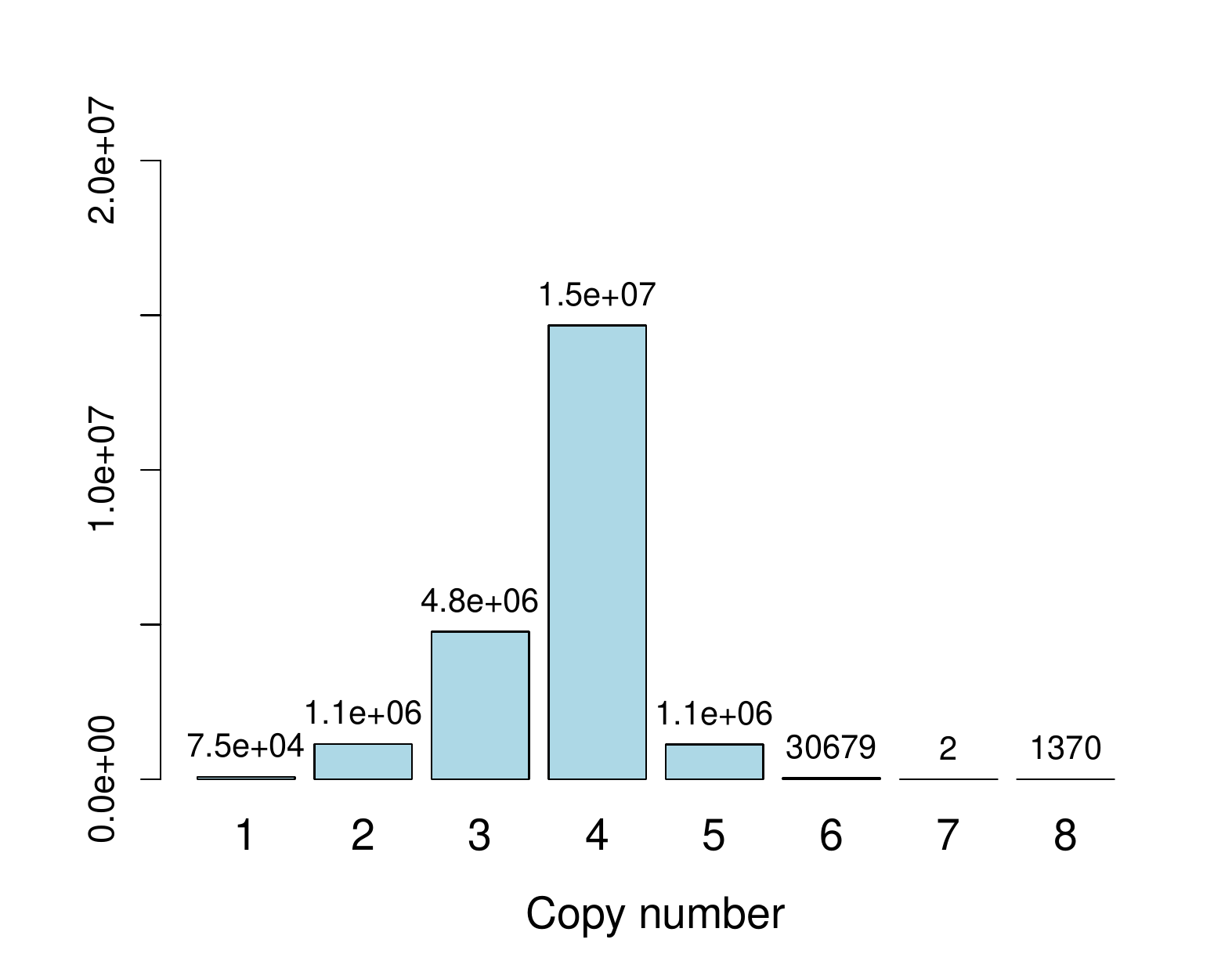}
\end{center}
Copy number frequencies in LNCaP for $100$bp-bins with a mappability
larger than $0.75$.

\subsection*{Figure S4 - Read depth of LNCaP MBD-seq by copy number}
\begin{center}
\includegraphics[width=\textwidth]{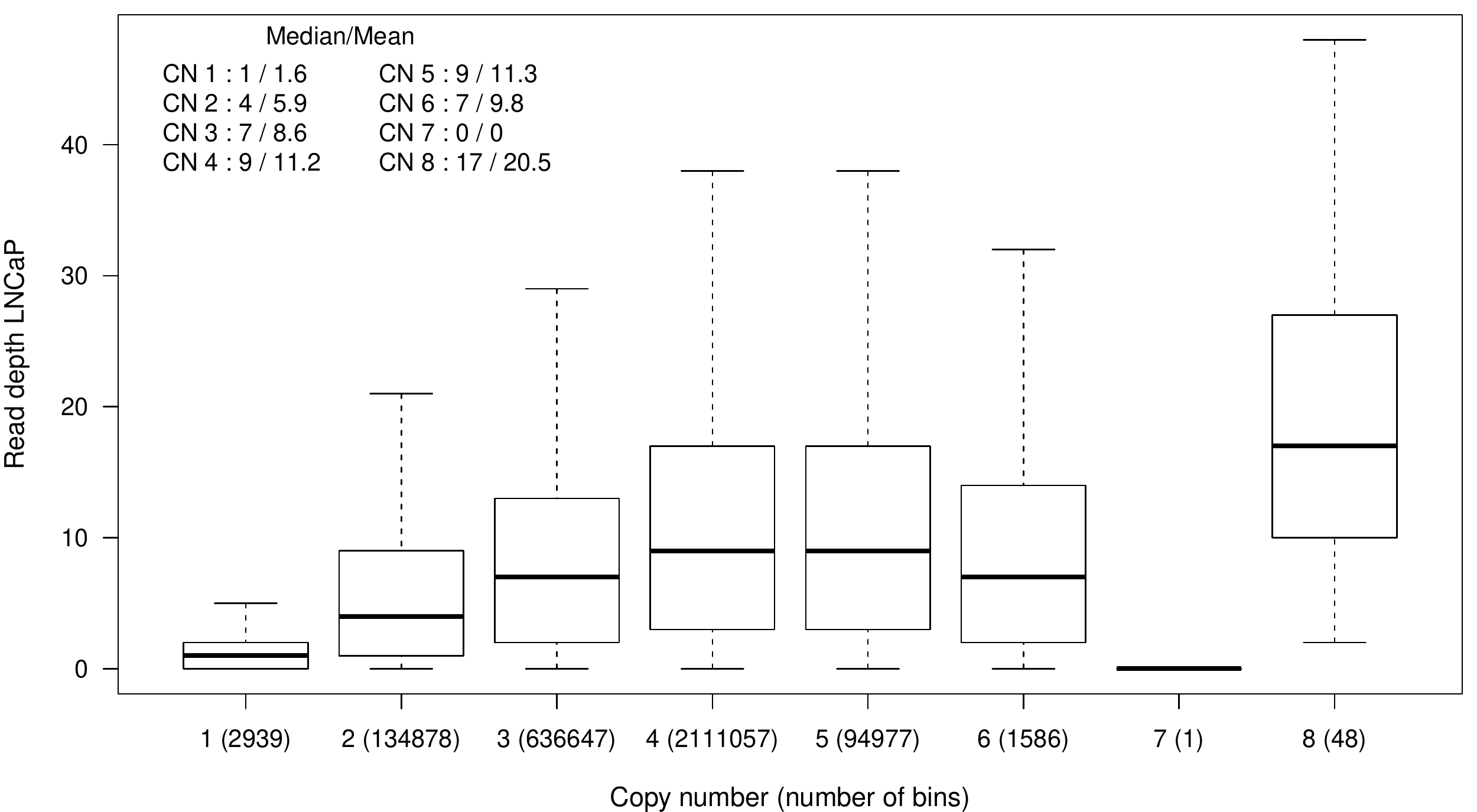}
\end{center}
Read depth stratified by copy number is shown for $100$bp-bins with a mappability larger than $0.75$
and with a SssI depth larger than four. Median and mean read depth are given per copy number state.

\subsection*{Figure S5 - Varying normalizing offsets between methylation kits}

\begin{center}
\includegraphics[width=\textwidth]{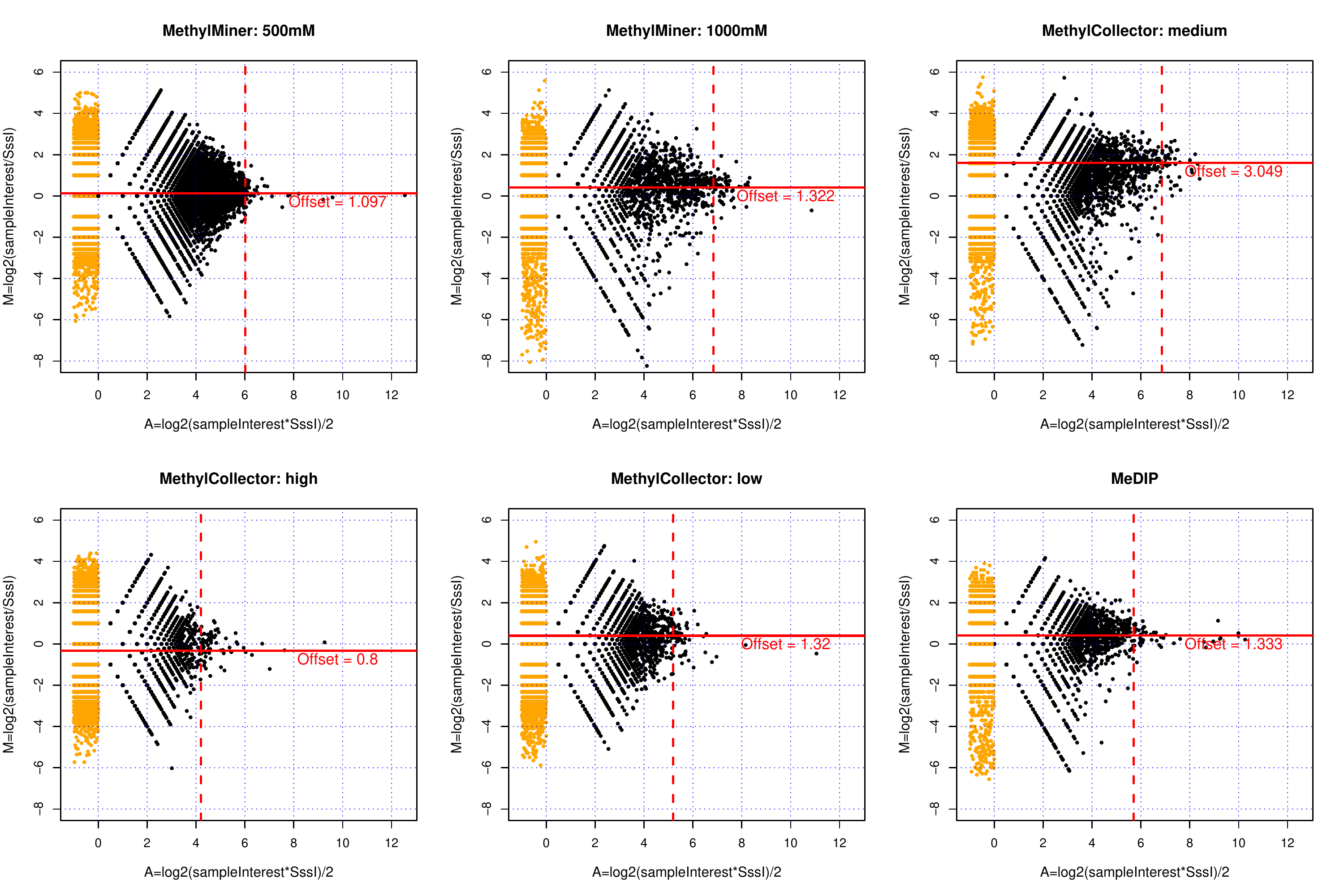}
\end{center}

Log-fold change ($M$) versus log-concentration ($A$) illustrated for
$25000$ randomly chosen bins for IMR-90 data derived using different
methylation kits. The red dotted line shows the $0.998$
quantile $q$ of $A$ determined from all bins. The red straight line
shows the estimated normalization offset $f=2^{median(M_{A > q})}$.
A 'smear' of yellow points at a low $A$ value represents counts that
are low in either of the two samples.

\subsection*{Figure S6 - Distribution of estimated methylation levels for SssI sample using Illumina HumanMethylation450 arrays}
\begin{center}
\includegraphics[width=0.5\textwidth]{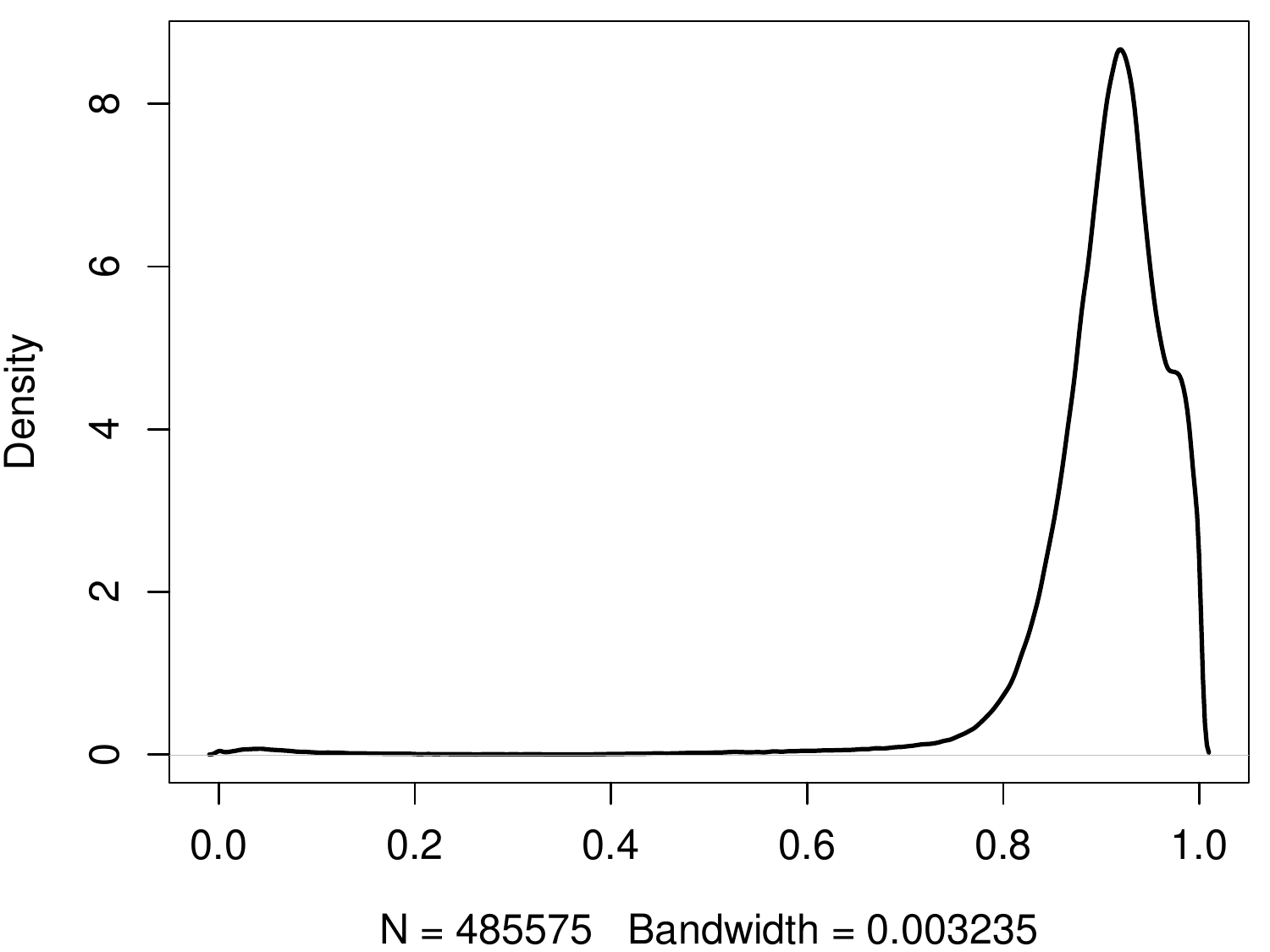}
\end{center}
Density plot of 450k beta values.

\section*{Supplementary Table}
\subsection*{Table S1 - Mean bias stratified by CpG island status and true methylation level}

 \par
    \mbox{
\begin{tabular}{rrrr|rrr}
  & \multicolumn{3}{c|}{No CpG-islands} & \multicolumn{3}{c}{CpG-islands}\\
  Method &  $[0,0.2]$ &   $(0.2,0.8]$ & $(0.8,1]$ &  $[0,0.2]$ &   $(0.2,0.8]$ & $(0.8,1]$\\ 
  \hline
BayMeth & 0.41 & 0.06 & -0.22 & 0.09 & 0.14 & -0.07 \\ 
  Batman & 0.69 & 0.29 & -0.01 & 0.11 & 0.15 & -0.02 \\ 
  MEDIPS & 0.05 & -0.27 & -0.54 & 0.05 & -0.16 & -0.47 \\ 
  BALM & -0.03 & -0.37 & -0.57 & -0.01 & -0.16 & -0.31 \\ 
   \hline
\end{tabular}
}

\clearpage

\section*{Additional file 3 --- BayMeth analysis of ``Bock'' data}

We applied (default) BayMeth to the MethylCap sequencing data of 
\cite{bock-etal-2010}, provided at
 \url{http://www.broadinstitute.org/labs/meissner/mirror/papers/meth-benchmark/index.html}, 
and denoted as the ``Bock'' data below. 
Absolute read densities are available for four samples: HUES6 ES cell line, 
HUES8 ES cell line, colon tumor tissue, colon normal tissue 
(same donor as for colon tumor tissue), based on hg18 and given for (non-overlapping) 50bp bins. 
There is no matched SssI sample available for these data. To take 
advantage of BayMeth in analyzing these data,  we use a non-matching 
SssI sample, but one chosen to be maximally compatible to the preparation 
conditions of Bock data\cite{bock-etal-2010} 
(i.e. MethylCap at low salt concentration: 200mM NaCl). 
Furthermore, RRBS data are available for each sample representing 
absolute DNA methylation levels at single CpGs.

In section 1, we outline all data preparation steps. First, all samples of interest are saved in a single GRanges object based on genome-wide non-overlapping 50bp bins. RRBS information is loaded and saved in the same object. Since the read density for the fully methylated sample is based on hg19, the Bock data are lifted over. Based on hg19 we derive CpG density and mappability estimates. Finally, all information is stored in a BayMethList data object. Section 2 describes the BayMeth analysis applied
on the former created BayMethList data object. Normalizing offsets are derived for all samples, before the empirical Bayes approach is used to get suitable prior parameters. Finally region-specific methylation estimates are computed.

\section*{Data preparation}

\subsection*{Samples of interest}
We applied BayMeth to the MethylCap sequencing data of \cite{bock-etal-2010}.
Data are available for four samples: 1) HUES6 ES cell line, 2) HUES8 ES cell line, 3) Colon tumor tissue, 4) Colon normal tissue (same donor as (3)).
Absolute read densities provided as bigwig files were downloaded, converted to {\tt GRanges} objects and saved in a {\tt GRangesList}:\\[0.2cm]
\begin{Schunk}
\begin{Sinput}
 setwd("./4_bock/")
 library(rtracklayer)
 data_names <- c("HUES6", "HUES8", "Colon_normal", "Colon_tumor")
 grl_bock_methylCap <- GRangesList()
 for(i in 1:length(data_names)){
     print(data_names[i])
     # import the data and convert to GRanges
     data_tmp <- import(paste("data/ChIP_absReadFreqW50_MethylCap_", data_names[i], "_all.bw", sep=""),"bw")
     data_tmp <- as(data_tmp,"GRanges")
     grl_bock_methylCap <- c(grl_bock_methylCap, GRangesList(data_tmp))
 }
\end{Sinput}
\end{Schunk}
Read densities are based on (non-overlapping) 50bp bins. Summary 
information for each sample is shown in Table~\ref{tab:sum}.\\[0.2cm]

\begin{Schunk}
\begin{Sinput}
 sumTab <- cbind(summary(values(grl_bock_methylCap[[1]])$score),
 summary(values(grl_bock_methylCap[[2]])$score),
 summary(values(grl_bock_methylCap[[3]])$score),
 summary(values(grl_bock_methylCap[[4]])$score))
\end{Sinput}
\end{Schunk}

\begin{table}[t]
\begin{center}
\begin{tabular}{rrrrr}
  \hline
 & HUES6 & HUES8 & Colon\_normal & Colon\_tumor \\
  \hline
Min. & 0.00 & 0.00 & 0.00 & 0.00 \\
  1st Qu. & 0.00 & 0.00 & 0.00 & 0.00 \\
  Median & 0.00 & 0.00 & 0.00 & 0.00 \\
  Mean & 2.00 & 1.81 & 1.96 & 1.99 \\
  3rd Qu. & 2.00 & 2.00 & 2.00 & 2.00 \\
  Max. & 374.00 & 400.00 & 407.00 & 400.00 \\
   \hline
\end{tabular}
\caption{Summary information for absolute read counts for each sample.}\label{tab:sum}
\end{center}
\end{table}
Of note, read density information for the different samples is not given for the same bins. To save all data in one {\tt GRanges} object, a genome-wide GRanges object for hg18 based on non-overlapping 50bp was created.\\[0.2cm]

\begin{Schunk}
\begin{Sinput}
 library(BSgenome.Hsapiens.UCSC.hg18)
 library(Repitools)
 library(GenomicRanges)
 # save all datasets in one GRanges object
 gb_hg18 <- genomeBlocks(Hsapiens, 1:24, width=50)
 #
 tumor <- normal <- hues6 <- hues8 <- rep(NA, length(gb_hg18))
 #
 fo_hues6 <- findOverlaps(gb_hg18, grl_bock_methylCap[[1]])
 fo_hues8 <- findOverlaps(gb_hg18, grl_bock_methylCap[[2]])
 fo_normal <- findOverlaps(gb_hg18, grl_bock_methylCap[[3]])
 fo_tumor <- findOverlaps(gb_hg18, grl_bock_methylCap[[4]])
 #
 inds_hues6 <- split(fo_hues6@subjectHits, fo_hues6@queryHits)
 ind_hues6 <- as.integer(names(inds_hues6))
 hues6[ind_hues6] <- values(grl_bock_methylCap[[1]])$score[fo_hues6@subjectHits]
 #
 inds_hues8 <- split(fo_hues8@subjectHits, fo_hues8@queryHits)
 ind_hues8 <- as.integer(names(inds_hues8))
 hues8[ind_hues8] <- values(grl_bock_methylCap[[2]])$score[fo_hues8@subjectHits]
 #
 # ... analogously for normal and tumor
 
 df <- DataFrame("hues6"=hues6, "hues8"=hues8, "normal"=normal, "tumor"=tumor)
 values(gb_hg18) <- df
\end{Sinput}
\end{Schunk}

To do this properly we have to ensure that the bins of \cite{bock-etal-2010} start at $1, 51, 101, 151, \ldots$ and have a width of 50bp. We have proved this using a modulo operation {\tt table(start(grl\_bock\_methylCap[[i]]) \%\% 50)} which resulted in $1$ for all bins, and {\tt table(width(grl\_bock\_methylCap[[i]]))}, which resulted in $50$ for all bins.
Using the function {\tt findOverlaps} the different read counts are saved as metadata at the corresponding positions in the object {\tt gb\_hg18}. If no information is provided for a bin, the read density is set to {\tt NA}.

\subsection*{Reduced representation bisulphite sequencing (RRBS) information}

Information on RRBS data are available on  \url{http://www.broadinstitute.org/labs/meissner/mirror/papers/meth-benchmark/RRBS/},
and used as gold standard in the following analysis. In the RRBS data for HUES6 and HUES8 we
removed lines where the strand information is neither "+" , "-" nor "*", but "b", and saved
the data in RRBS\_cpgMethylation\_HUES6\_strandCleaned.RRBS.bed and RRBS\_cpgMethylation\_HUES8\_strandCleaned.RRBS.bed, respectively.

Both, the number of reads that overlay a cytosine (T) and
the number of cytosines that stay a cytosine (M), i.e. are methylated, are given. Note, that for one CpG site there is only
information from one strand available.\\[0.2cm]

\begin{Schunk}
\begin{Sinput}
 data_names <- c("HUES6_strandCleaned", "HUES8_strandCleaned", "Colon_normal", "Colon_tumor")
 # create container to save datasets
 grl_bock_rrbs <- GRangesList()
 for(i in 1:length(data_names)){
     # import the data and convert to GRanges
     data_tmp <- import(paste("data/RRBS_cpgMethylation_", data_names[i], ".RRBS.bed", sep=""),"BED")
     data_tmp <- as(data_tmp,"GRanges")
 
     # extract the number of reads that overlay a cytosine and the number
     # of cytosines that stay a cytosine i.e. are methylated
     name <- values(data_tmp)$name
     cpg <- strsplit(name, "/")
     cpg <- do.call(rbind, cpg)
     cpg <- sapply(1:ncol(cpg), function(u){as.numeric(cpg[,u])})
     colnames(cpg) <- c("numMeth", "total")
 
     # add the corresponding columns to the GRanges object
     # (meth correponds approximately to score/1000)
     values(data_tmp) <- cbind(values(data_tmp),
                             DataFrame(cpg, meth=cpg[,1]/cpg[,2]))
 
     grl_bock_rrbs <- c(grl_bock_rrbs, GRangesList(data_tmp))
 }
 names(grl_bock_rrbs) <- c("HUES6", "HUES8", "Colon_normal", "Colon_tumor")
\end{Sinput}
\end{Schunk}

To get smooth methylation estimates, we summarized CpG based RRBS data within 150bp bins (overlapping by 100bp). The methylation level for one 150bp bin $i$ is thereby derived as:

\begin{equation*}
	m_i= \frac{\sum M_{\in i}}{\sum T_{\in i}}.
\end{equation*}
That means using information for all CpG sites that fall into bin $i$.\\[0.2cm]

\begin{Schunk}
\begin{Sinput}
 gb_hg18_150 <- resize(gb_hg18, 150, fix="center")
 # get the corresponding rrbs estimates
 meth_names <- c("rrbs_hues6_meth", "rrbs_hues8_meth", "rrbs_normal_meth", "rrbs_tumor_meth")
 denom_names <- c("rrbs_hues6_denom", "rrbs_hues8_denom","rrbs_normal_denom", "rrbs_tumor_denom")
 for(i in 1:4){
     rrbs_tmp <- grl_bock_rrbs[[i]]
     fo_tmp <- findOverlaps(gb_hg18_150, rrbs_tmp)
     inds_tmp <- split(fo_tmp@subjectHits, fo_tmp@queryHits)
 
     nmeth <- values(rrbs_tmp)$numMeth
     total <- values(rrbs_tmp)$total
 
     methI <- sapply(inds_tmp, function(u) sum(nmeth[u])/sum(total[u]))
     denomI <- sapply(inds_tmp, function(u) sum(total[u]))
 
     denom <- meth <- rep(NA, length(gb_hg18))
     # assign the derived estimates to the corresponding genomic bins
     ind_tmp <- as.integer(names(inds_tmp))
     meth[ind_tmp] <- methI
     denom[ind_tmp] <- denomI
     tmp_df <- DataFrame(meth, denom)
     colnames(tmp_df) <- c(meth_names[i], denom_names[i])
     values(gb_hg18) <- cbind(values(gb_hg18), tmp_df)
 }
\end{Sinput}
\end{Schunk}

Figure~\ref{fig:rrbs_methylcap_normal_hg18} shows a smooth density representation of the RRBS methylation estimates versus the MethylCap read density after filtering bins where no truth exists and only taking a minimum depth of 20 in RRBS.\\


\begin{figure}
\begin{center}
	\includegraphics[width=.6\textwidth]{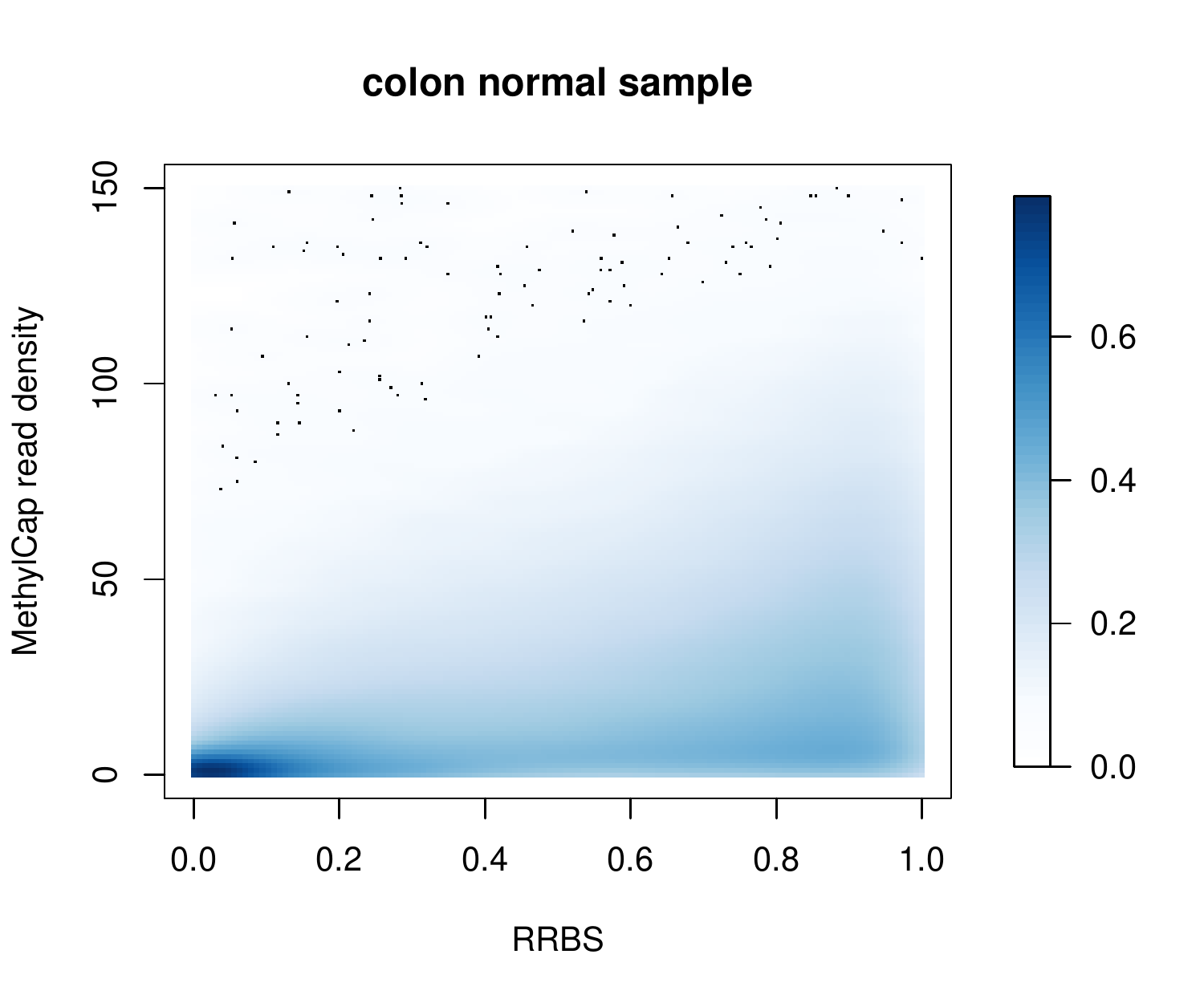}
	\caption{Comparison between read frequencies and DNA methylation levels derived from RRBS for the colon normal sample. Unprocessed read frequencies for MethylCap were correlated with DNA methylation levels as determined by RRBS. 
}\label{fig:rrbs_methylcap_normal_hg18}
\end{center}
\end{figure}

\subsection*{Lift-over to hg19}

Since the data for the fully  methylated (SssI treated) sample are based on hg19, the bin coordinates of hg18 are transferred to the corresponding position on hg19.\\[0.2cm]

\begin{Schunk}
\begin{Sinput}
 chain <- import.chain("data/hg18ToHg19.over.chain")
 gb_hg19 <- liftOver(gb_hg18, chain)
 gb_hg19 <- unlist(gb_hg19)
\end{Sinput}
\end{Schunk}

We remove all bins with a width unequal to 50bp.\\[0.2cm]
\begin{Schunk}
\begin{Sinput}
 library(BSgenome.Hsapiens.UCSC.hg19)
 w.idx <- which(width(gb_hg19) != 50)
 gb_hg19r <- gb_hg19[-w.idx]
\end{Sinput}
\end{Schunk}
Lifting the bins over to hg19 caused overlapping bins. Hence, we remove all bins that have more than one overlap (namely with itself).\\[0.2cm]

\begin{Schunk}
\begin{Sinput}
 fo <- findOverlaps(gb_hg19r, gb_hg19r)
 inds <- split(fo@subjectHits, fo@queryHits)
 len <- unlist(lapply(inds, length))
 w2.idx <- which(len != 1)
 gb_hg19r <- gb_hg19r[-w2.idx]
\end{Sinput}
\end{Schunk}

\subsection*{SssI sample, CpG density and mappability information}

BayMeth quantifies methylation of an affinity-enrichment sequencing dataset best by taking advantage of a full methylated control data set. Here, we use a sample treated with SssI and analysed using MethylCap at low salt concentration, i.e., $200$ mM NaCl,
to be maximally compatible to the preparation conditions of \cite{bock-etal-2010}.\\[0.2cm]

\begin{Schunk}
\begin{Sinput}
 library(BSgenome.Hsapiens.UCSC.hg19)
 f <- "data/SSSl_low.bam"
 names(f) <- "SssI_low"
 counts <- annotationBlocksCounts(f, gb_hg19r, seq.len=150)
\end{Sinput}
\end{Schunk}

The CpG density is calculated by symmetrically extending the bins around the bin center to a length of 700bp and linear weighting the CpG sites falling into this range.\\[0.2cm]

\begin{Schunk}
\begin{Sinput}
 gbA <- resize(gb_hg19r, 1, fix="center")
 cpgdens <- cpgDensityCalc(gbA, organism=Hsapiens, w.function="linear", window=700)
\end{Sinput}
\end{Schunk}

Mappability probabilities are derived from \url{http://hgdownload.cse.ucsc.edu/goldenPath/hg19/encodeDCC/wgEncodeMapability/wgEncodeCrgMapabilityAlign50mer.bigWig}.\\[0.2cm]

\begin{Schunk}
\begin{Sinput}
 library(rtracklayer)
 bw <- BigWigFile("data/wgEncodeCrgMapabilityAlign50mer_hg19.bigWig")
 map <- import(bw)
 score <- score(map)
 wd <- width(map)
 fo <- findOverlaps(gb_hg19r, map)
 ind <- split(fo@subjectHits,fo@queryHits)
 mapv <- numeric(length(gb_hg19r)) # default of 0
 w <- as.numeric(names(ind))
 # take weighted mean
 mapv[w] <- sapply(ind, function(u) sum( wd[u]*score[u] ) / sum(wd[u]) )
 values(gb_hg19r) <- cbind(values(gb_hg19r), DataFrame("cpgdens"=cpgdens, "map_ucsc"=mapv, "SssI-low"=counts))
 save(gb_hg19r, file="data/bock_data_prepared.Rdata")
\end{Sinput}
\end{Schunk}

SssI read densities, CpG density and mappability are saved as further metadata columns in gb\_h19r.

\clearpage
\section*{BayMeth Analysis}

Here, my session info:\\[0.2cm]

\begin{Schunk}
\begin{Sinput}
 sessionInfo()
 #R Under development (unstable) (2013-07-03 r63169)
 #Platform: x86_64-unknown-linux-gnu (64-bit)
 #
 #locale:
 # [1] LC_CTYPE=en_CA.UTF-8       LC_NUMERIC=C              
 # [3] LC_TIME=en_US.UTF-8        LC_COLLATE=en_CA.UTF-8    
 # [5] LC_MONETARY=en_US.UTF-8    LC_MESSAGES=en_CA.UTF-8   
 # [7] LC_PAPER=en_US.UTF-8       LC_NAME=C                 
 # [9] LC_ADDRESS=C               LC_TELEPHONE=C            
 #[11] LC_MEASUREMENT=en_US.UTF-8 LC_IDENTIFICATION=C       
 #
 #attached base packages:
 #[1] parallel  stats     graphics  grDevices utils     datasets  methods  
 #[8] base     
 #
 #other attached packages:
 # [1] lattice_0.20-15                    fields_6.7                        
 # [3] spam_0.29-3                        Repitools_1.7.13                  
 # [5] BSgenome.Hsapiens.UCSC.hg18_1.3.19 BSgenome_1.29.1                   
 # [7] Biostrings_2.29.15                 rtracklayer_1.21.9                
 # [9] GenomicRanges_1.13.36              XVector_0.1.0                     
 #[11] IRanges_1.19.24                    BiocGenerics_0.7.4                
 #
 #loaded via a namespace (and not attached):
 # [1] bitops_1.0-6       edgeR_3.3.7        grid_3.1.0         KernSmooth_2.23-10
 # [5] limma_3.17.21      RCurl_1.95-4.1     Rsamtools_1.13.29  Rsolnp_1.14       
 # [9] stats4_3.1.0       tools_3.1.0        truncnorm_1.0-6    XML_3.98-1.1      
 #[13] zlibbioc_1.7.0 
\end{Sinput}
\end{Schunk}

We start the analysis by loading the data. We remove bins with zero reads in all 
four samples and in the control, and generate a {\tt BayMethList} object. 
This object is initialized with four entries:
\begin{itemize}
  \item {\tt windows}: A {\tt GRanges} object representing the genomic bins of interest.
  \item {\tt control}:  A {\tt matrix} of read counts obtained by an affinity
        enrichment sequencing experiment for the fully methylated
        (SssI) treated sample. The number of rows must be equal
        to ‘length(windows)’. Each column contains the counts of
        one sample. The number of columns must be either one or
        equal to the number of columns of ‘sampleInterest’.
  \item {\tt sampleInterest}: A matrix of read counts obtained by an
	affinity enrichment sequencing experiment for the samples
	of interest. The number of rows must be equal to
	‘length(windows)’. Each column contains the counts of one
	sample.
  \item {\tt cpgDens}: A numeric vector containing the CpG density for
        ‘windows’. The length must be equal to ‘length(windows)’
\end{itemize}

\begin{Schunk}
\begin{Sinput}
 library(Repitools)
 # load the prepared data object
 load("data/bock_data_prepared.Rdata")
 metDat <- as.matrix(values(gb_hg19r))
 # remove bins where we have no read depth in none of the samples
 rs <- rowSums(metDat[, c("hues6", "hues8", "normal", "tumor", "SssI.low.SssI_low")])
 wr <-  which(rs == 0)
 gb_hg19_noZero <- gb_hg19r[-wr]
 metDat <- metDat[-wr,]
 map <- metDat[, "map_ucsc"]
 sssI <- matrix(metDat[,"SssI.low.SssI_low"], ncol=1)
 colnames(sssI) <- "sssI"
 bockBL <- BayMethList(
     window=window(gb_hg19_noZero),
     control=sssI,
     sampleInterest=cbind(hues6=metDat[,"hues6"], hues8=metDat[,"hues8"],
 		    normal=metDat[,"normal"], tumor=metDat[,"tumor"]),
     cpgDens=metDat[,"cpgdens"])
\end{Sinput}
\end{Schunk}

We only include autosomes in the analysis and concentrate on bins with 
with at least $75\%$ mappable bases.\\[0.2cm]

\begin{Schunk}
\begin{Sinput}
 # only consider autosomes
 as.idx <- !(seqnames(windows(bockBL)) %in% c("chrX", "chrY"))
 as.idx <- as.vector(as.idx)
 bockBL <- bockBL[as.idx]
 map <- map[as.idx]
 bockBL <- bockBL[map > 0.75] 
\end{Sinput}
\end{Schunk}

Next, we determine the normalizing constant for each sample.
The normalizing factor $f$ is essentially a scaling factor between highly methylated regions in 
the corresponding sample relative to the SssI control, see Figure~\ref{fig:offset}.\\[0.2cm]
\begin{Schunk}
\begin{Sinput}
 bockBL <- determineOffset(bockBL, q=0.998, controlPlot=list(show=TRUE, mfrow=c(2,2), nsamp=100000, 
    main=colnames(sampleInterest(bockBL)), ask=FALSE))
 fOffset(bockBL)
 #        hues6 hues8   normal    tumor
 #[1,] 2.289898  2.75 1.285714 1.272727
\end{Sinput}
\end{Schunk}

\begin{figure}
	\begin{center}
		\includegraphics[width=0.7\textwidth]{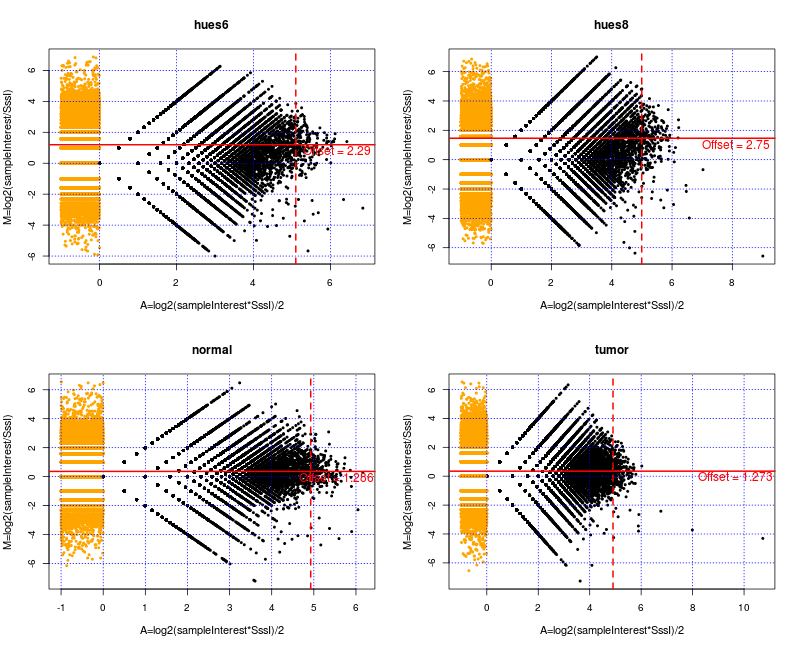}
		\caption{Log-fold change ($M$) versus log-concentration ($A$) illustrated for all four samples randomly sampling data of $100000$ bins in each case. 
		The red dotted line shows the $0.998$ quantile $q$ of $A$ determined from all bins. The red straight line shows the estimated normalization offset $f=2^{median(M_{A > q})}$.  
		A 'smear' of yellow points at a low $A$ value represents counts that are low in either of the two samples.}\label{fig:offset}
	\end{center}
\end{figure}

Using the empirical Bayes approach we have to be aware of bins with unusual high 
counts of reads. These might cause problems  in the optimization routine as they 
can cause {\tt NA} or {\tt Inf} values returned by the hypergeometric function. 
Some of these high read counts can be explained by unannotated high copy number 
regions, see~\cite{pickrell.etal2011}. We mask these bins out for the empirical 
Bayes procedure to avoid numerical problems. However, note that we will finally
obtain methylation estimates for almost all of these bins.\\[.3cm]
\begin{Schunk}
\begin{Sinput}
 ## mask suspicious regions
 #wget http://eqtl.uchicago.edu/Masking/seq.cov1.ONHG19.bed.gz
 library(rtracklayer)
 hcRegions <- import("data/seq.cov1.ONHG19.bed", asRangedData=FALSE)
 bockBL <- maskOut(bockBL, hcRegions)
\end{Sinput}
\end{Schunk}

Using this reduced dataset we derive the prior parameters based on empirical Bayes.
We use a uniform prior distribution for the methylation level
and consider $K=100$ separate CpG groups. The algorithm is 
run on four CPUs in parallel.\\[0.2cm]

\begin{Schunk}
\begin{Sinput}
 ## find prior parameters using empirical Bayes
 bockBL <- empBayes(bockBL, ngroups = 100, ncomp = 1, maxBins = 50000, 
   method="beta", ncpu=4, verbose=FALSE)
\end{Sinput}
\end{Schunk}

The prior parameters for all samples are saved in a list, which can be accessed 
using the function {\tt priorTab(.)}. The first list element contains a vector with
the assigned CpG density group for each bin. Of note, the length of this vector
is equal to the numbers of bins used in the analysis. The second list 
element saves the number of
mixture components used and the third contains a string indicating the type of
prior ("beta" or "DBD"). The following entries contain the prior parameters for
each sample. One list element corresponds thereby to one sample.
Figure~\ref{fig:empBayes} shows the mean of the obtained prior predictive distribution 
of the SssI sample depending on CpG density group for all four samples.\\[0.2cm]

\begin{Schunk}
\begin{Sinput}
 plot(priorTab(bockBL)[[4]][1,]/priorTab(bockBL)[[4]][2,], 
   type="l", xlab="CpG group", ylab="Mean (a/b)", xaxt="n")
 axis(1, at=seq(1,100,10), labels=levels(priorTab(bockBL)[[1]])[seq(1,100,10)])
 for(i in 2:4){
     lines(priorTab(bockBL)[[3+i]][1,]/priorTab(bockBL)[2,], type="l", col=i)
 }
 legend("topright", c("HUES6", "HUES8", "Normal", "Tumor"), lty=1, col=1:4)
\end{Sinput}
\end{Schunk}
\begin{figure}
\begin{center}
	\includegraphics[width=.6\textwidth]{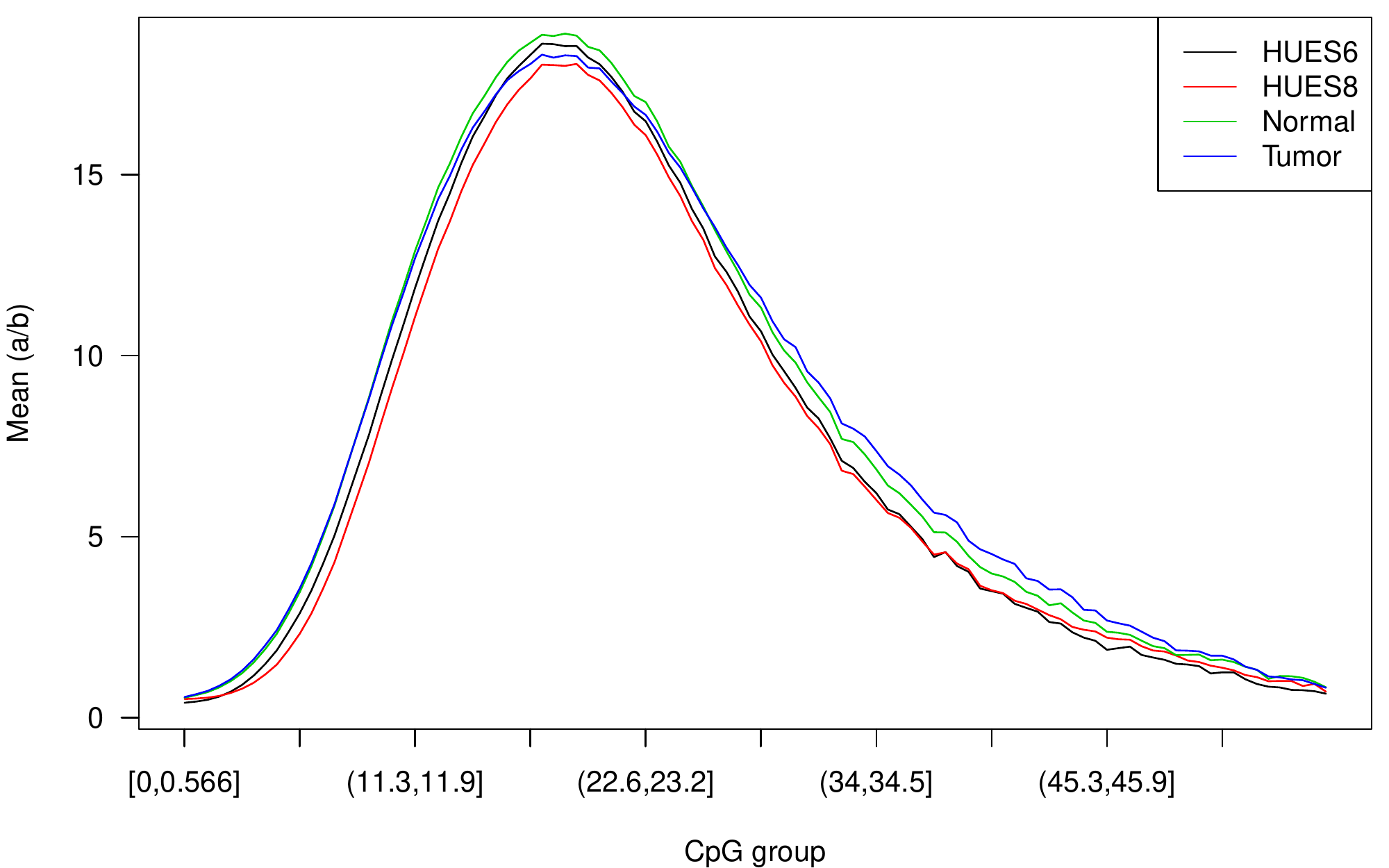}
\caption{Mean of the prior predictive distribution depending on CpG density group for all four samples.
}
\label{fig:empBayes}
\end{center}
\end{figure}
To get methylation estimates we call:\\

\begin{Schunk}
\begin{Sinput}
 bockBL <- methylEst(bockBL, verbose=TRUE, controlCI = list(compute = FALSE))
\end{Sinput}
\end{Schunk}

This function assigns a list to the slot {\tt methEst} in our {\tt BayMethList} object. 
Here, the mean, variance and potential credible intervals are saved for each sample.
The mean and variance can be accessed using {\tt methEst(bockBL)\$mean} and {\tt methEst(bockBL)\$var} .

Figure~\ref{fig:BayMeth_rrbs} shows regional methylation estimates of 
BayMeth compared to RRBS for all samples.\\[0.2cm]
\clearpage
\begin{Schunk}
\begin{Sinput}
 mE <- methEst(bockBL)$mean   
 mV <- methEst(bockBL)$var
 cP <- cpgDens(bockBL)
 sssI <- control(bockBL)
 sI <- sampleInterest(bockBL)
 ## get the truth for all samples
 rrBS <- as.matrix(values(windows(bockBL))[,5:12])
 rrBS <- as.matrix(rrBS)
 #
 # combine everything in one matrix to facilitate plotting
 all <- cbind(mE, rrBS, cP, sssI, sI, mV)
 colnames(all) <- c("bayMeth_hues6", "bayMeth_hues8", "bayMeth_normal", "bayMeth_tumor",
   colnames(rrBS), "cpgDens", "sssI", "hues6", "hues8", "normal", "tumor", 
   "bayMeth_varHues6", "bayMeth_varHues8", "bayMeth_varNormal", "bayMeth_varTumor")
 #
 sNames <- c("a) HUES6", "b) HUES8", "c) Colon normal", "d) Colon tumor")
 #
 alls <- all
 #
 col <- "dodgerblue4"
 Lab.palette <- colorRampPalette(c("blue", "orange", "red"), space = "Lab")
 par(mfrow=c(2,2), mar=c(3.5,4, 3, 4.5), mgp=c(2.5,1,0), cex.lab=.85, cex.main=1, cex.axis=.75, pty="s", las=1)
 zlim <- c(0,2.34)
 lim <- c(0,1)
 for(i in 1:4){
   all <- alls
   all <- all[!is.na(all[,5+2*(i-1)]),]
   all <- all[!is.na(all[,i]),]
   #
   ## define a limit for the truth
   limit_truth <- 20
   all <- all[all[,6+2*(i-1)] > limit_truth,]
   #
   ## separation by variance
   limit_var <- 0.0225 
   all <- all[all[,19+(i-1)] < limit_var,]
   #
   ## separation by SssI control
   limit_control <- 9
   all <- all[all[,"sssI"] > limit_control,]
   #
   ## smooth density representation
   mysmoothScatter(all[,5+2*(i-1)], all[,i], pch=".",  
     col=col, colramp=Lab.palette, xlab="RRBS", ylab="BayMeth", 
     main=sNames[i], xlim=lim, ylim=lim,
     cex=0.05, horizontal=F, zlim=zlim, 
     axis.args=list(at=zlim, labels=c("low", "high")))
   text(0.5, 0.05, sum(!is.na(all[,i])), col="white", cex=0.85)
   abline(c(0,0), c(1,1), col="green", lwd=1.3, lty=2)
 }
\end{Sinput}
\end{Schunk}

Here, {\tt mysmoothScatter} represents an adaptation of the function
{\tt smoothScatter} to get a color key next to the figures.
\begin{figure}
\begin{center}
	\includegraphics[width=.6\textwidth]{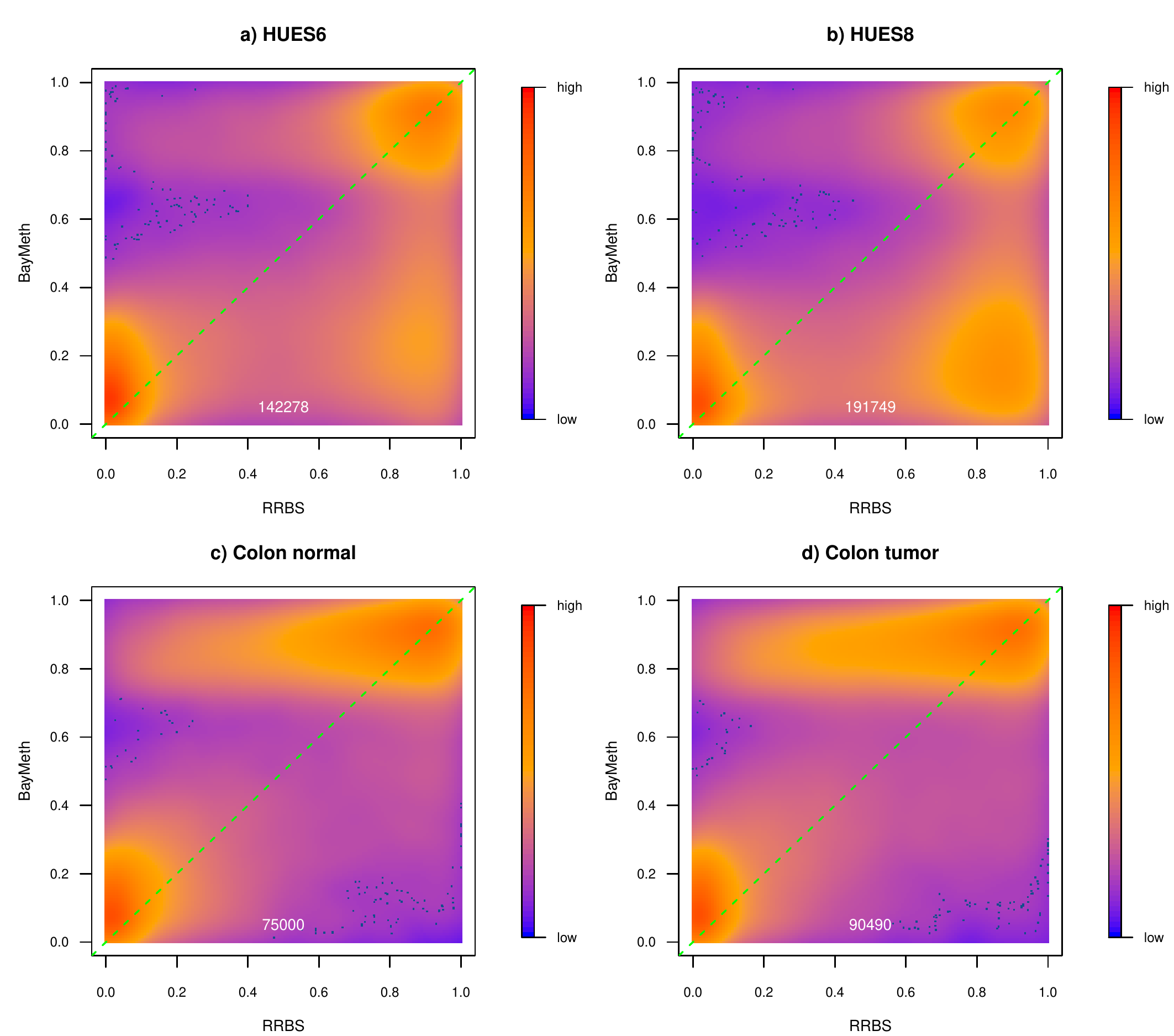}
\caption{Smooth color density representation of variance 
estimates obtained by BayMeth versus number of reads in the SssI control for a read depth 
larger than $20$ in RRBS. The red box contains the bins used in Figure 9 having at least
a depth of 10 in SssI and a standard deviation smaller than $0.15$, i.e. a variance smaller
than $0.025$.} \label{fig:BayMeth_rrbs}
\end{center}
\end{figure}

\end{bmcformat}
\end{document}